\documentclass[hyper,letterpaper,12pt]{article}
\pdfoutput=1
\usepackage{a4wide}
\usepackage{amsmath}
\usepackage[
      colorlinks=true,
      linkcolor=blue,
      urlcolor=blue,
      filecolor=black,
      citecolor=blue,
            ]{hyperref}
\usepackage{color}
\usepackage{graphicx}
\usepackage{amsfonts}
\usepackage{amssymb}
\usepackage{epstopdf}
\usepackage{hyperref}

\def\beq{\begin{equation}}
\def\eeq{\end{equation}}

\begin{document}

\title{\bf \Large Introduction to Holographic  Superconductor Models}

\author{\large
~Rong-Gen Cai$^1$\footnote{E-mail: cairg@itp.ac.cn}~,
~~Li Li$^2$\footnote{E-mail: lili@physics.uoc.gr}~,
~~Li-Fang Li$^3$\footnote{E-mail: lilf@itp.ac.cn}~,
~~Run-Qiu Yang$^1$\footnote{E-mail: aqiu@itp.ac.cn}\\
\\
\small $^1$State Key Laboratory of Theoretical Physics,\\
\small Institute of Theoretical Physics, Chinese Academy of Sciences,\\
\small Beijing 100190,  China.\\
\small $^2$ Crete Center for Theoretical Physics, Department of Physics, \\
\small University of Crete, 71003 Heraklion, Greece.\\
\small $^3$State Key Laboratory of Space Weather, \\
\small Center for Space Science and Applied Research, Chinese Academy of Sciences,\\
\small Beijing 100190, China.}
\date{\today}
\maketitle

\vspace{-12cm}\hspace{12cm}{CCTP-2015-06}
\vspace{10cm}

\vspace{-10cm}\hspace{11.7cm}{CCQCN-2015-65}
\vspace{10cm}

\begin{abstract}
\normalsize In the last years it has been shown that some properties of strongly coupled superconductors can be potentially described by classical general relativity living in one higher dimension, which is known as holographic superconductors. This paper gives a quick and introductory overview of some holographic  superconductor models  with s-wave, p-wave and d-wave orders in the literature from point of view of bottom-up, and summarizes some basic properties of these holographic models in various regimes. The competition and coexistence of these
superconductivity orders are also studied in these superconductor models.
\end{abstract}

\tableofcontents

\section{ Introduction}
The phenomenon of superconductivity was discovered in the early part of the last century that the electrical resistivity of a material  suddenly drops to zero below a critical temperature $T_c$. More importantly, the so-called Meissner effect tells us that the magnetic field is expelled in the superconducting phase,  which is distinguished from the perfect conductivity.
In the latter case a pre-existing magnetic field will be trapped inside the sample.
Conventional superconductors are well described by BCS theory~\cite{Bardeen:1957mv}, where the condensate is a Cooper pair of electrons bounded together by phonons. According to the symmetry of the spatial part of wave function of the Cooper pair, superconductors can be classified as the s-wave, p-wave, d-wave, f-wave superconductors, etc. However, some materials of significant theoretical and practical interest, such as high temperature cuprates and heavy fermion compounds, are beyond BCS theory. There are indications that the  involving physics is in strongly coupled regime, so one needs a departure from the quasi-particle paradigm of Fermi liquid theory~\cite{Carlson:2002}. Condensed matter theories have very few tools to do this.

On the other hand, although some of the deeper questions arising from the Anti-de Sitter/Conformal Field Theory(AdS/CFT) correspondence~\cite{Maldacena:1997re,Gubser:1998bc,Witten:1998qj} remain to be understood from first principles, this duality creating an interface between gravitational theory and dynamics of quantum field theory provides an invaluable source of physical intuition as well as computational power. In particular, in a ``large $N$ and large $\lambda$" limit, the gravity side can be well described by classical general relativity, while the dual field theory involves the dynamics with strong interaction.\footnote{Loosely speaking,  $N^2$ can be considered as the degrees of freedom in the dual field theory, and $\lambda$ as the characteristic strength of interactions. An elementary introduction to this correspondence can be found in the next section.} It is often referred to as ``holography" since a higher dimensional gravity system is described by a lower dimensional field theory without gravity, which is very reminiscent of an optical hologram.
There are indeed many physical motivations that lead to this amazing holographic duality in the literature, such as renomalization group flow and black hole membrane paradigm. A very new perspective was proposed, which is called the exact holographic mapping~\cite{Qi:2013caa}. By constructing a unitary mapping from the Hilbert space of a lattice system in flat space (boundary) to that of another lattice system in one higher dimension (bulk), it  provides a more explicit and complete understanding of the bulk theory for a given boundary theory and can be compared with AdS/CFT correspondence.

It has been shown that the AdS/CFT correspondence can indeed provide solvable models of strong coupling superconductivity, see refs.~\cite{Horowitz:2010gk,Herzog:2009xv,Iqbal:2011ae,Musso:2014efa} for reviews. The physical picture is that some gravity background would become unstable as one tunes some parameter, such as temperature for black hole and chemical potential for AdS soliton, to developing some kind of hair. The emergency of the hair in the bulk corresponds to the condensation of a composite charged operator in the dual field theory. More precisely, the dual operator acquires a non-vanishing vacuum expectation value breaking the U(1) symmetry spontaneously. It has been uncovered that this simple holographic setup shows similar properties with real superconductors.

The holographic s-wave superconductor model known as Abelian-Higgs model was first realized in refs.~\cite{Hartnoll:2008vx,Hartnoll:2008kx}. According to the AdS/CFT correspondence, in the gravity side, a Maxwell field and a charged scalar field are introduced to describe the U(1) symmetry and the scalar operator in the dual field theory, respectively. This holographic model undergoes a phase transition from black hole with no hair (normal phase/conductor phase) to the case with scalar hair at low temperatures (superconducting phase). The holographic model for the insulator/superconductor
phase transition has been realized in ref.~\cite{Taka} at zero temperature by taking the AdS soliton as the gravitational background in the Abelian-Higgs model.  Holographic d-wave model was constructed by introducing a charged massive spin two field propagating in the bulk~\cite{Chen:2010mk,Benini:2010pr}. The superconductivity in the high temperature cuprates is well known to be of d-wave type.

In recent years, evidence from several materials suggests that we now have examples of p-wave superconductivity, providing us new insights into the understanding of unconventional superconductivity in strongly correlated electron systems~\cite{Mackenzie}.
To realize a holographic p-wave model, one needs to introduce a charged vector field in the bulk as a vector order parameter. Ref.~\cite{Gubser:2008wv} presented a holographic p-wave model by introducing an SU(2) Yang-Mills field into the bulk, where a gauge boson generated by one SU(2) generator is dual to the vector order parameter. The authors of refs.~\cite{Cai:2013pda,Cai:2013aca} constructed a holographic p-wave model by adopting a complex vector field charged under a U(1) gauge field, which is dual to a strongly coupled system involving a charged vector operator with a global U(1) symmetry. An alternative holographic realization of p-wave superconductivity emerges from the condensation of a two-form field in the bulk~\cite{Aprile:2010ge,Donos:2011ff,Donos:2012gg}.

The philosophy for holographic setups is that even though the underlying microscopic description of the theory with a gravity dual is quite likely to be different form that arising in materials of experimental interest, it may uncover some universal aspects of the strongly coupled dynamics and kinematics, thus would help the development of new theories of superconductivity. By mapping the quantum physics of strongly correlated many body systems to the classical dynamics of black hole physics in one higher dimension, the holographic approach provides explicit examples of theories without a quasi-particle picture in which computations are nevertheless feasible.

The models studied in the literature can be roughly divided into two classes, i.e., the bottom-up and top-down models. In the former approach the holographic model is constructed phenomenologically by picking relevant bulk fields corresponding to the most important operators on the dual field theory and then writing down a natural bulk action considering general symmetries and other features of dual system. Thus the holographic description is necessarily effective and can be used to describe a wide class of dual theories instead of a definite single theory.  In the top-down approach, the construction of a model is uniquely determined by  a consistent truncation from string theory or supergravity. One can usually have a much better control over the dual field theory, nonetheless, the resulted models are  much more complicated.

This paper aims at providing a quick and introductory overview of those three kinds of holographic  superconductor models from the point of view of bottom-up.\footnote{Holographic superconductor models constructed in the top-down approach can be found, for example, in refs.~\cite{Denef:2009tp,Gubser:2009qm,Gauntlett:2009dn,Gubser:2009gp,Gauntlett:2009bh,KalyanaRama:2011ny,Bobev:2011rv}.} The organization of the paper is as follows. In the next section, we review basic elements of Ginzburg-Landau theory of superconductivity and holographic duality, as a warm up. A brief introduction to the Abelian-Higgs model is presented in section~\ref{sect:higsmodel}. In section~\ref{sect:p-wave}, we first introduce the SU(2) Yang-Mill model, focusing on its condensate and conductivity,   then study the Maxwell-vector model, paying more attention to the vector condensate induced by magnetic fields and its complete phase diagram in terms of temperature and chemical potential, and finally discuss the third p-wave model by introducing a two-form  in the bulk. This model can exhibit a novel helical superconducting phase. Section~\ref{sect:dwave} is devoted to holographic d-wave models. In the next two sections, we pay attention to the competition and coexistence among different orders, including different superconducting orders in section~\ref{sect:competition} as well as superconducting order and magnetic order in section~\ref{sect:M&S}. The conclusion and some discussions are included in section~\ref{sect:conclusion}.

%
\section{Preliminary}
\label{sect:prelimi}
\subsection{Ginzburg-Landau theory}
The microscopic origin of traditional superconductivity is well understood by BCS theory, which explained the superconducting current as a superfluid of pairs of electrons interacting through the exchange of phonons. However, at phenomenological level, the Ginzburg-Landau theory of superconductivity~\cite{Ginzburg:1950} had great success in explaining the macroscopic properties of superconductors.~\footnote{To have a wide description of the Ginzburg-Landau theory see ref.~\cite{cyrot:1973} and references therein.}

In this phenomenological theory, the free energy of a superconductor can be expressed in terms of a complex order parameter field, $\Psi$, which is directly related to the density of the superconducting component.  By assuming smallness of $|\Psi|$ and smallness of its gradients, the free energy near the superconducting critical temperature $T_c$ has the following form,

\begin{equation}
\label{GL}
F=F_n+\alpha|\Psi|^2+\frac{\beta}{2}|\Psi|^4+\frac{1}{2m^*}|(-i\nabla-e^*\vec{A})\Psi|^2+\frac{|\vec{B}|^2}{2\mu_0},
\end{equation}
where $F_n$ is the free energy in the normal phase, $\vec{A}$ is the vector potential and $B=\nabla\times\vec{A}$ is the magnetic field. $m^*$ and $e^*$ are effective mass and charge of condensate. If one considers the BCS theory, $m^*=2m$ and $e^*=2e$ with $m$ and $e$ the mass and charge of electrons forming Copper pairs. $\alpha$ and $\beta$ are two phenomenological parameters, which behave as $\alpha=\alpha_0(T_c-T)$ with $\alpha_0$ and $\beta$ two positive constants. Note that we work in the unites $\hbar=c=1$. It is obvious that the free energy is invariant under the following transformation,
\begin{equation}\label{uone}
\Psi\rightarrow\Psi e^{i\theta(x)},\quad \vec{A}\rightarrow\vec{A}+\frac{1}{e^*}\nabla\theta(x),
\end{equation}
which is known as the U(1) gauge symmetry.

Minimising the free energy with respect to the order parameter and the vector potential, one obtains the Ginzburg-Landau equations,
\begin{eqnarray}
&&\alpha\Psi+\beta|\Psi|^2\Psi+\frac{1}{2m^*}(-i\nabla-e^*\vec{A})^2\Psi= 0, \\
&& \vec{J_s}=-\frac{ie^*}{2m^*}(\Psi^*\nabla\Psi-\Psi\nabla\Psi^*)-\frac{e^{*2}}{m^*}\Psi^*\Psi\vec{A}.\label{london}
\end{eqnarray}
The first equation determines the order parameter $\Psi$, and the second one provides the superconducting current $\vec{J_s}$ which is the dissipation-less electrical current.  The Ginzburg-Landau theory actually can be derived from the BCS microscopic theory. Thus, the electrons that contribute to superconductivity would form a superfluid and $|\Psi|^2$ indicates the fraction of electrons condensed into a superfluid.

This phenomenological theory can give many useful information even in homogeneous case. Let us consider a homogeneous superconductor with no superconducting current, so the equation for $|\Psi|$ simplifies to
\begin{equation}
\alpha\Psi+\beta|\Psi|^2\Psi=0.
\end{equation}
Above the superconducting transition temperature, $T>T_c$, one only gets a trivial solution $\Psi=0$, which corresponds to the normal state of the superconductor. Below the critical temperature, $T<T_c$, apart from the trivial solution, there are a series of non-trivial solutions which read
\begin{equation}\label{break}
|\Psi_0|=\sqrt{-\frac{\alpha}{\beta}}=\sqrt{\frac{\alpha_0(T_c-T)}{\beta}}.
\end{equation}
Furthermore, compared with $\Psi=0$, those solutions have lower potential energy, thus are dominant. Note that there are infinite solutions giving the ground state of superconducting phase. However, the true ground state can only choose one solution from them. Therefore, the ground state will change under the U(1) transformation~\eqref{uone}. In such case, we call that the U(1) symmetry is spontaneously broken.  From~\eqref{break} one sees that $\Psi$ approaches zero as $T$ gets closer to $T_c$ from below, which is a typical behaviour of a second order phase transition. One can find from~\eqref{london} that in the homogeneous case one can neglect the contribution from the first term and thus the superconducting current is proportional to the vector potential, i.e., $\vec{J_s}\propto\vec{A}$. If one takes a time derivative on both sides, one will obtain $\vec{E}=-\partial_t\vec{A}\propto\partial_t\vec{J_s}$. This means that the electric fields accelerate superconducting electrons resulting in the infinite DC conductivity. If one takes the curl and combines with Maxwell's equations, one will find $\nabla^2\vec{B}\propto\vec{B}$ indicating the decay of magnetic fields inside a superconductor, i.e., the Meissner effect.

The Ginzburg-Landau equations predict two characteristic lengths in a superconductor. The first one is  the coherence length $\xi$ which is given by
\begin{equation}\label{glcoherence}
\xi=\sqrt{\frac{1}{2m^*|\alpha|}}.
\end{equation}
It is the characteristic exponent of the variations of the density of superconducting component. In the BCS theory $\xi$ denotes the characteristic Cooper pair size. The other one is the penetration length $\lambda$ which reads
\begin{equation}\label{glpentration}
\lambda=\sqrt{\frac{m^*}{\mu_0 e^{*2}|\Psi_0|^2}}=\sqrt{\frac{m^* \beta}{\mu_0 e^{*2}|\alpha|}},
\end{equation}
where $\Psi_0$ is the equilibrium value of the order parameter in the absence of electromagnetic fields. This length characterises the speed of exponential decay of the magnetic field at the surface of a superconductor.

Note that from definitions~\eqref{glcoherence} and~\eqref{glpentration} the temperature dependences near $T_c$ behave as
\begin{eqnarray}
\xi & \propto& (T_c-T) ^{-1/2},\\
\lambda &\propto& (T_c-T) ^{-1/2}.
\end{eqnarray}
Both diverge as $T\rightarrow T_c$ from below with the critical exponent $1/2$. Nevertheless, the ratio $\kappa=\lambda/\xi$ known as the Ginzburg-Landau parameter is temperature independent. Type-I superconductors correspond to cases with $0<\kappa<1/\sqrt{2}$, and type-II superconductors correspond to cases with $\kappa>1/\sqrt{2}$. One of the most important findings from the Ginzburg-Landau theory was that in a type-II superconductor, strong enough magnetic fields can penetrate the superconductor by forming the hexagonal lattice of quantised tubes of flux, called the Abrikosov vortex lattice.

Finally one point we would like to emphasize is that in the Ginzburg-Landau theory the U(1) symmetry is broken spontaneously in the superconducting phase transition. Actually, only the spontaneous symmetry breaking feature itself can lead to many fundamental phenomenological properties of superconductivity, without any precise detail of the breaking mechanism specified~\cite{Weinberg:1986}. In this review, we will show how a similar effective approach constitutes the basis of superconductivity in terms of holographic description.

\subsection{Holographic duality}
The original conjecture proposed by Maldacena~\cite{Maldacena:1997re} was that type-IIB string theory on the product spacetime $AdS_5 \times S^5$ should be equivalent to $\mathcal{N}= 4$ SU(N) supersymmetric Yang-Mills theory on the 3+1 dimensional boundary. This super-Yang-Mills theory is a conformal field theory, so this duality is named AdS/CFT correspondence. Later, this conjecture has been generalized to more general gravitational backgrounds and cases without supersymmetry and conformal symmetry.\footnote{The simple examples are Lifshitz symmetry~\cite{Kachru:2008yh} and Schr\"odinger symmetry~\cite{Son:2008ye,Balasubramanian:2008dm}, while more generic cases are those with generalized Lifshitz invariance and hyperscaling violation~\cite{Charmousis:2010zz,Gouteraux:2011ce,Huijse:2011ef,Dong:2012se}, and the associated Schr\"odinger cousins~\cite{Kim:2012nb}. However, those take us outside the best understood AdS/CFT framework. We shall focus on the most well defined case involving the bulk geometry with the asymptotically AdS behaviour.} From a modern perspective, the correspondence is an equality between a quantum field theory (QFT) in d dimensional spacetime and a (quantum)  gravity theory in d+1 spacetime dimensions. This correspondence is also sometimes called gauge/gravity duality, gauge/string duality or holographic correspondence (or duality).

A remarkable usefulness of the correspondence comes from the fact that it is a strong-weak duality: when the quantum field theory is strongly coupled, the dual gravitational theory is in a weakly interacting regime and thus more mathematically tractable, and vice versa. So the holographic duality provides us a powerful toolkit for studying strongly interacting systems. The backbone of the correspondence was elaborated by the authors of refs.~\cite{Gubser:1998bc,Witten:1998qj}. For every gauge invariant operator $\mathcal{O}$ in the QFT, there is a corresponding dynamical field $\Phi$ in the bulk gravitational theory. The partition function in gravity side is equal to the generating functional of the dual boundary field theory. More specifically,  adding a source $J$ for $\mathcal{O}$ in the QFT is equivalent to impose a boundary condition for the dual field $\Phi$ at the boundary of the gravity manifold (say at $z\rightarrow 0$), i.e., the field $\Phi$ tends towards the value $\Phi\rightarrow\phi_0= J$ at the boundary up to an overall power of $z$. The formula reads
\begin{equation}
\label{gkpw}
Z_{{\rm bulk}}[\Phi\rightarrow \phi_0=J]=\left<{\rm exp}\left(i\int\sqrt{-g_0}\;d^dx\; J\mathcal{O}\right)\right>_{\rm QFT},
\end{equation}
where $g_0$ is the determinant of the background metric of dual field theory.
If we want to study a strongly coupled field theory, we can translate it into a weakly coupled gravity system. In the semiclassical limit, the partition function is equal to the on-shell action of the bulk theory and thus one only needs to solve particular differential equations of motion. Therefore we can compute expectation values and correlation functions of the operator $\mathcal{O}$ in the (strongly coupled) QFT by differentiating the left side with respect to $J=\phi_0$.

In order to get familiar with the calculation by holography, let us consider $\mathcal{O}$ as a scalar operator which is dual to the bulk scalar field also denoted as $\Phi$. The minimal bulk action for $\Phi$ is given by
\begin{equation}\label{acscalar}
S_0=\int d^{d+1}x\sqrt{-g}\left[-\frac{1}{2}(\partial\Phi)^2-\frac{1}{2}m^2\Phi^2\right].
\end{equation}
For illustration the gravity background is fixed as the pure $AdS_{d+1}$ in Poinc\'are coordinates
\begin{equation}
ds^2=\frac{L^2}{z^2}(dz^2-dt^2+d\vec{x}\cdot d\vec{x}),
\end{equation}
together with a profile for the scalar field $\Phi=\Phi(z,t,\vec{x})$.  Here $\vec{x}$ are $d-1$ spatial coordinates, $t$ is a timelike coordinate and $z$ is  the radial spatial coordinate. $L$ is known as AdS radius and the conformal boundary of AdS is located at $z\rightarrow 0$. Note that the geometry is invariant under the scaling transformation $(z, t, \vec{x})\rightarrow(\lambda z,\lambda t, \lambda \vec{x})$. Actually, the full isometry group of $AdS_{d+1}$ is identical to the conformal group in d dimensional boundary spacetime.

To calculate the on-shell action for $\Phi$,  we need to solve the equation of motion derived from action~\eqref{acscalar}. Working in Fourier space $\Phi(z, t, \vec{x})\rightarrow\Phi(z, \omega, \vec{k})=\Phi(z) e^{-i\omega t+i\vec{k}\cdot\vec{x}}$, we obtain
\begin{equation}
\partial^2_z\Phi(z)-\frac{d-1}{z}\partial_z\Phi(z)-(k^2+\frac{m^2L^2}{z^2})\Phi(z)=0,\quad k^2=-\omega^2+\vec{k}^2.
\end{equation}
Near the boundary $z\rightarrow 0$, the above equation admits the general asymptotic solution,
\begin{equation}
\Phi(z, \omega, \vec{k})\sim A(k)\; z^{d-\Delta}+B(k)\; z^\Delta, \quad z\rightarrow 0,
\end{equation}
with $\Delta=d/2+\sqrt{m^2L^2+d^2/4}$.\footnote{Note that $AdS_{d+1}$ spacetime is stable even when the mass squared $m^2$ of scalar field is negative provided $m^2L^2\geq m^2_{BF}L^2=-d^2/4$~\cite{Breitenlohner:1982jf}. The lower bound $m^2_{BF}=-d^2/4L^2$ is often called Breitenloner-Freedman (BF) bound.} The relation between A and B is determined by the interior of AdS. Making Fourier transformation back into real space, we then obtain
\begin{equation}\label{asypsa}
\Phi(z, t, \vec{x})\sim A(t,\vec{x})\; z^{d-\Delta}+B(t,\vec{x})\; z^\Delta, \quad z\rightarrow 0,
\end{equation}

We now try to identify which term can be considered as the source $J$ of the dual operator $\mathcal{O}$. It turns out that, as long as $m^2L^2>-d^2/4+1$, the mode $A$ is non-normalizable with respect to the inner product
\begin{equation}\label{innerp}
(\Phi_1,\Phi_2)=-i\int_{\Sigma_t}dz d\vec{x}\sqrt{-g}g^{tt}(\Phi^*_1\partial_t\Phi_2-\Phi_2\partial_t\Phi^*_1),
\end{equation}
with $\Sigma_t$ a constant-$t$ slice. The $B$ mode in this case is normalizable. We identify the coefficient $A$ as the source term, i.e.,
\begin{equation}
J(t, \vec{x})=\phi_0(t, \vec{x})=A(t, \vec{x})=\lim_{z\rightarrow 0} z^{\Delta-d}\;\Phi (z, t, \vec{x}).
\end{equation}
This means that the on-shell action, and thus the partition function in AdS is a functional of $J(t, \vec{x})$. We can now argue that the scaling dimension of $\mathcal{O}$ is $\Delta$ without further calculation. Consider the scale transformation $(z, t, \vec{x})\rightarrow\lambda(z, t, \vec{x})$, the scalar $\Phi$ under such operation transforms as $\tilde{\Phi}(\lambda z,\lambda t,\lambda\vec{x})=\Phi(z,t,\vec{x})$. So the source term $J=A$ must transform as $\tilde{J}(\lambda t,\lambda\vec{x})=\lambda^{\Delta-d}J(t,\vec{x})$, and thus according to~\eqref{gkpw} $\mathcal{O}$ transforms as $\mathcal{\tilde{O}}=\lambda^{-\Delta} \mathcal{O}$ which suggests the dimension of $\mathcal{O}$ should be $\Delta$.

Calculating the on-shell bulk action in terms of the solution with asymptotic expansion~\eqref{asypsa}, one will find that the on-shell action will diverge near the boundary $z\rightarrow 0$. This divergence is interpreted as
dual to UV divergences of the boundary field theory. Actually, the infrared (IR) physics of the bulk near the boundary corresponds to the ultraviolet (UV) physics of dual QFT, and vice versa. This is called UV/IR relation~\cite{Peet:1998wn} and the radial direction $z$ plays the role of energy scale in the dual boundary theory. Physical processes in the bulk occurring at different radial positions correspond to different field theory processed with energies which scale as $E\sim 1/z$.

The divergence can be cured by adding local counter terms at the boundary, known as holographic renormalization~\cite{Balasubramanian:1999re,Bianchi:2001de,Bianchi:2001kw}. For the present case the counter term to be introduced is
\begin{equation}
S_{ct}=\frac{\Delta-d}{2L}\int_{z\rightarrow 0} d^dx\sqrt{-\gamma}\; \Phi^2,
\end{equation}
where $\gamma$ is the determinant of the induced metric at the boundary. So the renormalized on-shell action should be $S^{ren}=S_0+S_{ct}$. We can then compute the expectation value by using the basic formula~\eqref{gkpw}. That relation implies
\begin{equation}
\left<\mathcal{O}\right>=-i\frac{\delta Z_{bulk}[J]}{\delta J}\sim\frac{\delta S^{ren}[J]}{\delta J},
\end{equation}
where we have taken the semiclassical limit $Z_{bulk}=e^{i S^{ren}}$. Straightforward calculation shows that
\begin{equation}
\left<\mathcal{O}\right>(t,\vec{x})=\frac{2\Delta-d}{L} B(t,\vec{x}).
\end{equation}
This is often summarised as saying that the ``non-normalizable"  mode $A$ gives the source in the dual field theory, whereas the ``normalizable" mode $B$ encodes the response.

In the real world, many important experimental processes such as transport and spectroscopy involve small time dependent perturbations about equilibrium. Those phenomena can be described by linear response theory, in which the basic quantity is the retarded Green's function.
The retarded Green's function is defined to linearly relate sources and corresponding expectation values. In frequency space, it can be written as
\begin{equation}
\delta \left<\mathcal{O}\right>(\omega,\vec{k})=G^{R}(\omega,\vec{k})\; \delta J(\omega,\vec{k}).
\end{equation}
Using above formula one can continue to compute  $G^R(\omega,\vec{k})$ which is given by
\begin{equation}
G^R(\omega,\vec{k})=\frac{2\Delta-d}{L} \frac{B(\omega,\vec{k})}{J(\omega,\vec{k})}.
\end{equation}
However, there is something subtle we shall discuss here. Different from the case in Euclidean signature where the bulk solution can be uniquely determined by additional requirement of regularity in the IR, while in the real-time Lorentzian signature, we must choose an appropriate boundary condition in far IR region of the geometry. This ambiguity reflects multitude of real-time Green's functions  (Feynman, retarded, advanced) in the QFT. Since the retarded Green's function describes causal response of the system to a perturbation, we involve an in-going condition describing stuff falling into the IR, i.e., moving towards larger $z$ as time passes. The advanced Green's function corresponds to the choice of out-going condition  enforced in the IR region. \footnote{An intrinsically real-time
holographic prescription was first proposed by the authors of ref.~\cite{Son:2002sd} by essentially analytically continuing the Euclidean prescription. It has been justified by a holographic version of the Schwinger-Keldysh formalism~\cite{Herzog:2002pc,Skenderis:2008dh}.}

Let us briefly consider the case with $-d^2/4< m^2L^2<-d^2/4+1$ where the second restriction comes from the unitary bound. One can easily check that both terms in~\eqref{asypsa} are normalizable with the inner product~\eqref{innerp}. So either one can be considered as a source, and the other one as a response. These two ways to quantize a scalar field in the bulk by imposing Dirichlet or Neumann like boundary conditions correspond to two different dual field theories~\cite{Klebanov:1999tb}, respectively. In the standard quantization, the corresponding operator has dimension $\Delta$, while in the alternative quantization, the corresponding operator has dimension $d -\Delta$.~\footnote{In fact, it has been shown that even more general quantisations are possible, like double trace deformation, see, for example, refs.~\cite{Witten:2001ua,Berkooz:2002ug}.}

The above  discussion only uses the near boundary expansion~\eqref{asypsa} and thus applies to generic asymptotically AdS geometries. It can also be applied to other fields such as components of the metric and Maxwell fields. To sum up, we first obtain a solution which satisfies appropriate boundary conditions, especially the condition in the deep IR. Then we compute the properly renormalized  on-shell action, identify the source and response from the asymptotic behaviour of the solution near the boundary, and compute the Green's function through linear response. In the next section we will use this procedure to compute the optical conductivity.

Another essential entry in the holographic dictionary is that the thermodynamic data of the QFT is entirely encoded in the thermodynamics of the black hole in the dual geometry.  QFT states with finite temperature are dual to black hole geometries, where the Hawking temperature of the black hole is identified with the temperature in the QFT. Turning on a chemical potential in this QFT corresponds to gravity with a conserved charge. The thermal entropy of QFT is identified as the area of black hole horizon and the free energy is related to  the Euclidian on-shell bulk action.  As space is limited, we only introduce essential issues which will be needed to discuss holographic superconductors.

Before the end of this subsection, let us point out that the holographic duality can be used to understand some hard nuts in quantum gravity from dual field theory side. A typical example is the black hole information paradox. It was first suggested by Hawking~\cite{Hawking:1976} that black holes destroy information which seemed to conflict with the unitarity postulate of quantum mechanics. The black hole information paradox can be resolved, at least to some extent, by holography, because it shows how a black hole can evolve in a manner consistent with quantum mechanics in some contexts, i.e.,  evolves in a unitary fashion~\cite{Hawking:2005kf,Hawking:2014tga}. There are some excellent papers talking about aspects of holographic duality, see, for example, refs.~\cite{Aharony:1999ti,Hartnoll:2009sz,McGreevy:2009xe,CasalderreySolana:2011us,Adams:2012th,Sachdev:2011wg} for more details.

\section{Holographic S-wave  Models}
\label{sect:higsmodel}

\subsection{The Abelian-Higgs model}

In this subsection, we begin with the Abelian-Higgs model~\cite{Hartnoll:2008vx} by introducing a complex scalar field $\Psi$, with mass $m$ and charge $q$, into the $(3+1)$ dimensional Einstein-Maxwell theory with a negative cosmological constant. The complete action can be written down as~\footnote{The model is a s-wave one since the condensed field is a scalar field dual to a scalar operator in the field theory side. This model can be straightforwardly generalized to other spacetime dimensions. Holographic s-wave superconductors with generalised couplings have also been considered in a number of works~\cite{Franco:2009yz,Aprile:2009ai,Aprile:2010yb,Liu:2010ka,Chen:2010hi,Peng:2011gh,Bigazzi:2011ak,Dey:2014xxa,Arean:2015wea}.}
\begin{equation}
\label{higsaction}
S =\frac{1}{2\kappa^2} \int d^4 x
\sqrt{-g} \left ( \mathcal{R}+\frac{6}{L^2}-\frac{1}{4}F_{\mu\nu}F^{\mu\nu}-|\nabla\Psi-i qA\Psi|^2-m^2 |\Psi|^2 \right),
\end{equation}
where $ 2 \kappa^2=16\pi G $ with $G$ being the Newtonian gravitational constant,  $\mathcal{R}$ is the scalar curvature of spacetime, $L$ is the AdS radius and Maxwell field strength $F_{\mu\nu}=\nabla_\mu A_\nu-\nabla_\nu A_\mu$. If one rescales $A_\mu\rightarrow A_\mu/q$ and $\Psi\rightarrow \Psi/q$, then the matter part has an overall factor $1/q^2$ in front of its Lagrangian, thus the back reaction of the matter fields on the metric becomes negligible when $q$ is large. The limit $q\rightarrow\infty$ with $q A_\mu$ and $q\Psi$ fixed is called the probe limit.  Here we will review the results obtained in this probe approximation, which can simplify the problem while retains most of the interesting physics. The study including the back reaction of matter fields can be found in ref.~\cite{Hartnoll:2008kx}.

The background metric is the AdS-Schwarzschild black hole with planar horizon
\begin{equation}
\label{AdSswtz}
 ds^2=-f(r)dt^2+\frac{dr^2}{f(r)}+r^2(dx^2+dy^2),~~~f(r)=r^2(1-r_h^3/r^3),
\end{equation}
where we have set the AdS radius $L$ to be unity. The conformal boundary is located at $r\rightarrow\infty$. The Hawking temperature of the black hole is determined by the horizon radius $r_h$: $T=3r_h/4\pi$.  The solution describes a thermal state of dual field theory in (2+1)-dimensions with temperature $T$. In addition, it is clear that the AdS-Schwarzschild black hole is an exact solution of the
action (\ref{higsaction}) when the matter sector is negligible.  It will be seen shortly  that when the temperature is lowered enough, the black hole solution will become unstable and a new stable black hole
solution appears with nontrivial scalar field.

To see the formation of scalar hair, we are interested in static, translationally invariant solutions, thus we consider the ansatz~\cite{Hartnoll:2008vx}
\begin{eqnarray}
\Psi=\psi(r)\;, \ \ \ A=\phi(r)\,dt\;.
\end{eqnarray}
The $r$ component of Maxwell equations implies that the phase of $\psi$ must be constant. Therefore, for convenience, one can take $\psi$ to be real.
This leads to the equations of motion~\footnote{In the probe limit, the concrete value of the charge $q$ does not play an essential role. Without loss of generality, we take $q$ to be one.}
\begin{equation}\label{higsEOMs}
\begin{split}
\psi''+(\frac{f'}{f}+\frac{2}{r})\psi'+\frac{\phi^2}{f^2}\psi-\frac{m^2}{f}\psi=0,\\
\phi''+\frac{2}{r}\phi'-\frac{2\psi^2}{f}\phi=0.
\end{split}
\end{equation}
As pointed out in ref.~\cite{Gubser:2008px}, the coupling of the scalar to the Maxwell field produces a negative effective mass for $\psi$ (see the third term in the first equation). Since this term becomes more important at low temperatures, we expect an instability towards forming nontrivial scalar hair.

The asymptotic behaviours of scalar field and gauge field near the AdS boundary are
\begin{eqnarray}
\psi=\frac{\psi_-}{r^{\Delta_-}}+\frac{\psi_+}{r^{\Delta_+}}+\cdots,\ \ \ \phi=\mu-\frac{\rho}{r}+\cdots,
\end{eqnarray}
where $\Delta_{\pm}=(3\pm\sqrt{3^2+4m^2})/2$, $\mu$ is the chemical potential and $\rho$ is the charge density in the dual field theory.
According to the AdS/CFT dictionary, the leading coefficient $\psi_-$ is regarded as the source of the dual scalar operator $\mathcal{O}$ with scaling dimension $\Delta_+$. Since we want the U(1)  symmetry to be broken spontaneously, we should turn off the source, i.e., $\psi_-=0$. Therefore the subleading term $\psi_+$ provides the vacuum expectation value $\langle\mathcal{O}\rangle$ in the absence of any source.~\footnote{We only consider the standard quantization here, regarding the leading coefficient as the source of dual operator. An alternative way inducing spontaneous symmetry breaking in holographic superconductors is to introduce double trace deformation~\cite{Faulkner:2010gj}.}

\begin{figure}[h]
\centering
\includegraphics[scale=0.95]{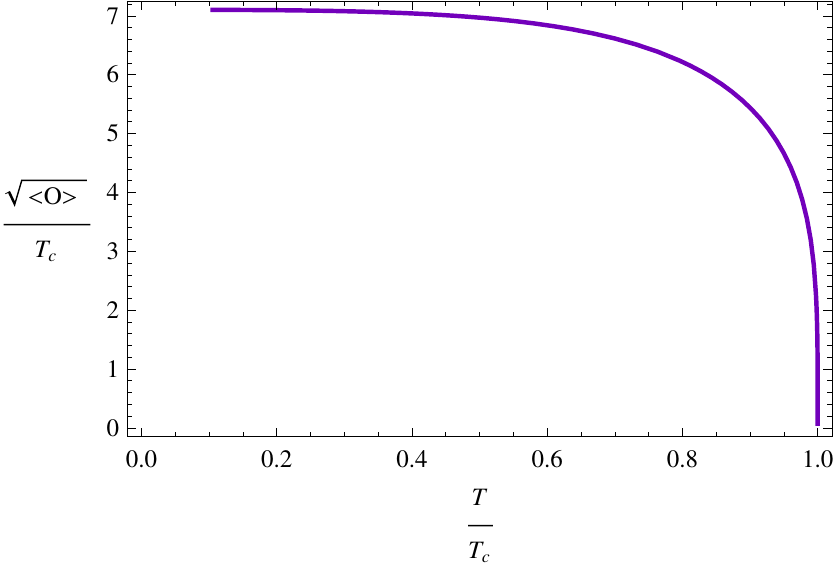}
\caption{\label{sconden} The condensate as a function of temperature. The critical temperature $T_c$ is proportional to $\sqrt{\rho}$. We choose $m^2=-2$.}
\end{figure}
Figure~\ref{sconden} shows how the condensate $\langle\mathcal{O}\rangle$ behaves as a function of temperature in a canonical ensemble with $\rho$ fixed to be one.  As one can see that there is a critical temperature $T_c$ below which the condensate appears, then rises quickly as the system is cooled and finally goes to a constant for sufficiently low temperatures. This behaviour is qualitatively similar to that obtained in BCS theory and observed in many materials. Near the critical temperature $T_c$, $\langle\mathcal{O}\rangle\sim (T_c-T)^{1/2}$, which is the typical result predicated by Ginzburg-Landau theory, see equation~\eqref{break}. By comparing the free energy of these hairy configurations to the solution $\psi=0, \phi=\rho(1/r_h-1/r)$ with no scalar hair, one finds that the hairy phase is thermodynamically favoured and the difference of free energies behaves like $(T_c-T)^2$ near the critical point, indicating a second order phase transition.

We now compute the optical conductivity, i.e., the conductivity as a function of frequency $\omega$, which is related to the retarded current-current two-point function for the U(1) symmetry, $\sigma(\omega)=\frac{1}{i\omega}G^{R}(\omega,\vec{k}=0)$. According to the holographic duality, this can be obtained by calculating electromagnetic fluctuations in the bulk. By symmetry, it is sufficient  to turn on the perturbation $\delta A=A_x(r)e^{-i\omega t}dx$, then the linearized equation of motion for $A_x$ is
\begin{equation}
A_x''+\frac{f'}{f}A_x'+(\frac{\omega^2}{f^2}-\frac{2\psi^2}{f})A_x=0\;.
\end{equation}

To obtain the real time correlation functions for the dual boundary theory, the holographic description associates in-going and out-going boundary conditions at the black hole horizon to retarded and advanced boundary correlators respectively~\cite{Son:2002sd}. To consider  causal behaviour, one should impose the in-going wave condition at the horizon: $A_x\sim f^{-i\omega/3r_h}$. Near the AdS boundary, the asymptotic behaviour of $A_x$ is given by
\begin{equation}
A_x=A^{(0)}+\frac{A^{(1)}}{r}+\cdots.
\end{equation}
According to the AdS/CFT correspondence, $A^{(0)}$ is the source, while $A^{(1)}$ is  dual to the current. Thus one can obtain
\begin{equation}
\sigma(\omega)=\frac{1}{i\omega}G^R(\omega)=\frac{1}{i\omega}\frac{A^{(1)}}{ A^{(0)}}\;.
\end{equation}
\begin{figure}[h!]
\centering
\includegraphics[scale=0.9]{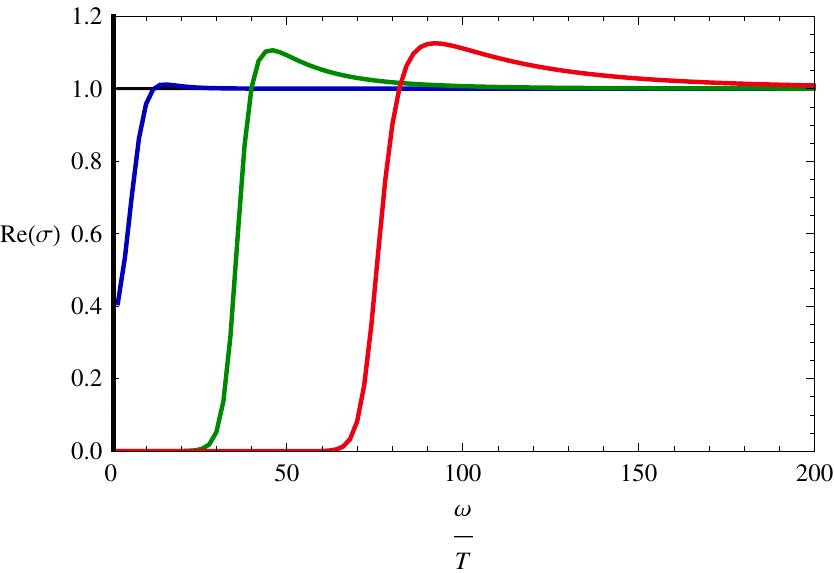}\ \ \ \
\includegraphics[scale=0.93]{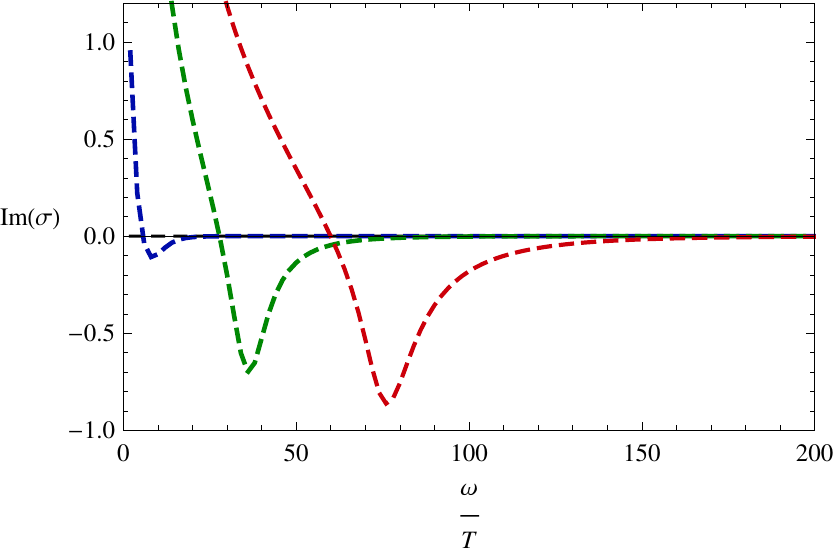} \caption{\label{sconduc} The optical conductivity as a function of frequence. The solid lines in the left plot are the real part of the conductivity, while the  dashed lines in the right plot are the imaginary part of the conductivity. We choose $m^2=-2$. The horizontal lines correspond to temperatures above $T_c$. Other curves from the left to right correspond to $T/T_c\simeq0.888$ (blue), $T/T_c\simeq0.222$ (green) and $T/T_c\simeq0.105$ (red), respectively. There is a delta function at the origin for the real part of the conductivity in the condensed phase.}
\end{figure}

The AC conductivity as a function of frequency is presented in figure~\ref{sconduc}. Above the critical temperature, the conductivity is a constant. As the temperature is lowered below $T_c$, the optical conductivity develops a gap at some special frequency $\omega_g$ known as gap frequency. As suggested in ref.~\cite{Horowitz:2008bn}, it can be identified with the one at the minimum of the imaginary part of the AC conductivity. Re$[\sigma(\omega)]$ is very small in the infrared and rises quickly at $\omega_g$.\footnote{It has been shown that the conductivity is directly related to the reflection coefficient with the frequency given the incident energy~\cite{Horowitz:2009ij}. The key point is that even as $T\rightarrow0$ there is still tunneling through the barrier provided by the effective potential. Therefore, a nonzero conductivity at small frequencies will always exist, and hence there is no hard gap in the optical conductivity at zero temperature. To obtain a superconductor with a hard gap, one might consider non-minimally coupled scalars in the bulk.} There also exists a small ``bump" slightly above $\omega_g$, which is reminiscent of the behaviour due to fermionic pairing~\cite{Gubser:2008wv}. For different choice of parameters, one can obtain a robust feature $\omega_g\simeq8T_c$ with deviations of less than $10\%$. Compared to the corresponding BCS value $\omega_g\simeq3.5T_c$, the result shown here is consistent with the fact that the holographic model describes a system at strong coupling. There is also a delta function at $\omega=0$ appearing as soon as $T<T_c$. This can be seen from the imaginary part of the conductivity.  According to the Kramers-Kronig relation
\begin{equation}\label{kkrelation}
\text{Im}[\sigma(\omega)] = - \frac{1}{\pi} {\mathcal{P}}\int_{-\infty}^{\infty} \frac{\text{Re}[\sigma(\omega')] d\omega'}{\omega'-\omega},
\end{equation}
one can conclude that the real part of the conductivity contains a Dirac delta function at $\omega=0$ if and only if the imaginary part has a pole, i.e., Im$(\sigma)\sim1/\omega$.

From above discussion, we see that this simple model can provide a holographically dual description of a superconductor. It predicts that a charged condensate emerges below a critical temperature via a second order transition, that the DC conductivity becomes infinite, and that the optical conductivity develops a gap at low frequency.  The temperature dependences of the coherence length $\xi$ as well as the penetration length $\lambda$ in the holographic model are both proportional to $(T_c-T)^{-1/2}$ near the critical temperature~\cite{Horowitz:2008bn,Maeda:2008ir}.
It has been shown that this holographic superconductor is type-II~\cite{Hartnoll:2008kx}. The condensate can form a lattice of vortices and the minimum of the free energy at long wavelength corresponds to a triangular array~\cite{Maeda:2009vf}. The effects of a superconducting condensate on holographic Fermi surfaces have been studied~\cite{Faulkner:2009am,Bagrov:2014mqa}. All these features are very reminiscent of real superconductors. Although the holographic model is very simple, it indeed captures some significant characteristics for superconductivity, thus helping us to understand real, strongly coupled superconductors.

\subsection{Holographic insulator/superconductor phase transition}

In  this subsection, let us consider a five-dimensional Einstein-Abelian-Higgs theory with following action
\begin{equation}
\label{5dhigsaction}
S= \frac{1}{2\kappa^2} \int d^5x \sqrt{-g} \left( {\cal R} +\frac{12}{L^2} -\frac{1}{4}F_{\mu\nu}F^{\mu\nu}-|\nabla_{\mu}\Psi -i q A_{\mu} \Psi |^2 -m^2 |\Psi|^2\right).
\end{equation}
When one does not include the matter sector,  the theory has a five-dimensional AdS-Schwarzschild black hole solution.  It is interesting to note that there also exists another exact solution, so-called AdS soliton,  in the theory (\ref{5dhigsaction}). The AdS soliton solution can be obtained by double Wick rotation from the AdS-Schwarzschild black hole as
\begin{eqnarray}\label{soliton}
ds^2 = f(r)d\chi^2+\frac{dr^2}{f(r)}+r^2(-dt^2+dx^2+ dy^2), \quad f(r) = r^2(1-r_0^4/r^4).
\end{eqnarray}
To remove the potential conical singularity, the spatial coordinate $\chi$ has to be periodic with a period $\pi/r_0$. If one considers the coordinates $(\chi, r)$, the geometry looks like a cigar and the tip is given by $r=r_0$.  The AdS soliton has no horizon, and therefore no entropy is associated with this solution.
Due to the existence of an IR cutoff at $r=r_0$ for the soliton solution, the field theory dual to this gravity background turns out to be in confined phase at zero temperature. Furthermore, this solution can be explained as a gravity dual to an insulator in condensed matter theory.  If one increases the chemical potential to a critical value, the AdS soliton solution becomes unstable to developing a scalar hair with nontrivial scalar profile. It is shown that the new solution can describe a superconducting phase~\cite{Taka}. In this way, the holographic insultor/superconductor phase transition at zero temperature can  be realized in the Abelian-Higgs model (\ref{5dhigsaction}).

\begin{figure}[h]
 \centering
  \includegraphics[scale=0.84]{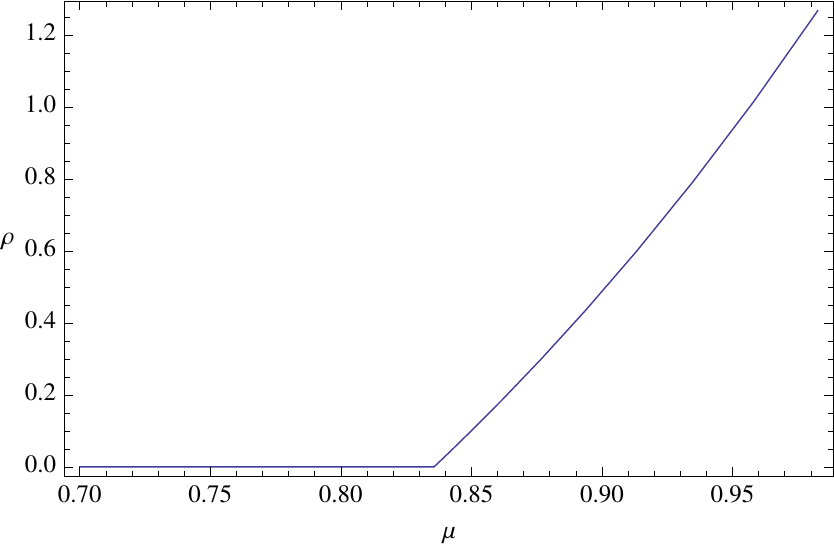}
    \hspace{0.5cm}
  \includegraphics[scale=0.85]{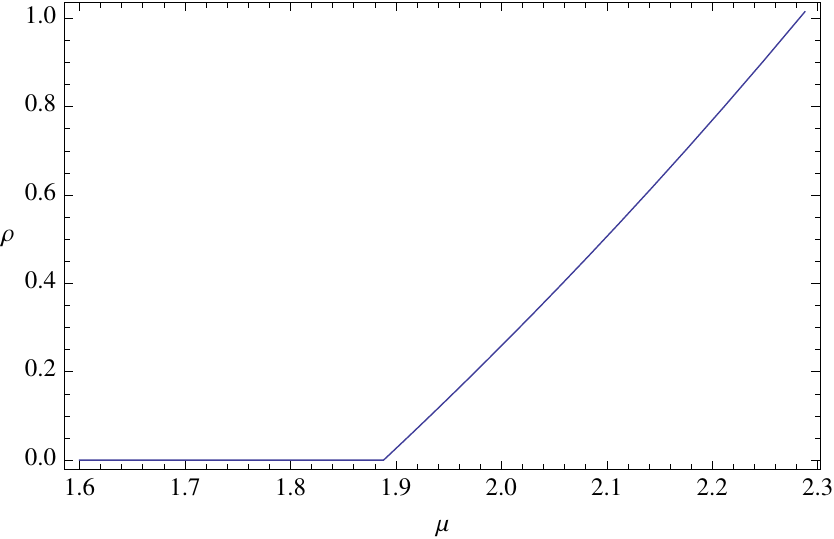}
 \caption{ \label{rhomu} The charge density $\rho$  as a function of $\mu$ when $\langle  {\cal O}_1\rangle \neq 0$ (left) and $\langle {\cal O}_2\rangle \neq 0$ (right). Its derivative jumps at the phase transition point. Used with permission from ref.~\cite{Taka}.}
\end{figure}

 More precisely, let us also consider the following ansatz in the probe limit
\begin{equation}
\Psi = \psi (r), \ \ \  A_{\mu}= \phi (r) dt.
\end{equation}
In the AdS soliton (\ref{soliton}) background, the equations of motions turn out  to be
\begin{eqnarray}
\begin{split}
& \psi'' +\left( \frac{f'}{f}+\frac{3}{r}\right) \psi' -\left( \frac{m^2}{f}-\frac{q^2 \phi^2}{r^2 f}\right) \psi =0, \\
& \phi'' +\left( \frac{f'}{f}+\frac{1}{r}\right) \phi' -\frac{2 q^2 \Psi^2}{f}\phi =0.
\end{split}
\end{eqnarray}
In the five-dimensional case, the BF bound is $m^2_{BF}=-4$. For simplicity, let us consider the case with $m^2=-15/4$.  To solve the equations of motion, we have to specify the boundary conditions both at the tip and the AdS boundary.  Near the AdS boundary, we have the following asymptotical form
\begin{equation}
\psi = \frac{\psi_-}{r^{3/2}} +\frac{\psi_+}{r^{5/2}}+ \cdots, \quad \phi = \mu -\frac{\rho}{r^2}+ \cdots.
\end{equation}
Note that in this case, both terms proportional to $\psi_-$ and $\psi_+$ are normalizable,  so the corresponding operators ${\cal O}_1$ and ${\cal O}_2$ have dimensions $\triangle =3/2$ and $\triangle =5/2$, respectively.
On the other hand, near the tip of the soliton, these fields behave like
\begin{eqnarray}\label{Newmann}
\begin{split}
&\psi= a +b \log (r-r_0)+ c (r-r_0) +\cdots, \\
&\phi= A +B \log (r-r_0) +C (r-r_0)+ \cdots,
\end{split}
\end{eqnarray}
where $a, b, c$ and $A, B, C$ are all constants. The field regularity at the tip requires us to take $b=B=0$. As in the previous subsection we can set $q=1$ and further set $r_0=1$ without loss of
generality.
\begin{figure}[h]
\centering
\includegraphics[scale=0.85]{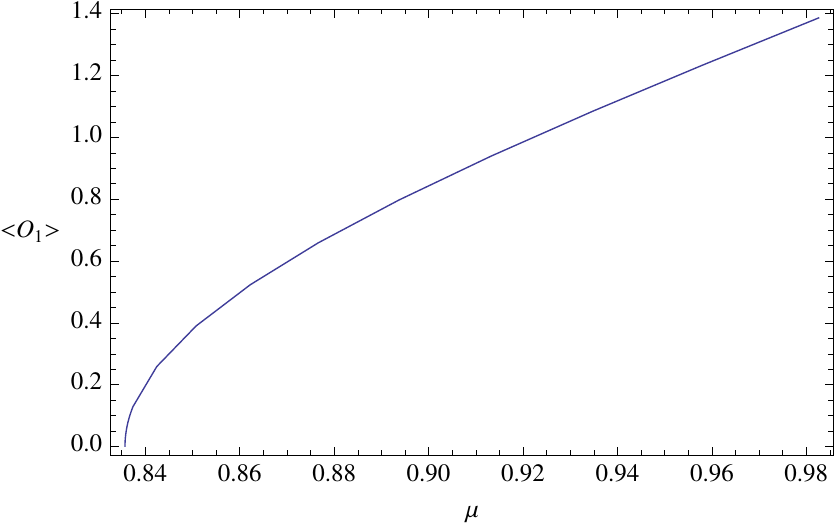}
 \hspace{0.3cm}
\includegraphics[scale=0.85]{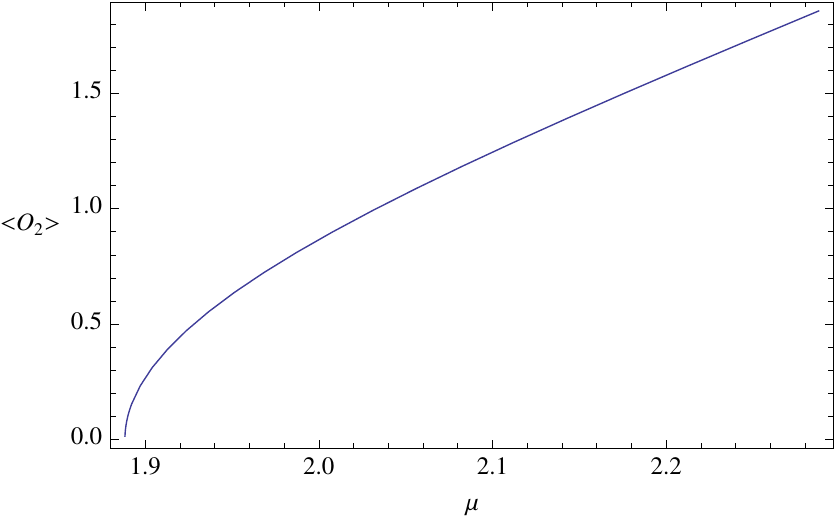}
  \caption{\label{Condensate} The behaviour of condensation for the operator $\langle {\cal O}_1 \rangle$ (left) and $\langle {\cal O}_2 \rangle$ (right) with respect to chemical potential.
  Used with permission from ref.~\cite{Taka}.}
\end{figure}
With the boundary conditions, solving the equations of motion, one can find that when the chemical potential $\mu$ is beyond some critical value, the condensation happens. Concretely,
for the operator ${\cal O}_1$, the critical chemical potential is $\mu_1=0.84$, while the critical chemical potential $\mu_2= 1.88$ for the operator ${\cal O}_2$. The behaviour of condensation is plotted in figure~\ref{Condensate} with respect to chemical potential.  In figure~\ref{rhomu} the charge density $\rho$ with respect to chemical potential is plotted. We can see that at the phase transition
point, its derivative is discontinuous, which verifies that the phase transition is indeed second order, since one has $\rho =\partial \Omega/\partial \mu$, where $\Omega$ is the Gibbs free energy density.

\begin{figure}[h]
 \centering
 \includegraphics[scale=0.88]{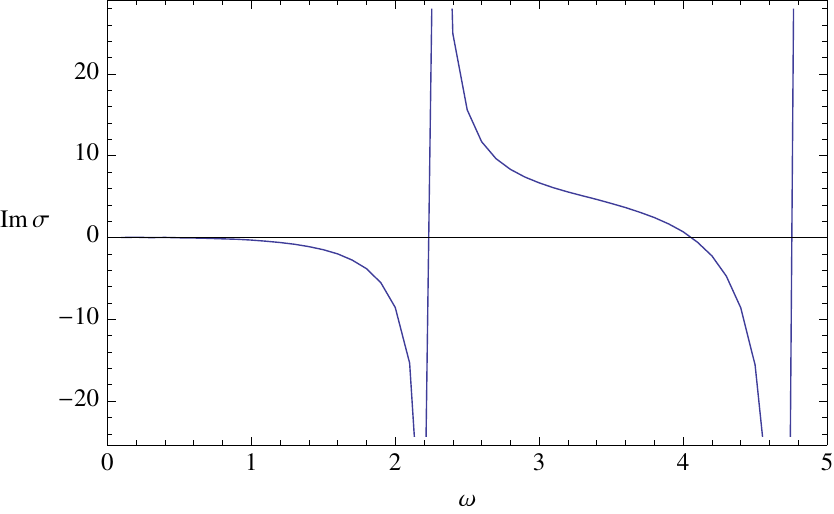}
  \hspace{0.3cm}
  \includegraphics[scale=0.90]{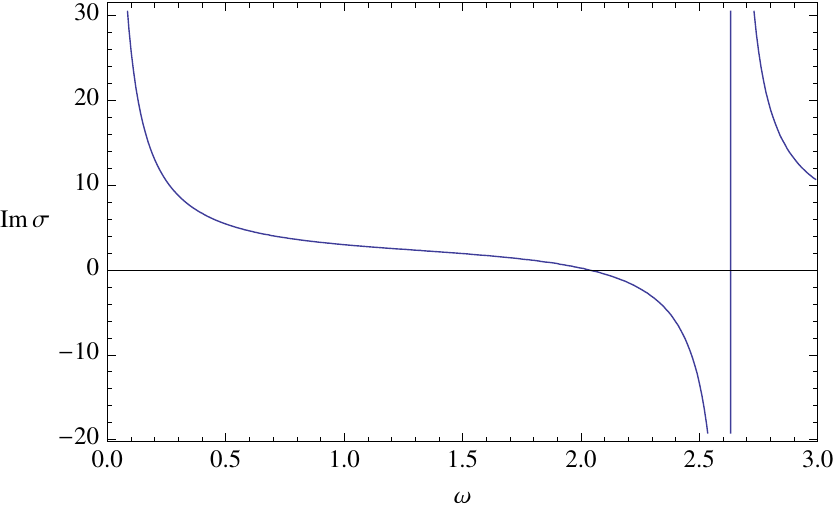}
 \caption{\label{BubbleSC} The imaginary part of the conductivity for the AdS soliton without condensation $\langle {\cal O}_{1,2} \rangle =0$  (left) and with condensation $\langle  {\cal O}_1\rangle \neq 0$ (right).  $\rho=0.0094$ and $\mu=0.84$ are taken in the right plot. Used with permission from ref.~\cite{Taka}.}
\end{figure}

To calculate conductivity $\sigma (\omega)$  we can consider the perturbation of the component $A_x$ in the soliton background. Assuming it has the form $A_x \sim e^{-i\omega t}$, its equation then turns out to be
\begin{equation}
A''_x +\left ( \frac{f'}{f}+\frac{1}{r}\right) A'_x +\left (\frac{\omega^2}{r^2f}- \frac{2 q^2 \Psi^2}{f}\right) A_x=0.
\end{equation}
At the tip one takes the Newmann boundary condition as in (\ref{Newmann}). Near the AdS boundary, one has the asymptotical form as
\begin{equation}
 A_x = A_x^{(0)}+ \frac{A_x^{(1)}}{r^2}+\frac{A_x^{(0)}\omega^2}{2}\frac{ \log \Lambda r}{r^2}+ \cdots,
 \end{equation}
 where $\Lambda$ is a cutoff.  The holographic conductivity can be obtained as
 \begin{equation}
 \sigma(\omega)= -\frac{2 i A_x^{(1)}}{\omega A_x^{(0)}}+ \frac{i \omega}{2}.
 \end{equation}
 Since the background  has no horizon, the real part of the conductivity always vanishes. This means that there is no dissipation. The imaginary part is plotted in figure~\ref{BubbleSC}: The left plot corresponds to the case of pure AdS solution without scalar hair, while the right one to the case with nontrivial scalar hair.  There exist poles periodically at the points where $A_x^{(0)}$ vanishes. These correspond to normalized modes dual to vector operators. One can see that when $\omega$ is large, both case are similar, while when $\omega \to 0$, they are quite different.  In the case without condensation, the imaginary part goes to zero when $\omega \to 0$, while it diverges in the case with condensation. According to the Kramers-Kronig relation~\eqref{kkrelation}, it shows that there is a delta functional
 support for the real part of conductivity at $\omega=0$.  Therefore the AdS soliton background with nontrivial scalar hair should be identified with  the superconductivity.

\begin{figure}[htbp]
 \centering
  \includegraphics[scale=0.43]{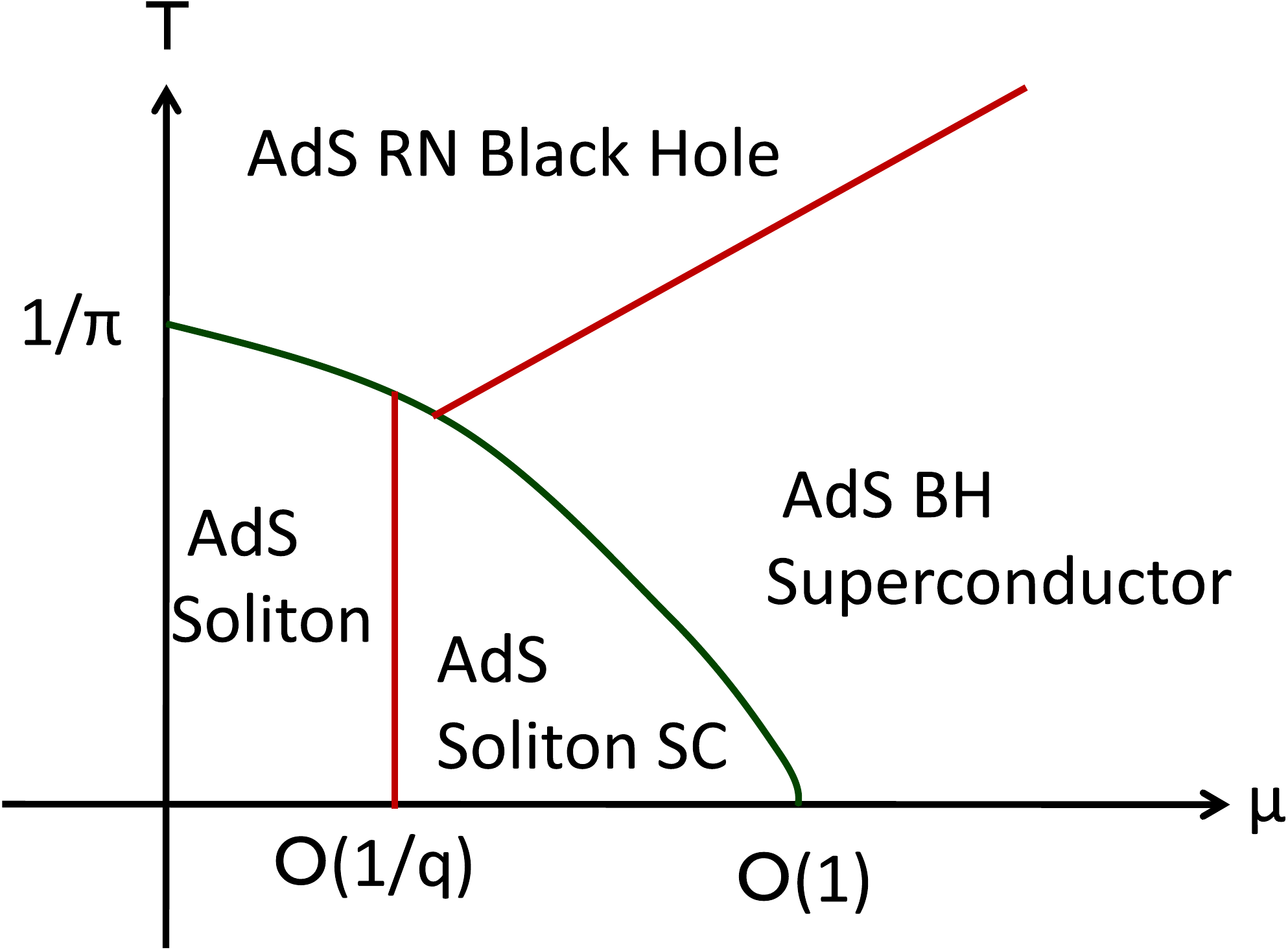}
 \caption{\label{schematicPhase} The phase diagram of AdS soliton and AdS black hole with a charged scalar field obtained in the large $q$ limit. Used with permission from ref.~\cite{Taka}.}
\end{figure}

In the Einstein-Abelian-Higgs theory (\ref{5dhigsaction}), besides the two phases described above, as in the four dimensional case, there exist another two solutions: AdS Reissner Nordstr\"om (AdS RN) black hole\footnote{Its precise form can be found in~\eqref{RNmetric} below.} without scalar hair and AdS RN black hole with scalar hair, the latter can be identified with a superconductivity phase, while the former is dual to a conductor phase. Combining the four phases together, one could have the phase digram of the theory, which is schematically plotted in figure~\ref{schematicPhase}. The green line in the figure denotes the first order phase transition, while two red lines represent second order phase transition.  Considering back reaction of matter sector, the complete phase diagrams in terms of temperature and chemical potential for the Abelian-Higgs model have been constructed in ref.~\cite{Horowitz:2010jq}. It is interesting to note that the behaviour of the entanglement entropy with respect to chemical potential is non-monotonic and seems to be universal in this kind of insulator/superconductor models~\cite{Cai:2012sk,Cai:2012es,Yao:2014fua}.

\section{Holographic P-wave Models}
\label{sect:p-wave}

\subsection{The SU(2) Yang-Mills P-wave model}
\label{sect:yangmills}
The first holographic p-wave model is constructed by introducing a SU(2) Yang-Mills field in asymptotically AdS spacetime. One of three U(1) subgroups is regarded as the gauge group of electromagnetism and the off-diagonal gauge bosons which are charged under this U(1) gauge field are supposed to condense outside the horizon. The full action is given by~\cite{Gubser:2008wv}
\begin{equation}\label{su2action}
S =\int d^4 x
\sqrt{-g}[\frac{1}{2\kappa^2}(\mathcal{R}+\frac{6}{L^2})-\frac{1}{4\hat{g}^2}
F^a_{\mu\nu} F^{a\mu \nu}],
\end{equation}
where $\kappa$ is the four dimensional gravitational constant, $\hat{g}$ is the Yang-Mills coupling constant and $L$ is the AdS radius.
The field strength for the SU(2) gauge field $A=A^a_{\mu}\tau^a dx^{\mu}$ is
\begin{equation}\label{supstrength}
 F^a_{\mu\nu}=\partial_\mu A^a_\nu-\partial_\nu A^a_\mu + \epsilon^{abc}A^b_\mu
 A^c_\nu,
\end{equation}
where $\mu,\nu=(t,r,x,y)$ denote the indices of spacetime and $a,b,c=(1,2,3)$ are the indices of the SU(2) group generators $\tau^a=\sigma^a/2i$ ($\sigma^a$ are Pauli matrices).
$\epsilon^{abc}$ is the totally antisymmetric tensor with $\epsilon^{123}=+1$.

Note that the ratio $\kappa/\hat{g}$ measures the influence of Yang-Mills field on the background geometry. For the case $\kappa/\hat{g}\ll 1$ with $A^a_\mu$ fixed, the back reaction of the matter field can be ignored, thus the metric is simply $AdS_4$ Schwarzschild black hole
\begin{equation}
 ds^2=\frac{r^2}{L^2}\left[-\left(1-\frac{r_h^3}{r^3}\right)dt^2+dx^2+dy^2\right]+\frac{L^2}{r^2}\frac{dr^2}{1-r_h^3/r^3},
\end{equation}
with the temperature given by $T=\frac{3r_h}{4\pi L^2}$. Without loss of generality, we shall choose $L=1$, and we also fix a scale by setting $r_h=1$.

\subsubsection{Vector condensate}

To realize the p-wave condensate, one takes the ansatz
\begin{equation}
A=\phi(r)\tau^3 dt+\varpi(r)\tau^1 dx\;.
\end{equation}
It is clear that the non-trivial profile of $\varpi(r)$ picks out the $x$ direction as special, thus the condensed phase breaks the gauge group $\text{U}(1)_3$ generated by $\tau^3$ and $SO(2)$ rotational symmetry in $x-y$ plane. The relevant equations are~\cite{Gubser:2008wv}
\begin{equation}\label{su2eoms}
\begin{split}
\phi'' + {2 \over r} \phi' -{1 \over r(r^3-1)} \varpi^2 \phi &= 0,\\
\varpi'' + {1+2r^3 \over r(r^3-1)} \varpi'+{r^2 \over (r^3-1)^2} \phi^2 \varpi &= 0,
\end{split}
\end{equation}
with primes representing the derivative with respect to $r$.

The regularity at the horizon $r=1$ demands the behaviour like
\begin{equation}
\phi=\phi_1(r-1)+\cdots,\quad \varpi=w_0+w_2(r-1)^2+\cdots,
\end{equation}
while the asymptotical expansion near the boundary $r\rightarrow\infty$ takes the form
\begin{equation}
\phi=\mu-{\rho\over r}+\cdots,\quad \varpi=W_0+{W_1\over r}+\cdots.
\end{equation}
According to the holographic dictionary, $\mu$ is regarded as chemical potential and $\rho$ is the total charged density, and $W_0$ is the source of the dual operator $J^x$. To spontaneously
break the U(1) symmetry, we should impose $W_0=0$, then the coefficient $W_1$ gives the vacuum expectation value of $J^x$. According to the two-fluid model, the total charge density $\rho$ can be divided into two components $\rho=\rho_n+\rho_s$, where $\rho_n$ is the normal component, while $\rho_s$ is the superconducting component. In the holographic setup, the normal charge density $\rho_n$ is proportional to the $\tau^3$ part of the electric field at the horizon, i.e., $\rho_n=\phi_1$. Therefore the superconducting charge density is $\rho_s=\rho-\rho_n$.
\begin{figure}[h]
\centering
\includegraphics[scale=0.65]{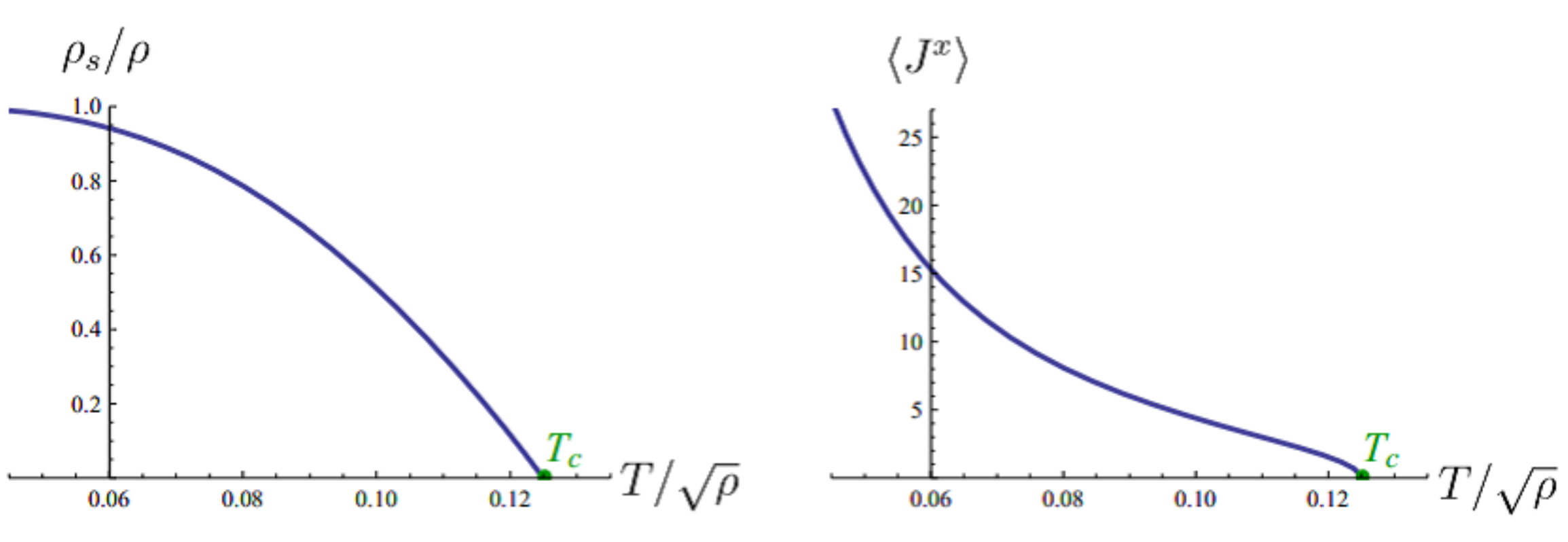}
\caption{The fraction $\rho_s/\rho$ (right) and the condensate (left) as a function of temperature. Adapted with permission from ref.~\cite{Gubser:2008wv}.}\label{sucondensate}
\end{figure}

By numerically solving the equations~\eqref{su2eoms}, one finds that the condensate is non-vanishing only when the rescaled temperature $T/\sqrt{\rho}$ is small enough, i.e., lower than $T_c$ at which the condensate first turns on. As one can see in the right plot of figure~\ref{sucondensate}, as the temperature is lowered, $\langle J^x\rangle$ increases continuously.
Near $T_c$, $\langle J^x\rangle$ vanishes as $\sqrt{T_c-T}$, which is the typical behaviour predicted by Ginzburg-Landau theory. The fraction $\rho_s/\rho$ of the charge carried by the superconducting condensate goes to zero linearly near $T_c$.~\footnote{The ration $\rho_s/\rho$ versus temperature in the left plot is reminiscent of the temperature dependence of the superfluid of liquid He II as measured from in the torsional oscillation disk stack experiment. However, we find $\rho_s$ goes to zero linearly here, while the experiment gives a critical exponent about $0.67$.}

We have interpreted $\text{U}(1)_3$ generated by $\tau^3$ as the gauge group of electromagnetism. The condensate of $\langle J^x\rangle$ spontaneously breaks this U(1) symmetry as well as the rotational symmetry, thus resulting in an anisotropic superconducting phase. To see this much more clearly, we shall calculate the optical conductivity, which can be deduced by the retarded Green's function of the $\text{U}(1)_3$ current. Similar to the previous section, in gravity side the linear response to electromagnetic probes is turned out to study how linear perturbations of the $\tau^3$ component of the gauge field propagate.

\subsubsection{Conductivity}
\label{sect:suconducticity}
In the presence of the condensate $\varpi\tau^1 dx$, the $x$ direction is special, so the conductivity $\sigma_{xx}$ along the $x$ direction is expected to be different from $\sigma_{yy}$ along the $y$ direction. To obtain consistent linearized equations, we can turn on the perturbation~\cite{Gubser:2008wv}
\begin{equation}\label{supertubation}
\delta A=e^{-i\omega t} \left[(a_t^1 \tau^1 + a_t^2 \tau^2) dt + a_x^3 \tau^3 dx + a_y^3 \tau^3 dy \right],
\end{equation}
where all the functions depend on $r$ only. By plugging the perturbation~\eqref{supertubation} into the linearized Yang-Mills equation, one finally obtains four second order equations
\begin{equation}\label{supertubationay}
{a_y^3}''+\frac{2r^3+1}{r(r^3-1)}{a_y^3}'+\left[\frac{\omega^2r^2}{(r^3-1)^2}-\frac{\varpi^2}{r(r^3-1)}\right]a_y^3=0,
\end{equation}
\begin{subequations}\label{supertubationax}
\begin{align}
{a_x^3}''+\frac{2r^3+1}{r(r^3-1)}{a_x^3}'+\frac{r^2}{(r^3-1)^2}\left(\omega^2a_x^3-\phi\varpi{a_t^1}-i\omega{a_t^2} \right)=0,\\
{a_t^1}''+\frac{2}{r}{a_t^1}'+\frac{\phi\varpi}{r(r^3-1)}{a_x^3}=0,\\
{a_t^2}''+\frac{2}{r}{a_t^2}'-\frac{\varpi}{r(r^3-1)}\left(\varpi {a_t^2}+i\omega {a_x^3}\right)=0,
\end{align}
\end{subequations}
and two first order constraint equations
\begin{equation}\label{supertubationax1}
\begin{split}
i\omega{a_t^1}'+\phi {a_t^2}'-\phi'{a_t^2}=0,\\
i\omega{a_t^2}'-\phi{a_t^1}'-(1-\frac{1}{r^3})\varpi{a_x^3}'+\phi'{a_t^1}+(1-\frac{1}{r^3})\varpi'{a_x^3}=0.
\end{split}
\end{equation}

It is clear that the equation of motion of the ${a_y^3}$ mode decouples from the others, and the conductivity $\sigma_{yy}$ exhibits similar ``soft gap" behaviour to the s-wave model~\cite{Hartnoll:2008vx}.\footnote{However, by considering the back reaction to the metric in the SU(2) model~\eqref{su2action}, it has been shown that the conductivity in the $y$ direction has a ``hard gap" at zero temperature, i.e., the real part of the conductivity is zero for an excitation frequency less than the gap frequency~\cite{Basu:2009vv}.} What we are interested in is the conductivity $\sigma_{xx}$ in the $x$ direction. The conductivity $\sigma_{xx}$ can be determined by solving the coupled equations~\eqref{supertubationax} with the constraints given by~\eqref{supertubationax1}. More precisely, we impose the ingoing wave condition at the horizon, which corresponds to a retarded Green's function,
\begin{equation}\label{suingoing}
\begin{split}
a_x^3 &= (r-1)^{-i\omega/4\pi T} \left[ 1 +a_x^{3(1)} (r-1) + a_x^{3(2)} (r-1)^2 + \cdots \right],\\
a_t^1 &= (r-1)^{-i\omega/4\pi T} \left[ a_t^{1(2)} (r-1)^2 + a_t^{1(3)} (r-1)^3 + \cdots \right],\\
a_t^2 &= (r-1)^{-i\omega/4\pi T} \left[ a_t^{2(1)} (r-1) + a_t^{2(2)} (r-1)^2 + \cdots \right],
\end{split}
\end{equation}
where all the coefficients can be fixed once $w_0$, $\phi_1$ and $\omega$ are specified. Near the conformal boundary $r\rightarrow\infty$, one has a generic solution to the equations of motion
\begin{equation}\label{suboundaty}
\begin{split}
a_x^3 &= A_x^{3(0)} + {A_x^{3(1)} \over r} + \cdots,\\
a_t^1 &= A_t^{1(0)} + {A_t^{1(1)} \over r} + \cdots,\quad a_t^2= A_t^{2(0)} + {A_t^{2(1)} \over r} + \cdots.
\end{split}
\end{equation}
As pointed out in ref.~\cite{Gubser:2008wv}, there exists a residual gauge invariance. After fixing this residual gauge freedom, one can finally obtain the gauge invariant conductivity along $x$ direction
\begin{equation}
\sigma_{xx} = -{i\over\omega A_x^{3(0)}}\left(A_x^{3(1)}+W_1 {i\omega A_t^{2(0)}+\mu A_t^{1(0)} \over \mu^2 - \omega^2}\right)\;.
\end{equation}
Numerical calculation can only display the continuous part of $\sigma_{xx}(\omega)$. One can reveal the non-analytic behaviour by virtue of the Kramers-Krong relations, which tells us that a simple pole in $\text{Im}[\sigma_{xx}(\omega)]$ at $\omega_0$ implies a delta function $\delta(\omega-\omega_0)$ to $\text{Re}[\sigma_{xx}(\omega)]$. Further more, the positivity constraint on the real part of conductivities requires any pole of $\text{Im}[\sigma_{xx}(\omega)]$ on the real axis to have a positive residue.

\begin{figure}[h!]
\centering
\includegraphics[scale=0.58]{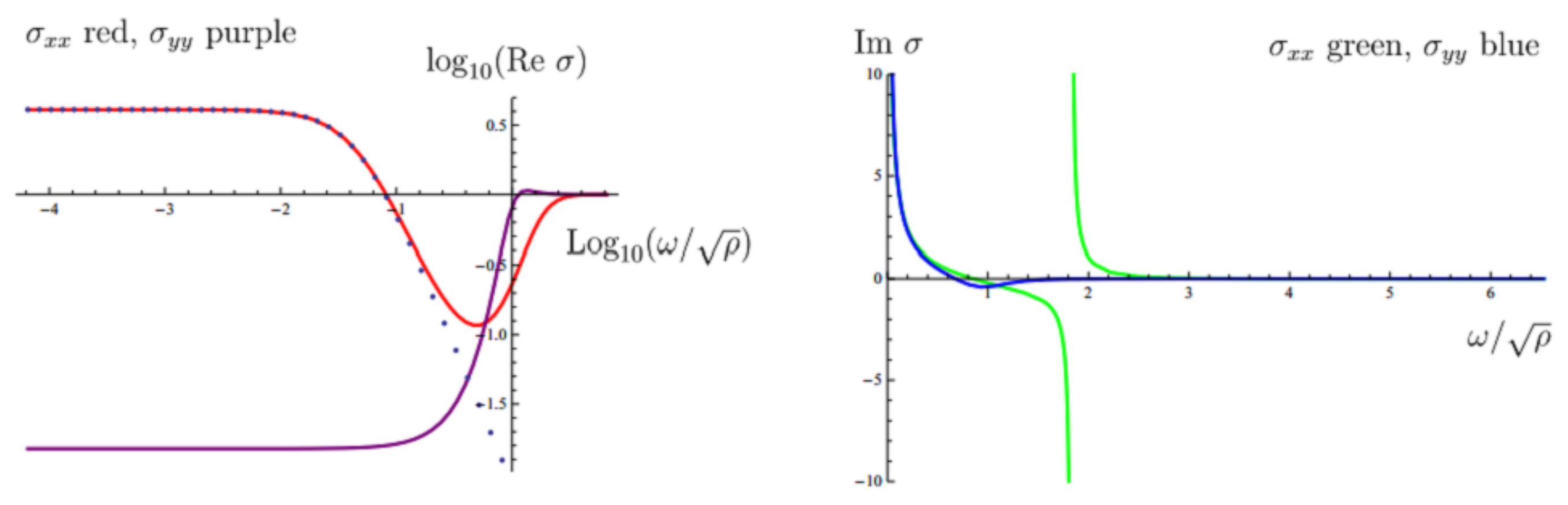}
\caption{Conductivities $\sigma_{xx}$ and $\sigma_{yy}$ with respect to frequency at $T/\sqrt{\rho}\simeq0.0779$. The dotted curves are the best fits of the Drude model prediction to $\text{Re}~\sigma_{xx}(\omega)$. Adapted with permission from ref.~\cite{Gubser:2008wv}.}\label{suconductivity}
\end{figure}

The behaviour of conductivities as a function of frequency $\omega$ is shown in figure~\ref{suconductivity}, from which one can see the following features~\cite{Gubser:2008wv}. First, both $\sigma_{xx}$ and $\sigma_{yy}$ approach constant for sufficiently large $\omega$. This is because the condensate involves dynamics with a characteristic energy scale set by $\sqrt{\rho}$. If $\omega\gg\sqrt{\rho}$, the propagation of the gauge boson should become insensitive to the condensate and can be approximated by the case in pure $AdS_4$, thus is a constant. Second, $\sigma_{yy}$ exhibits gapped dependence similar to the Abelian-Higgs model in figure~\ref{sconduc}. $\text{Re}(\sigma)$ is very small in the infrared, then rises quickly at $\omega =\omega_g \simeq \sqrt{\rho}$. There is a slight ``bump'' a little above $\omega_g$ which is reminiscent of the behaviour expected for fermionic pairing. Third, there is a pole in $\text{Im}[\sigma_{xx}]$ at $\omega=\omega_0\simeq1.8\sqrt{\rho}$. Therefore, there is a delta function contribution to $\text{Re}[\sigma_{xx}]$ at $\omega=\omega_0$. Finally, in the small $\omega$ region, $\text{Re}[\sigma_{xx}]$ can be well parameterized in terms of the Drude model
\begin{equation}
\text{Re}[\sigma_{\text{Drude}}] = {\sigma_0 \over 1 + \omega^2 \tau^2}\;,
\end{equation}
where $\sigma_0$ gives the DC conductivity and $\tau$ is the scattering time. The best fit gives a narrow Drude peak in $\sigma_{xx}$ and suggests conductivity due to quasi-particles with scattering time to diverge as $T\rightarrow0$.

We do not have a microscopic description of the condensate in the language of the dual theory without gravity. However, we know clearly that there is an SU(2) current algebra, and the component $J^x$ develops an expectation for sufficiently large chemical potential. Yet we only turn on $\tau^1 dx$ mode corresponding to the p-wave background. The $(p+ip)$-wave case can be realized by involving the combination $\tau^1dx+\tau^2dy$. This mode results in an isotropic superconducting phase which exhibits a pseudogap\footnote{The terminology ``pseudogap'' here is to denote a well defined gap in the dissipative conductivity at low frequencies in which the conductivity is not identically zero.} at low temperatures and a nonzero Hall conductivity with no external magnetic field~\cite{Roberts:2008ns}. However, it should be pointed out that $(p+ip)$ configurations are unstable against turning into pure p-wave background. The insulator/superconductor phase transition for the SU(2) p-wave model has been
studied in  ref.~\cite{Akhavan:2010bf}.

A new ground state can be found when a magnetic component of the gauge field is larger than a critical value, which forms a triangular Abrikosov lattice in the spatial directions perpendicular to the magnetic field~\cite{Ammon:2011je,Bu:2012mq}. In the same spirit, a p-wave superconductor for which the dual field is explicitly known has been constructed in refs.~\cite{Ammon:2008fc,Basu:2008bh,Ammon:2009fe} by embedding a probe of two coincident D7-branes in the AdS black hole background. From this top-down approach one can try to identify the SU(2) chemical potential as an isospin chemical potential and the condensate as a $\rho$ meson. The back reaction of the gauge field on the metric in the SU(2) Yang-Mills model has been considered in refs.~\cite{Ammon:2009xh}. It is interesting to note that when the back reaction is strong enough, the phase transition will be a first order one. The holographic SU(2) p-wave superconductor model has been extended to include, for example, the Gauss-Bonnet term~\cite{Cai:2010zm,CNZ} and Chern-Simons coupling~\cite{Zayas:2011dw}.
In addition, based on the backreacted metric, the behaviour of entanglement entropy in the holographic superconducting phase transitions has been studied in refs.~\cite{Cai:2012nm,Arias:2012py,Cai:2013oma}.

\subsection{The Maxwell-Vector P-wave model}
\label{sect:vector}
Let us introduce a charged vector field into the $(d+1)$ dimensional Einstein-Maxwell theory with a negative cosmological constant. The full action reads~\cite{Cai:2013pda,Cai:2013aca}
\begin{equation}\label{vecaction}
\begin{split}
S=\frac{1}{2\kappa^2}\int d^{d+1} x
\sqrt{-g}(\mathcal{R}+\frac{d(d-1)}{L^2}+\mathcal{L}_m),\\
\mathcal{L}_m=-\frac{1}{4}F_{\mu\nu} F^{\mu \nu}-\frac{1}{2}\rho_{\mu\nu}^\dagger\rho^{\mu\nu}-m^2\rho_\mu^\dagger\rho^\mu+iq\gamma \rho_\mu\rho_\nu^\dagger F^{\mu\nu},
\end{split}
\end{equation}
where  a dagger denotes complex conjugation and $\rho_\mu$ is a complex vector field with mass $m$ and charge $q$. We define $\rho_{\mu\nu}=D_\mu\rho_\nu-D_\nu\rho_\mu$ with the covariant derivative $D_\mu=\nabla_\mu-iqA_\mu$. The last non-minimal coupling term characterizes the magnetic moment of the vector field $\rho_\mu$.

Since $\rho_\mu$ is charged under the U(1) gauge field, according to AdS/CFT correspondence, its dual operator $\hat{J^\mu}$ will carry the same charge under this symmetry and a vacuum expectation value of this operator will then trigger the U(1) symmetry breaking spontaneously. Thus, the condensate of the dual vector operator will break the U(1) symmetry as well as the spatial rotational symmetry since the condensate will pick out one direction as special. Therefore, viewing this vector field as an order parameter, the holographic model can be used to mimic a p-wave superconductor (superfluid) phase transition. The gravity background without vector hair ($\rho_\mu=0$)/with vector hair ($\rho_\mu\neq0$) is used to mimic the normal phase/superconducting phase in the dual system.

Indeed, it was shown in ref.~\cite{Cai:2013pda} that working on the probe limit, as one lowers the temperature, the normal phase becomes unstable to developing nontrivial configuration of the vector field. The calculation of the optical conductivity reveals that there is a delta function at the origin for the real part of the conductivity, which means the condensed phase is indeed superconducting.
In this subsection, we shall review the effect of a background magnetic field on the model and its complete phase diagram in terms of temperature and chemical potential.

\subsubsection{Condensate induced by magnetic field}
Generally speaking, to consider the case with a magnetic field, one needs to solve coupled partial differential equations which is much more involved in practice. However, if one is interested in the instability induced by the magnetic field, one can overcome this difficulty by only focusing the dynamics near the critical point at which the condensate is very small. More precisely, one can introduce a deviation parameter $\epsilon$ from the critical point at which the condensate begins to appear. The coupled equations of motion can then be solved order by order in terms of the power of $\epsilon$.

Following the above procedure, we now turn on a magnetic field to study how the applied magnetic field influences the system. The background is taken to be a (3+1) dimensional AdS-Schwarzschild black hole~\eqref{AdSswtz}.
A consistent ansatz is as follows~\cite{Cai:2013pda}
\begin{equation}\label{vecmatterB}
\begin{split}
\rho_\nu dx^\nu=[\epsilon\rho_x(r,x)e^{ipy}+\mathcal{O}(\epsilon^3)]dx+[\epsilon\rho_y(r,x)e^{ipy}e^{i\theta}+\mathcal{O}(\epsilon^3)]dy,\\
A_\nu dx^\nu=[\phi(r)+\mathcal{O}(\epsilon^2)]dt+[Bx+\mathcal{O}(\epsilon^2)]dy,
\end{split}
\end{equation}
where $\rho_x(r,x)$, $\rho_y(r,x)$ are all real functions, $p$ is a real constant and the constant $\theta$ is the phase difference between the $x$ and $y$ components of the vector field $\rho_\mu$. The constant magnetic field $B$ is perpendicular to the $x-y$ plane.

The profile of $\phi$ can be uniquely determined at the zeroth order of $\epsilon$, which takes the form
\begin{equation}\label{vecat}
 \phi(r)=\mu(1-r_h/r)\;,
\end{equation}
with $\mu$ interpreted as the chemical potential. The equations of motion for $\rho_x$ and $\rho_y$ can be deduced from~\eqref{vecaction} at order $\mathcal{O}(\epsilon)$. We further separate the variables as $\rho_x(r,x)=\varphi_x(r)X(x)$ and $\rho_y(r,x)=\varphi_y(r)Y(x)$. Then one can get the following equations~\footnote{ In order to satisfy the equations of motion with the given ansatz, $\theta$ can only be chosen as $\theta_+=\frac{\pi}{2}+2n\pi$ or $\theta_-=-\frac{\pi}{2}+2n\pi$ with $n$ an arbitrary integer. Here and below the upper signs correspond to the $\theta_+$ case and the lower to the $\theta_-$ case.}
\begin{equation}\label{veceomabc}
\begin{split}
\varphi_x''+\frac{f'}{f}\varphi_x'+\frac{q^2\phi^2}{f^2}\varphi_x-\frac{m^2}{f}\varphi_x-\frac{E}{r^2f}\varphi_x=0,\\
-\ddot{X}\mp (1+\gamma)qBY+(qBx-p)^2X=EX,\\
-\ddot{Y}\mp (1+\gamma)qB X+(qBx-p)^2Y=EY,
\end{split}
\end{equation}
where the prime denotes the derivative with respect to $r$ and the dot denotes the derivative with respect to $x$. We have also made a consistent assumption $\varphi_x=\varphi_y$ and $E$ is a constant coming from variables separation. The last two equations for $X(x)$ and $Y(x)$ can be solved analytically and the eigenvalue is given by $E=(2n+1)|qB|\pm (1+\gamma)qB$ where $n$ can be chosen as a non-negative integer.

We are interested in how the applied magnetic field influences on the transition temperature from the normal phase to the condensed phase. The effective mass of the charged vector field in the lowest energy state, i.e., in the lowest Landau level $n=0$ depends on the magnetic field $B$ and the non-minimal coupling parameter $\gamma$ as
\begin{equation}\label{veclowestmass}
m_{\rm eff}^{2}=m^2-\frac{|\gamma qB|}{r^2}-\frac{q^2\phi^2}{f}\;.
\end{equation}
It is clear that the increase of the magnetic field $B$ decreases the effective mass and thus tends to raise the transition temperature, even in the case that the electric field is turned off. Only the magnetic field itself can trigger the phase transition. This result has an analogy to the QCD vacuum instability induced by a strong magnetic field to spontaneously developing the $\rho$-meson condensate. It is clear that the last term in~\eqref{vecaction} describing a non-minimal coupling of the vector field $\rho^\mu$ to the gauge field $A_\mu$ plays a crucial role in  the instability. Note that similar coupling can be found in many formalisms used to describe the coupling of magnetic moment to the background magnetic field for charged vector particles~\cite{Djukanovic:2005ag, Young:1963zza}.

\begin{figure}[h!]
\centering
\includegraphics[scale=1.0]{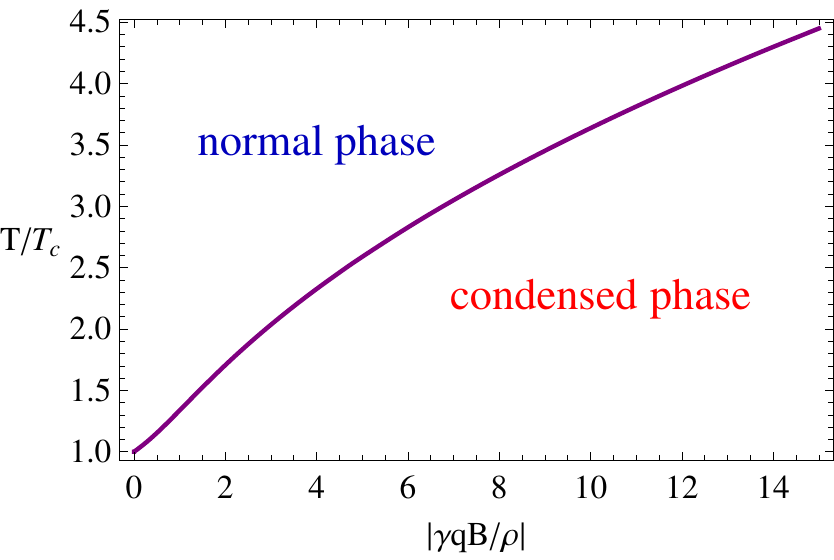}
\caption{\label{vecmagneticn}  The transition temperature from the normal phase to the condensed phase as a function of magnetic field. $T_c$ is the critical temperature in the case without magnetic field. The magnetic field raises the transition temperature. One has chosen $m^2=3/4$.  This figure was taken from ref.~\cite{Cai:2013pda}.}
\end{figure}
The $(T, B)$ phase diagram for the lowest Landau level is depicted in figure~\ref{vecmagneticn} in the case with fixed charge density $\rho=\mu r_h$. To determine which side of the phase transition line is the condensed phase, we can consider the equation~\eqref{veclowestmass}. It suggests that the magnetic field decreases the effective mass. So if we increase the magnetic field at a fixed temperature, the normal state will become unstable for sufficiently large magnetic field.

It is clear that the transition temperature increases with the applied magnetic field. In ordinary superconductors an external magnetic field suppresses superconductivity via diamagnetic and Pauli pair breaking effects. However, it has also been proposed that the magnetic field induced superconductivity can also be realized in type-II superconductors~\cite{PhysRevLett.58.1482,Rasolt:1992zz}, in which the Abrikosov flux lattice may enter a quantum limit of the low Landau level dominance with a spin-triplet pairing. And possible experimental evidence for the strong magnetic induced superconductivity can be found, for example, in refs.~\cite{levy2005,uji2010}.

\begin{figure}[h]
\centering
\includegraphics[scale=0.44]{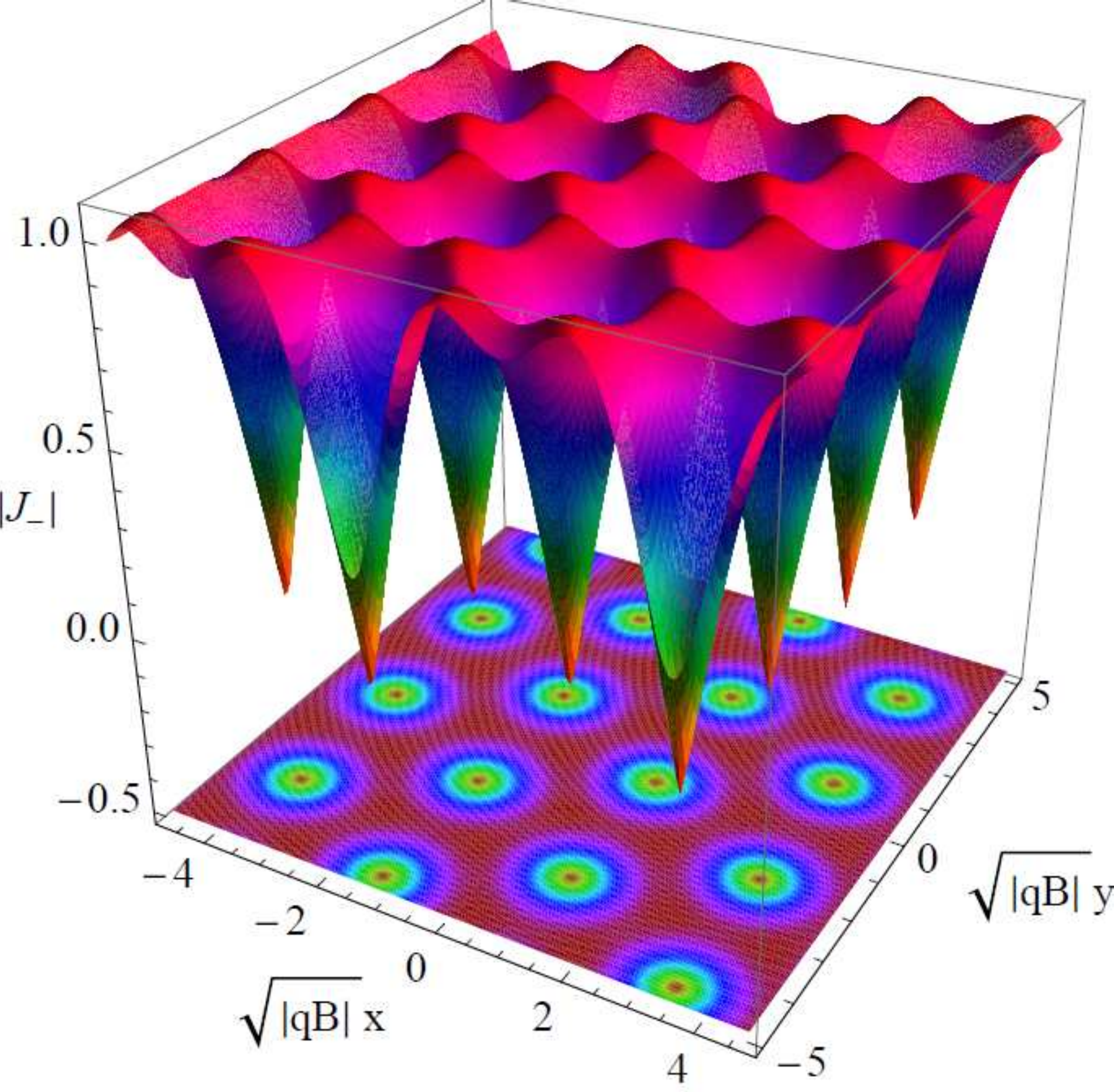}\caption{\label{veclattice} The vortex lattice structure for the triangular lattice in $x-y$ plane. The contour plot is also drawn in the bottom. In particular, the condensate vanishes in the core of each vortex. The figure was taken from ref.~\cite{Cai:2013pda}.}
\end{figure}

Due to the degeneracy in $p$, a linear superposition of the solutions with different $p$ is also a solution of the model at $\mathcal{O}(\epsilon)$. We can take this advantage to construct a class of vortex lattice solutions. As a typical example, the triangular lattice is shown in figure~\ref{veclattice}. It should be stressed that it is the special combinations $J_{\pm}=\langle\hat{J^x}\pm i\hat{J^y}\rangle$ which exhibit the vortex lattice structure. Strictly speaking, to obtain the true ground state, one should calculate the free energy of the solutions with different lattice structures from the action to find which configuration minimizes the free energy. It turns out that the linear analysis presented here is not sufficient to determine the most stable solution, thus should include higher order contributions.  Furthermore, it is worthwhile to mention that in the AdS soliton background, the external magnetic field triggered phase transition and vortex lattice structure also happen for the
vector field p-wave model~\cite{Cai:2013kaa}.

The response of this system to the magnetic field is quite different from the behaviour of ordinary superconductor where the magnetic field makes the transition more difficult.  But the result here is quite similar to the case of QCD vacuum instability induced by strong magnetic field to spontaneously developing the $\rho$-meson condensate~\cite{Chernodub:2010qx,Chernodub:2011mc}. Although so, it was shown that in model~\eqref{vecaction} the condensate of the vector operator forms a vortex lattice structure in the spatial directions perpendicular to the magnetic field. Of course, the non-minimal coupling term in the action plays a crucial role in both cases. Therefore in some sense, this model is a holographic setup of the study of $\rho$-meson condensate.
\subsubsection{The complete phase diagram}
The probe approximation neglecting the back reaction of the matter fields can indeed uncover many key properties. Nevertheless, it still loses some important information, such as the phase structure of the system. In the following paragraphs, we will discuss both the black hole background and soliton background in full back reaction case. Then a complete phase diagram in terms of temperature and chemical potential will be shown. We shall consider a $(4+1)$-dimensional bulk theory~\cite{Cai:2014ija}.

We would like to study a dual theory with finite chemical potential or charge density accompanied by a U(1) symmetry, so we turn on $A_t$ in the bulk. We want to allow for states with a non-trivial current $\langle\hat{J_x}\rangle$, for which we further introduce $\rho_x$ in the bulk. Because a non-vanishing $\langle\hat{J_x}\rangle$ picks out $x$ direction as special, which obviously breaks the rotational symmetry in spatial plane, thus we should introduce an additional function in the $xx$ component of the metric in order to describe the anisotropy. Therefore,
for the matter part, we consider the ansatz
\begin{equation}\label{vecmatter}
\rho_\nu dx^\nu=\rho_x(r)\,dx\;,\quad A_\nu dx^\nu=\phi(r)\,dt\;.
\end{equation}
We will consider black hole and soliton backgrounds separately.
\\
\\
{ \it (1)  AdS black hole with vector hair.}
\\
For the black hole background, we adopt the following metric ansatz
\begin{equation}\label{vecBHans}
ds^2=-a(r)e^{-b(r)}dt^2+\frac{dr^2}{a(r)}+r^2\,(c(r)dx^2+dy^2+dz^2)\;.
\end{equation}
The position of horizon is denoted as $r_h$ at which $a(r_h)=0$ and the conformal boundary is located at $r\rightarrow\infty$.
One finds that the $r$ component of Maxwell equations implies that the phase of $\rho_x$ must be constant. Without loss of generality, we can take $\rho_x$ to be real. Then, the independent equations of motion in terms of above ansatz are deduced as follows
\begin{equation}\label{vecBHeom}
\begin{split}
\phi''+(\frac{c'}{2c}+\frac{b'}{2}+\frac{3}{r})\phi'-\frac{2q^2\rho_x^2}{r^2ac}\phi=0,\\
\rho_x''+(\frac{a'}{a}-\frac{c'}{2c}-\frac{b'}{2}+\frac1r)\rho_x'+\frac{e^{b}q^2\phi^2}{a^2}\rho_x-\frac{m^2}{a}\rho_x=0,\\
b'-\frac{2a'}{a}-\frac{c'}{c}+\frac{2\rho_x'^2}{3rc}-\frac{re^b\phi'^2}{3a}-\frac{2e^b q^2\rho_x^2\phi^2}{3ra^2c}+\frac{8r}{a}-\frac{4}{r}=0,\\
c''+(\frac{a'}{a}-\frac{c'}{2c}-\frac{b'}{2}+\frac{3}{r})c'+\frac{2{\rho_x'}^2}{r^2}-\frac{2e^b q^2\rho_x^2\phi^2}{r^2a^2}+\frac{2m^2\rho_x^2}{r^2a}=0,\\
(\frac{3}{r}-\frac{c'}{2c})\frac{a'}{a}+(\frac{1}{r}+\frac{b'}{2})\frac{c'}{c}-\frac{\rho_x'^2}{r^2c}+\frac{e^b\phi'^2}{2a}+\frac{3e^b q^2\rho_x^2\phi^2}{r^2a^2c}-\frac{m^2\rho_x^2}{r^2ac}-\frac{12}{a}+\frac{6}{r^2}=0,
\end{split}
\end{equation}
where the prime denotes the derivative with respect to $r$.

When $\rho_x=0$, there exists an exactly analytical black hole solution, namely, AdS Reissner-Nordstr\"om black hole which reads
\begin{equation}\label{RNmetric}
\begin{split}
ds^2=-f(r)dt^2+\frac{dr^2}{f(r)}+r^2(dx^2+dy^2+dz^2),\\
f(r)=r^2\left[1-\left(1+\frac{\mu^2}{3r_h^2}\right)\left(\frac{r_h}{r}\right)^4+\frac{\mu^2}{3r_h^2}\left(\frac{r_h}{r}\right)^6\right], \quad \phi(r)=\mu\left[1-\left(\frac{r_h}{r}\right)^2\right].
\end{split}
\end{equation}
This solution is dual to a conductor phase in the dual field theory. However, the full coupled equations of motion do not admit an analytical solution with non-trivial $\rho_x$. Therefore, we have to solve them numerically. We will use shooting method to solve equations~\eqref{vecBHeom}. In order to find the solutions for all the five functions, i.e., $\rho_x(r),\phi(r), a(r), b(r)$ and $c(r)$ one must impose suitable boundary conditions  both at conformal boundary $r\rightarrow\infty$ and at the horizon $r=r_h$.

In order to match the asymptotical AdS boundary, the general falloff near the AdS boundary behaves as
\begin{equation} \label{vecboundarybh}
\begin{split}
\phi=\mu-\frac{\rho}{r^2}+\cdots,\quad \rho_x=\frac{{\rho_x}_-}{r^{{\Delta}_-}}+\frac{{\rho_x}_{+}}{r^{{\Delta}_+}}+\cdots,\\
a=r^2(1+\frac{a_4}{r^4})+\cdots,\quad c=1+\frac{c_4}{r^4}+\cdots,\quad b=0+\frac{b_4}{r^4}+\cdots,
\end{split}
\end{equation}
where the dots stand for the higher order terms in the expansion of $1/r$ and ${\Delta}_\pm=1\pm\sqrt{1+m^2}$.\footnote{The $m^2$ has a lower bound as $m^2=-1$ with ${\Delta}_+={\Delta}_-=1$. In that case, there exists a logarithmic term in the asymptotical expansion of $\rho_x$. One has to treat such a term as the source set to be zero to avoid the instability induced by this term~\cite{Horowitz:2008bn}. We will always consider the case with $m^2 >-1$.} In general, in the above expansion we must impose ${\rho_x}_-=0$, which  meets the requirement that the condensate appears spontaneously. According to the AdS/CFT dictionary, up to a normalization, the coefficients $\mu$, $\rho$, and ${\rho_x}_{+}$ are regarded as chemical potential, charge density and the $x$ component of the vacuum expectation value of the vector operator $\hat{J^\mu}$ in the dual field theory, respectively.

We focus on black hole configurations that have a regular event horizon located at $r_h$ and require the regularity conditions at the horizon $r=r_h$, which means that all five functions $\{\rho_x,\phi, a, b, c\}$ would have finite values at $r_h$ and admit a series expansion in terms of $(r-r_h)$. After substituting such series expansion into equations~\eqref{vecBHeom}, one finds there are only six independent parameters at the horizon, i.e.,  $\{r_h,\rho_x(r_h),\phi'_x(r_h),c(r_h),b(r_h)\}$ and other coefficients can be expressed in terms of those parameters.

Two free parameters $b(r_h)$ and $c(r_h)$ can be fixed by AdS boundary conditions that $b(r\rightarrow\infty)=0$ and $c(r\rightarrow\infty)=1$. Without loss of generality, the location of $r_h$ can be fixed to be one in our numerical calculation. We are then left with two independent parameters $\{\rho_x(r_h),\phi'(r_h)\}$. By choosing $\phi'(r_h)$ as the shooting parameter to match the source free condition, i.e., ${\rho_x}_-=0$, we finally have a one-parameter family of solutions labeled by the value of $\rho_x$ at the horizon. After solving the set of equations, we can read off the condensate $\langle \hat{J^x}\rangle$, chemical potential $\mu$ and charge density $\rho$ directly from the asymptotical expansion~\eqref{vecboundarybh}.
\\
\\
{\it (2) AdS soliton with vector hair.}

To construct homogeneous charged solutions with vector hair in the soliton background, we take the metric as
\begin{equation}\label{ansatz}
ds^2=\frac{dr^2}{r^2g(r)}+r^2(-f(r)dt^2+h(r)dx^2+dy^2+g(r)e^{-\chi(r)}d\eta^2)\;,
\end{equation}
where $g(r)$ vanishes at the tip $r=r_0$ of the soliton. The asymptotical AdS boundary is located at $r\rightarrow\infty$. Further, in order to obtain a smooth geometry at the tip $r_0$, $\eta$ should be made with an identification
\begin{equation}
\label{Gamma}
\eta\sim\eta+\Gamma\;,\qquad \Gamma=\frac{4\pi e^{\frac{\chi(r_0)}{2}}}{r_0^2 g'(r_0)}\;.
\end{equation}
This gives a dual picture of the boundary theory with a mass gap, which is reminiscent of an insulating phase.

The independent equations of
motion are deduced as follows
\begin{equation}\label{vecsleoms}
\begin{split}
\phi''-(\frac{f'}{2f}-\frac{g'}{g}-\frac{h'}{2h}+\frac{\chi'}{2}-\frac{3}{r})\phi'-\frac{2q^2\rho_x^2}{r^4gh}\phi=0,\\
\rho_x''+(\frac{f'}{2f}+\frac{g'}{g}-\frac{h'}{2h}-\frac{\chi'}{2}+\frac{3}{r})\rho_x'+\frac{q^2\phi^2}{r^4fg}\rho_x-\frac{m^2}{r^2g}\rho_x=0, \\
f''-(\frac{f'}{2f}-\frac{g'}{g}-\frac{h'}{2h}+\frac{\chi'}{2}-\frac{5}{r})f'-\frac{\phi'^2}{r^2}-\frac{2q^2\rho_x^2\phi^2}{r^6gh}=0,\\
\chi'-\frac{f'}{f}-\frac{2g'}{g}-\frac{h'}{h}+\frac{2{\rho_x'}^2}{3rh}-\frac{\phi'^2}{3rf}-\frac{2q^2\rho_x^2\phi^2}{3r^5fgh}+\frac{8}{rg}-\frac{8}{r}=0,\\
h''+(\frac{f'}{2f}+\frac{g'}{g}-\frac{h'}{2h}-\frac{\chi'}{2}+\frac{5}{r})h'+\frac{2{\rho_x'}^2}{r^2}-\frac{2q^2\rho_x^2\phi^2}{r^6fgh}+\frac{2m^2\rho_x^2}{r^4gh}=0,\\
(\frac{6}{r}-\frac{f'}{f}-\frac{h'}{h})\frac{g'}{g}+(\frac{f'}{f}+\frac{h'}{h})\chi'-\frac{f'h'}{fh}-\frac{2\rho_x'^2}{r^2h}+\frac{\phi'^2}{r^2f}+\frac{6q^2
\rho_x^2\phi^2}{r^6fgh}-\frac{2m^2\rho_x^2}{r^4gh}\\-\frac{24}{r^2g}+\frac{24}{r^2}=0,
\end{split}
\end{equation}
where the prime denotes the derivative with respect to $r$.
Similar to the black hole case, we will solve those coupled equations of motion numerically by use of shooting method. In order to find the solutions for all the six functions $\mathcal{F}=\{\rho_x,\phi,f,g,h,\chi\}$ one must impose suitable boundary conditions at both conformal boundary $r\rightarrow\infty$ and the tip $r=r_0$.

The asymptotical expansion for metric fields and matter fields near the boundary $r\rightarrow\infty$ is as follows
\begin{equation} \label{vecboundarysl}
\begin{split}
&\phi=\mu-\frac{\rho}{r^2}+\cdots,\quad \rho_x=\frac{{\rho_x}_-}{r^{{\Delta}_-}}+\frac{{\rho_x}_{+}}{r^{{\Delta}_+}}+\cdots,\quad f=1+\frac{f_4}{r^4}+\cdots\\
&g=1+\frac{g_4}{r^4}+\cdots,\quad h=1+\frac{h_4}{r^4}+\cdots,\quad \chi=0+\frac{\chi_4}{r^4}+\cdots,
\end{split}
\end{equation}
where the dots stand for the higher order terms of $1/r$. We choose the source free condition ${\rho_x}_-=0$ as before. The coefficients $\mu$, $\rho$, and ${\rho_x}_{+}$ are directly related to the chemical potential, charge density and $x$ component of the vacuum expectation value of the vector operator $\hat{J^\mu}$ in the dual system, respectively.

We impose the regularity conditions at the tip $r=r_0$, which means that all functions have finite values and admit a series expansion in terms of $(r-r_0)$ as
\begin{equation}\label{vecseries}
\mathcal{F}=\mathcal{F}(r_0)+\mathcal{F}'(r_0)(r-r_0)+\cdots.
\end{equation}
By plugging the expansion~\eqref{vecseries} into~\eqref{vecsleoms}, one can find that there are six independent parameters at the tip $\{r_0,\rho_x(r_0),\phi(r_0),f(r_0),h(r_0),\chi(r_0)\}$. However, there exist four useful scaling symmetries in the equations of motion, which read
\begin{equation} \label{vecscaling1}
\chi\rightarrow \chi+\lambda,\quad \eta\rightarrow e^{\lambda/2}\eta\;,
\end{equation}
\begin{equation} \label{vecscaling2}
\phi\rightarrow\lambda \phi,\quad t\rightarrow\lambda^{-1} t,\quad f\rightarrow\lambda^2 f\;,
\end{equation}
\begin{equation} \label{vecscaling3}
\rho_x\rightarrow\lambda \rho_x,\quad x\rightarrow\lambda^{-1} x,\quad h\rightarrow\lambda^2 h\;,
\end{equation}
and
\begin{equation} \label{vecscaling4}
r\rightarrow\lambda r,\quad (t,x,y,\eta)\rightarrow{\lambda^{-1}}(t,x,y,\eta),\quad(\phi,\rho_x)\rightarrow\lambda(\phi,\rho_x)\;,
\end{equation}
where in each case $\lambda$ is a real positive constant.

By using above four scaling symmetries, we can first set $\{r_0=1,f(r_0)=1,h(r_0)=1,\chi(r_0)=0\}$ for performing numerics. After solving the coupled differential equations, one should use the first three symmetries again to satisfy the asymptotic conditions $f(\infty)=1$, $h(\infty)=1$ and $\chi(\infty)=0$. We choose $\phi(r_0)$ as the shooting parameter to match the source free condition, i.e., ${\rho_x}_-=0$. Finally, for fixed $m^2$ and $q$, we have a one-parameter family of solutions labeled by $\rho_x(r_0)$. After solving the set of equations, we can read off the condensate $\langle \hat{J^x}\rangle$, chemical potential $\mu$ and charge density $\rho$ from the corresponding coefficients in~\eqref{vecboundarysl}. It should be noticed that different solutions obtained in this way will have different periods $\Gamma$ for $\eta$ direction. We should use the last scaling symmetry to set all of the periods $\Gamma$ equal in order to obtain same boundary geometry. We shall fix $\Gamma$ to be $\pi$ in this section.

\begin{figure}[h!]
\includegraphics[width=0.5\textwidth]{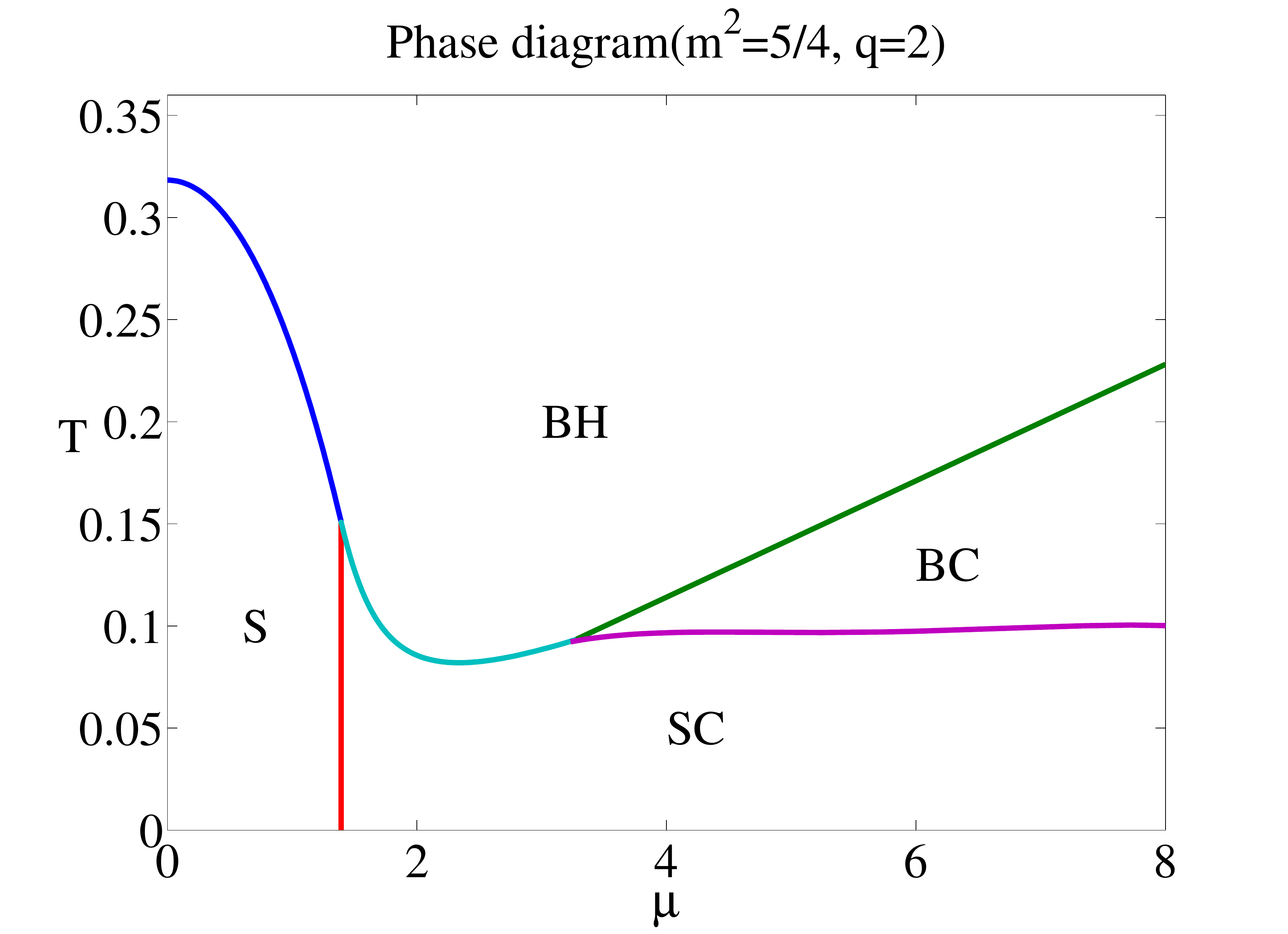}
\includegraphics[width=0.5\textwidth]{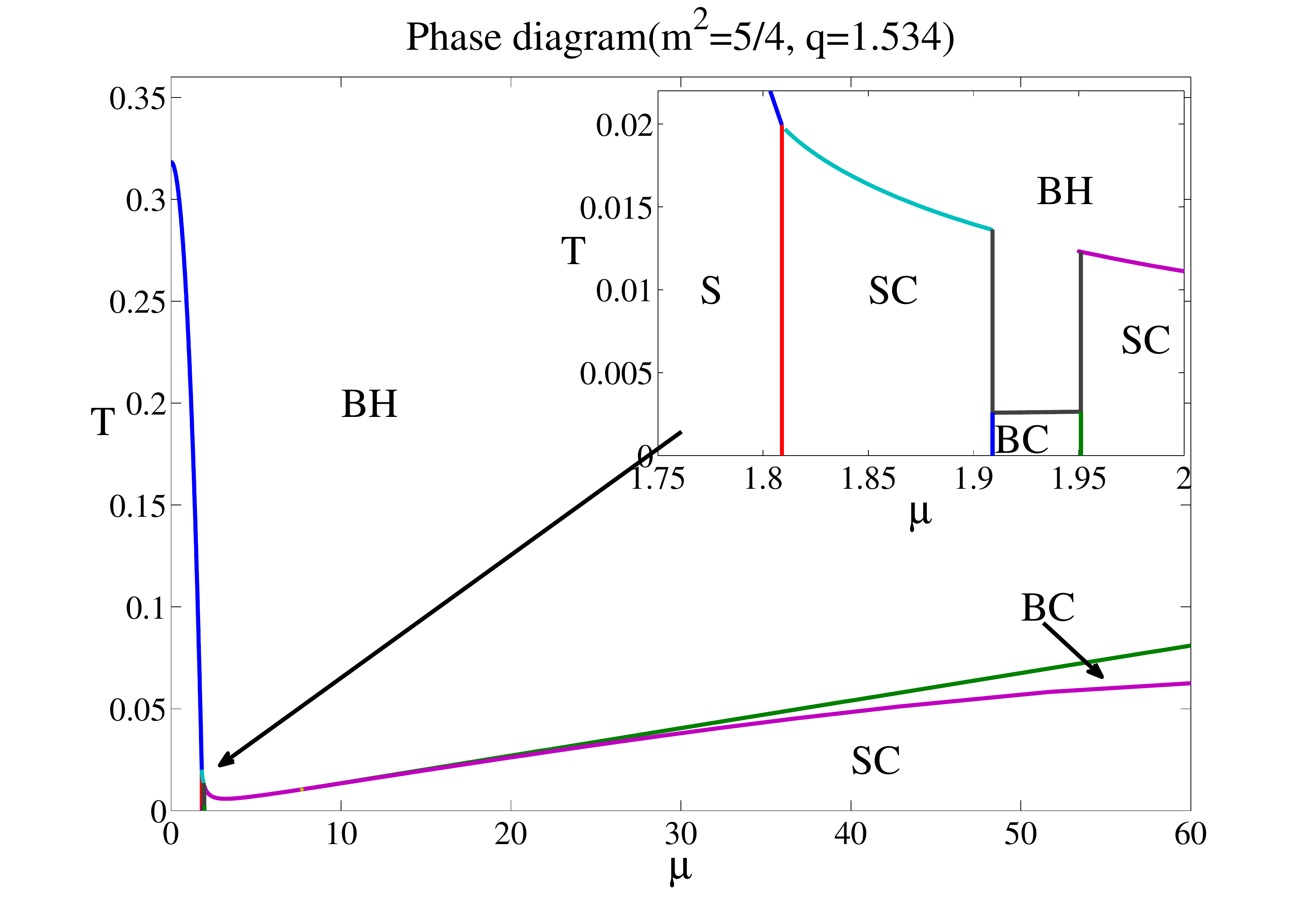}\\
\includegraphics[width=0.5\textwidth]{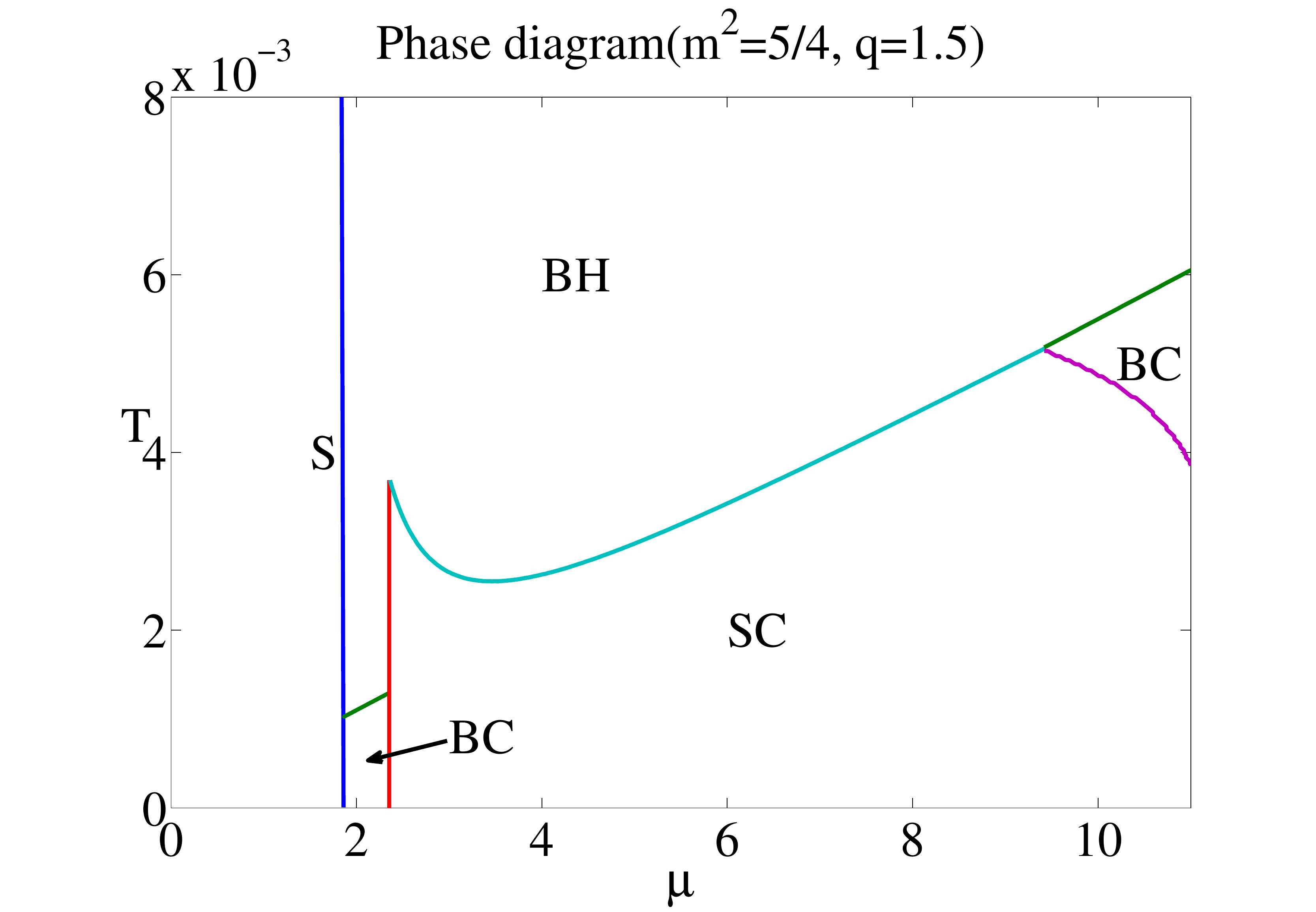}
\includegraphics[width=0.5\textwidth]{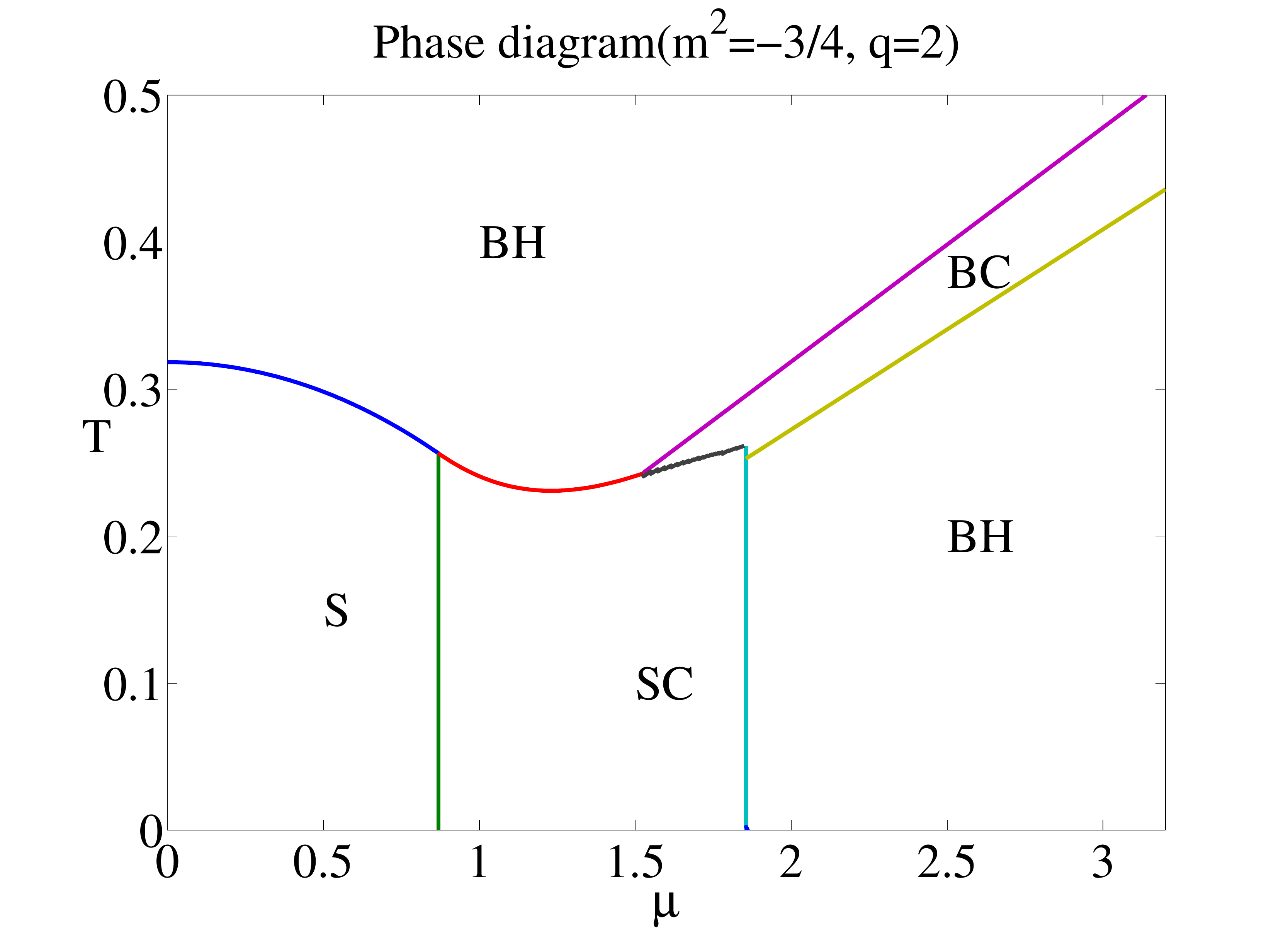}\\
\includegraphics[width=0.5\textwidth]{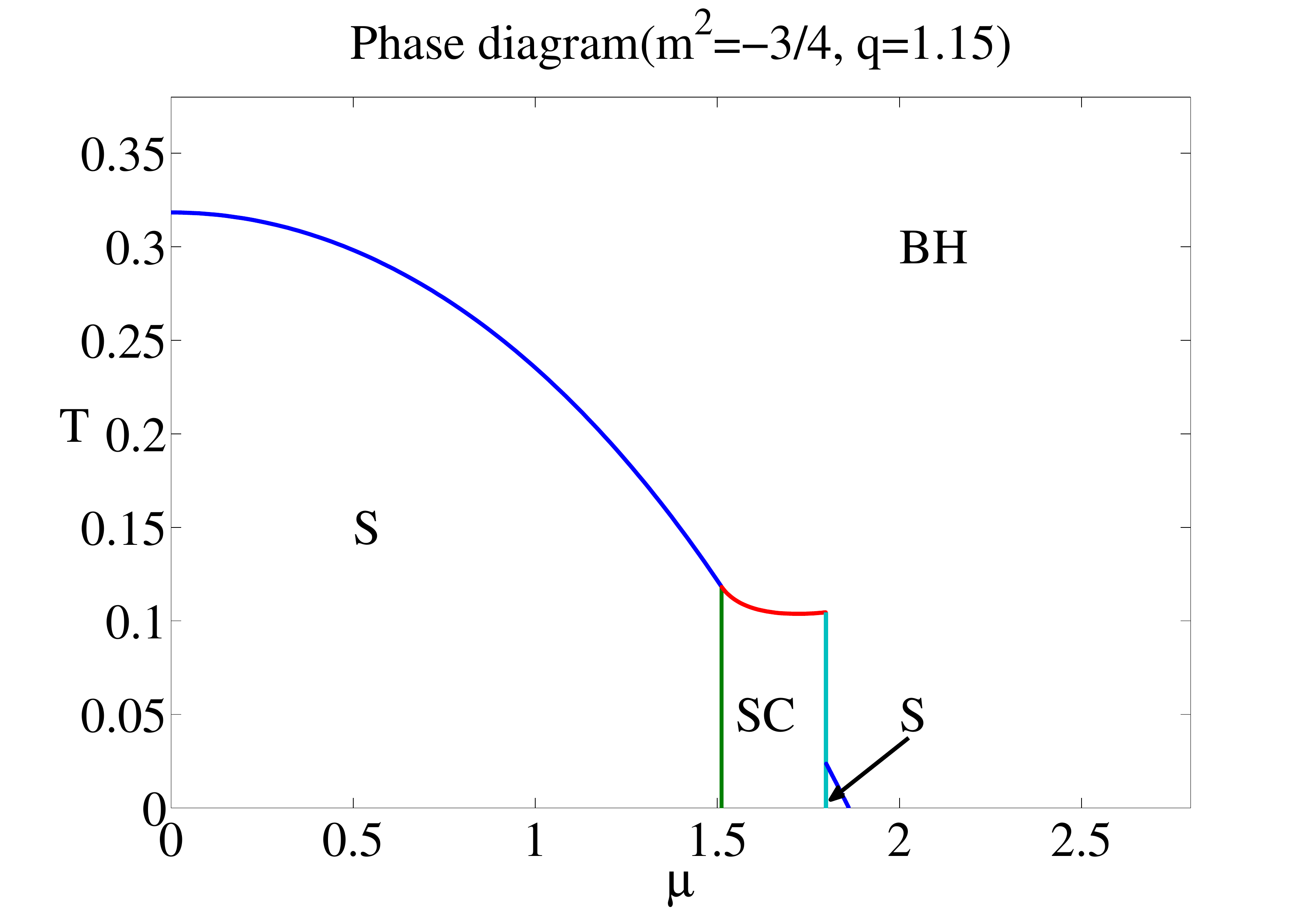}
\includegraphics[width=0.5\textwidth]{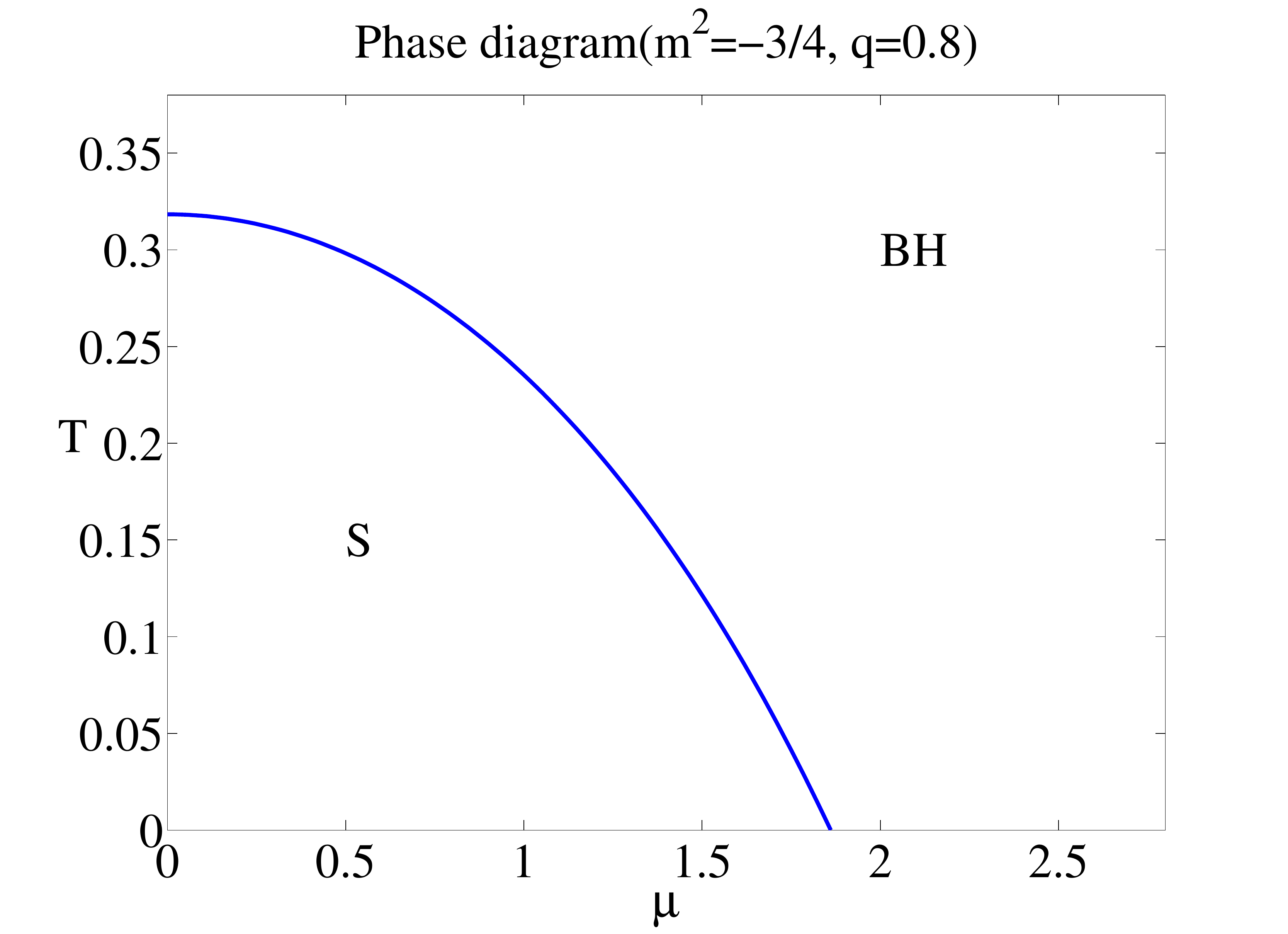}\\
\caption{\label{completephase} The complete phase diagrams of the Maxwell-vector model with S=pure AdS soliton, BH=AdS Reissner-Nordstr\"om black hole, SC=hairy soliton, and BC=hairy black hole. In each region the thermodynamically stable phase is labeled. As $m^2$ and $q$ are changed, the shape of each region gets modified. This figure is described in ref.~\cite{Cai:2014ija} in more detail.}
\end{figure}

These two kinds of situations have been well studied in ref.~\cite{Cai:2014ija}. There are four different bulk solutions given by the pure AdS soliton, AdS Reissner-Nordstr\"om black hole and their vector hairy counterparts. According to the AdS/CFT dictionary, the hairy solution is dual to a system with a non-zero vacuum expectation value of the charged vector operator which breaks the U(1) symmetry and the spatial rotation symmetry spontaneously. The above four solutions in the bulk correspond to an insulating phase, a conducting phase, a soliton superconducting phase and a black hole superconducting phase, respectively. Since we do not turn on magnetic field, the model is left with two independent parameters, i.e., the mass $m$ of the vector field giving the scaling dimension of the dual vector operator and its charge $q$ controlling the strength of the back reaction on the background geometry. The phase structure of the model heavily depends on those two parameters. There exist second order, first order and zeroth order\footnote{In the theory of superfluidity and superconductivity, a discontinuity of the free energy was discussed theoretically and an exactly solvable model for such phase transition was given
in ref.~\cite{Maslv:2004}. The zeroth order transition was also observed in holographic superconductors in refs.~\cite{Zeng:2013yoa,Zeng:2014dra,Chaturvedi:2015hra}.} phase transitions as well as the ``retrograde condensation" in which the hairy solutions exist only above a critical temperature or below a critical chemical potential with the free energy much larger than the solutions without hair.

With four kinds of phases at hand, the complete phase diagrams can be constructed in terms of temperature and chemical potential. At each point in $T$-$\mu$ plane, one should find the phase which has the lowest free energy. Since there are many types of phase transitions in both soliton and black hole backgrounds, the $T$-$\mu$ phase diagrams are expected to be much more complicated than the holographic s-wave model~\cite{Horowitz:2010jq} and the Yang-Mills p-wave model~\cite{Akhavan:2010bf}. Some typical examples are shown in figure~\ref{completephase}. We can see from the complete phase diagrams that in some cases, more than one superconducting phase appears in a phase diagram in the model. The phase diagrams for some realistic superconducting materials are usually complicated, and indeed, more than one superconducting phase can occur, for example, see refs.~\cite{Kordyuk,Chubukov,Yuan}. Definitely, it is of great interest to see whether this model is relevant to those superconducting materials.

\subsection{The Helical P-wave model}
\label{sect:helical}
The gravity solutions above mainly describe spatially homogeneous superconducting states. However, it has long been known that it is possible to have superconducting states that are spatially inhomogeneous. A well known example is the Fulde-Ferrell-Larkin-Ovchinnikov (FFLO) phase, for which a Cooper pair consisting of two fermions with different
Fermi momenta condenses leading to an order parameter with non-vanishing total momentum~\cite{Fulde:1964zz,larkin:1964zz}. In this section, we shall introduce a holographic model which can realize p-wave superconducting phase with a helical order. That is to say, the order parameter points in a given direction in a plane which then rotates as one moves along the direction orthogonal to the plane.

We consider a (4+1) dimensional model with  a gauge field $A_\mu$ and a charged two-form $C_{\mu\nu}$~\cite{Donos:2011ff,Donos:2012gg}
\begin{equation}\label{eomhelical}
\begin{split}
S=\int d^5 x\sqrt{-g}\big[\mathcal{R}+12&-\frac{1}{4}F_{\mu\nu}F^{\mu\nu}-\frac{1}{4}C^{\mu\nu}C^\dagger_{\mu\nu}+\frac{{i}}{24m}\epsilon^{\mu\nu\rho\sigma\lambda}C_{\mu\nu}
H^\dagger_{\rho\sigma\lambda}\big]\,,
\end{split}
\end{equation}
where we have chosen units where the AdS radius is unity, a dagger denotes complex conjugation and the field strengths read
\begin{equation}\label{ders}
F=dA\;,\qquad H=dC+{i} e\, A\wedge C\;.
\end{equation}

The gauge field $A_\mu$ is dual to a current in the dual theory and the two-form $C_{\mu\nu}$ corresponds to a self-dual rank two tensor operator with scaling dimension $\Delta=2+|m|$. In particular, this charged operator has angular momentum $l=1$ and thus can serve as an order parameter for $p$-wave superconductors. Since what we are interested in is a system at finite temperature and chemical potential with respect to the global U(1) symmetry, we will construct electrically charged asymptotically AdS black holes in gravity side. The normal phase with no condensate is described by the electrically charged Reissner-Nordstr\"om AdS black hole, which is spatially homogeneous and isotropic. This model is specified by two parameters $m$ and $e$. It was shown in ref.~\cite{Donos:2011ff} that when $e^2>m^2/2$ this black hole is unstable to developing non-trivial two-form hair that is dual to p-wave superconductors with helical order.

\subsubsection{Boundary conditions}
The helical black hole solution was constructed in ref.~\cite{Donos:2012gg} in which the authors adopted the ansatz
\begin{equation}\label{helicalansatz}
\begin{split}
ds^{2}&=-g\,f^{2}\,dt^{2}+g^{-1}{dr^{2}}+h^{2}\,\omega_{1}^{2}+r^{2}\,\left(e^{2\alpha}\,\omega_{2}^{2}+e^{-2\alpha}\,\omega_{3}^{2}\right),\\
C&=(i\, c_{1}\,dt+c_2 dr)\wedge\omega_{2}+c_{3}\,\omega_{1}\wedge\omega_{3}\;, \qquad A=a\, dt\;,
\end{split}
\end{equation}
where the one-forms $\omega_i$ are given by
\begin{equation}\label{helicalforms}
\begin{split}
&\omega_{1}=dx_{1}\;,\\
&\omega_{2}=\cos\left(kx_{1}\right)\,dx_{2}-\sin\left(kx_{1}\right)\,dx_{3}\;,\\
&\omega_{3}=\sin\left(kx_{1}\right)\,dx_{2}+\cos\left(kx_{1}\right)\,dx_{3}\;.
\end{split}
\end{equation}
Note that the constant $t$ and $r$ slices in the above metric are spatially homogeneous of Bianchi type $\text{VII}_0$. All eight functions in the ansatz  depend on the radial coordinate $r$ only and $k$ is a constant. After substituting the ansatz into the action~\eqref{eomhelical}, one finds that $c_1$ and $c_2$ can be determined by other functions, thus we are left with six independent functions including $f$, $g$, $h$, $\alpha$, $c_{3}$ and $a$. More precisely, $f$ and $g$ satisfy first order differential equations and other functions satisfy second order equations.

To solve the coupled equations of motion for above six functions, one needs to specify suitable boundary conditions in the horizon $r_h$ and the conformal boundary $r\rightarrow\infty$. Regularity at the horizon demands that $g(r_h)=a(r_h)=0$ and all of them have analytic expansion in terms of $(r-r_h)$. We then find that the full expansion at the horizon is fixed by six parameters, i.e., $r_h, f(r_h), h(r_h), \alpha(r_h), a'(r_h)$ and $c_3(r_h)$. Near the boundary $r\rightarrow\infty$, one demands asymptotically AdS geometry with the fall-off
\begin{eqnarray}\label{uvexpan}
\begin{split}
&g=r^{2}\,\left(1-{M}{r^{-4}}+\cdots \right),\quad f=f_{0}\left(1-{c_{h}}{r^{-4}}+\cdots\right),\\
&h=r\,\left(1+{c_{h}}{r^{-4}}+\cdots \right),\quad \alpha={c_{\alpha}}{r^{-4}}+\cdots,\\
&a=f_{0}\,\left(\mu-{\rho}{r^{-2}}/2+\cdots\right),\quad c_{3}={c_{v}}{r^{-\left|m\right|}}+\cdots,
\end{split}
\end{eqnarray}
which is determined by eight parameters $M, f_0, c_h, c_\alpha, \mu, \rho, c_v$ and $k$. One should note that the expansion of $c_3$ is chosen so that the charged operator dual to the two-form $C$ has no source, thus can spontaneously acquire an expectation value proportional to $c_v$ which is spatially modulated in the $x_1$ direction with period $2\pi/k$. $\mu$ and $\rho$ are regarded as the chemical potential and charge density in the dual system respectively. The holographic interpretation of the other UV parameters will be given below.
Observe that when $k\neq0$ the order parameter rotates in the $(x_2, x_3)$ plane as one moves along the $x_1$ direction thus there is a reduced helical symmetry.

There are two scaling symmetries of the coupled equations which can be used to set $\mu=f_0=1$. To solve the six differential equations, we need to specify ten integration constants. However, we have fourteen parameters in two boundaries minus two for the scaling symmetries. Therefore, we expect to leave with a two parameter family of black hole solutions
which can be selected as temperature $T$ and wave number $k$.
\subsubsection{Thermodynamics}
\label{sect:thermo}
We shall work in grand canonical ensemble with the chemical potential $\mu$ fixed. The thermodynamic potential of the boundary thermal state is identified with temperature $T$ times the on-shell bulk action in Euclidean signature. We denote $w$ as the density of thermodynamic potential per spatial volume in dual field theory. Then one can obtain the following expression for the free energy density~\footnote{For more details about this result, please see ref.~\cite{Donos:2013woa}.}
\begin{equation}
w=-M=3M+8c_h-\mu\rho-s\,T\;,
\end{equation}
where the entropy density $s=4\pi r_h^2 h(r_h)$ and $f_0$ is set to be one. From above equation one can immediately obtain the Smarr-type formula
\begin{equation}
4M+8c_h-\mu\rho-s\,T=0\;.
\end{equation}
An on-shell variation of the total action for fixed $k$ gives us the first law
\begin{equation}
\delta w=-s\delta T-\rho\delta\mu,
\end{equation}
and hence $w=w(T, \mu)$.

The expectation value of the dual stress-energy tensor is given, after setting $f_0=1$, by
\begin{equation}
\begin{split}
&T_{tt}=3M+8c_{h},\quad
T_{x_{1}x_{1}}=M+8c_{h},\\
&T_{x_{2}x_{2}}=M +8c_{\alpha}\,\cos\left(2kx_{1} \right) ,\\
&T_{x_{3}x_{3}}=M -8c_{\alpha}\,\cos\left(2kx_{1} \right) ,\\
&T_{x_{2}x_{3}}=-\,8\,c_{\alpha}\,\sin\left(2kx_{1}\right).
\end{split}
\end{equation}
Obviously the stress-energy tensor is traceless as a consequence of the underlying conformal symmetry. We further extract the energy density $\varepsilon=T_{tt}=3M+8c_{h}$ from which we can rewrite $w=\varepsilon-s T-\mu\rho$ and the first law takes the more familiar form $\delta \varepsilon=T \delta s+\mu\delta\rho$.
The average hydrostatic pressure $\bar p$ is defined as minus the average of the trace of the spatial components. We get $\bar p=M+8c_h/3$, and hence the system satisfies the thermodynamical relation $\varepsilon+\bar p=Ts+\mu\rho$.
\subsubsection{Helical p-wave solutions}
\label{sect:helicalsolu}
We focus on the specific case with $m=1.7$ and $e=1.88$~\footnote{The main reason for this choice is to obtain real scaling dimensions. For other values of $m, e$ which can avoid complex scaling dimensions will give similar results~\cite{Donos:2013woa}.} and set $\mu=f_0=1$. Starting from the AdS Reissner-Nordstr\"om black hole solution, as the temperature is lowered, the first instability appears at $T_c\simeq0.0265$ and $k=k_c\simeq0.550$. Below $T_c$, there is a continuum of hairy black hole solutions appearing with different values of $k$.

\begin{figure}[h!]
\centering
\includegraphics[scale=0.4]{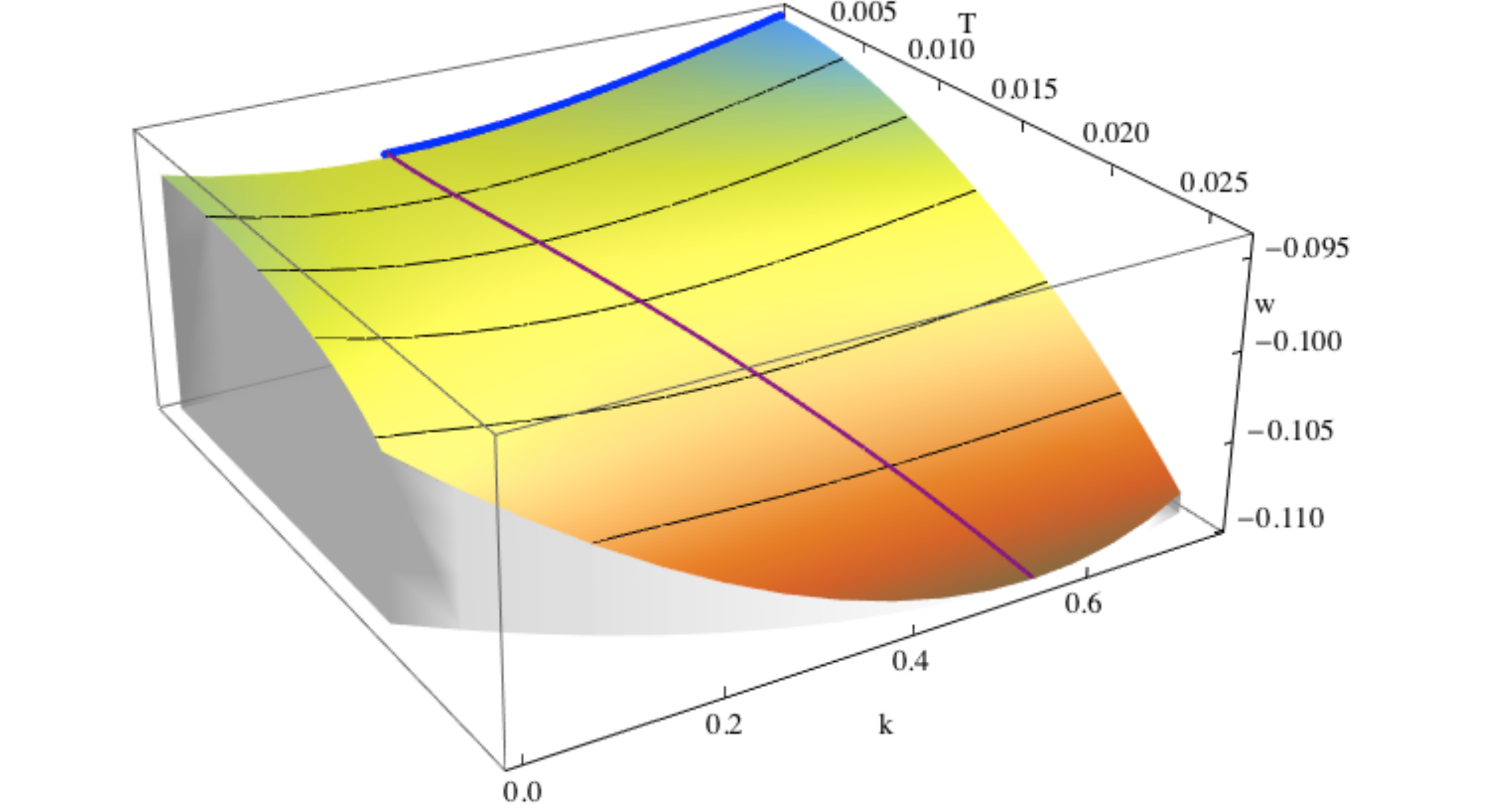}
\caption{The free energy density as a function of $T$ and $k$ for the case of $m=1.7$ and $e=1.88$. The red curve denotes the thermodynamically favored p-wave superconducting phase minimizing the free energy with respect to $k$ at fixed $T$. The black curves correspond to curves with constant $T$. The blue line is for the case of some domain wall solutions. Reprinted with kind permission from ref.~\cite{Donos:2012gg}.
\label{helicalfig1}}
\end{figure}

Figure~\ref{helicalfig1} summaries the free energy density $w$ as a function of temperature $T$ and wave number $k$. One can see that all hairy solutions have smaller free energy than the normal solutions at the same temperature and the transition to the p-wave preferred branch is second order. For a given temperature $T<T_c$, there is a one parameter family of solutions specified by $k$, and the most thermodynamically preferred solution is denoted by the red line. One can prove that while the general hairy solutions in figure~\ref{helicalfig1} have $c_h\neq0$, the solutions on the red line do have vanishing $c_h$~\cite{Donos:2013woa}.

\begin{figure}[h!]
\centering
\includegraphics[scale=0.85]{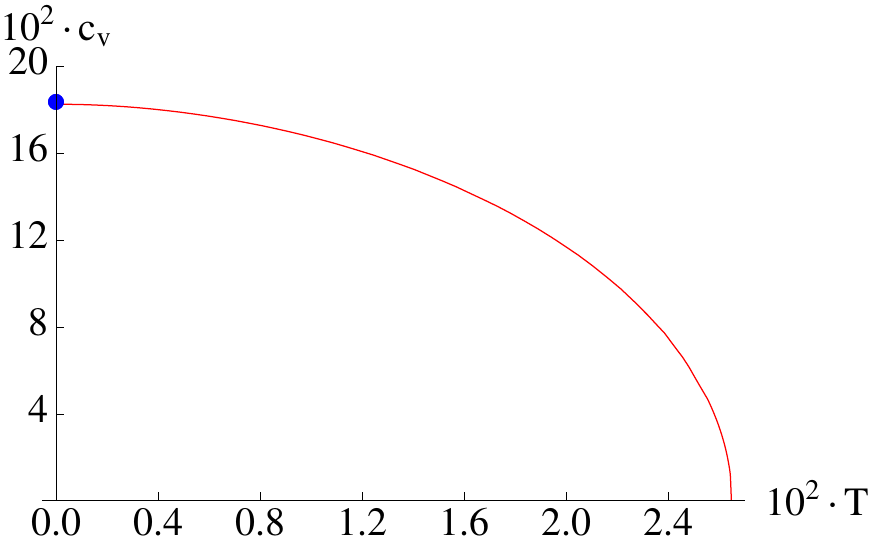}
\includegraphics[scale=0.85]{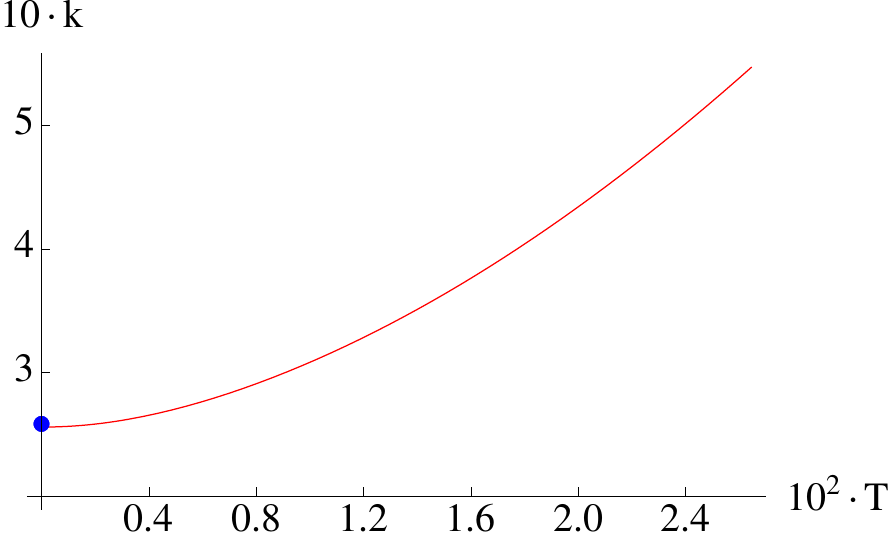}
\caption{The condensate $c_v$ and wave number $k$ as a function of $T$ for the thermodynamically preferred hairy solutions. The blue dots represent the quantities for the domain wall solutions. Used with permission from ref.~\cite{Donos:2012gg}.
\label{helicalfig2}}
\end{figure}

The helical superconducting order can be fixed by the condensate $c_v$ and wave number $k$, which are shown in figure~\ref{helicalfig2} with respect to $T$ for the red line in figure~\ref{helicalfig1}. Near $T_c$ one can find the critical phenomenon $c_{v}\simeq1.7\times 10^5 T_c^{3.7}\left(1-T/T_{c} \right)^{1/2}$, which is the famous mean field behaviour. As the temperature is lowered, the red line moves smoothly down to sufficiently low temperature at which $k\equiv k_0\simeq0.256$. In particular, the ground state at $T=0$ is also spatially modulated.

The $T=0$ limit of hairy solutions approach a smooth domain wall solution which interpolates between $AdS_5$ in the UV and a new IR fixed point with an anisotropic scaling. This fixed point in the IR reads
\begin{equation}
\begin{split}
&g=K\,r^{2},\quad f=\bar f_0r^{z-1},\quad h=k h_{0}\,,\quad  \alpha=\alpha_{0}\,,\\
&a=a_{0}r^{z},\quad
c_{3}=k c_{0}\,r\,,
\end{split}
\end{equation}
with $K, h_0, \alpha_0, a_0, c_0$ and $z$ all constants. This fixed point solution is invariant under the anisotropic scaling $r\rightarrow\lambda^{-1}r$, $t\rightarrow \lambda^{z}t$, $x_{2,3}\rightarrow\lambda x_{2,3}$ and $x_1\to x_1$. All those constants can be determined by the equations of motion.\footnote{Note that one can set $\bar f_0=k=1$ by scaling $t$ and $x_1$.} As a typical example, choosing $m=1.7$ and $e=1.88$, one can obtain
\begin{equation}
\begin{split}
&z\simeq1.65 ,\qquad K\simeq 0.995,\qquad h_{0}\simeq0.993,\\
&\alpha_{0}\simeq-0.380,\quad a_{0}\simeq0.265, \quad ~c_{0}\simeq 3.69.
\end{split}
\end{equation}

The domain wall solutions interpolating between the UV fixed point and the IR fixed point can be specified by the wave number $k$~\cite{Donos:2012gg}. One can see in figure~\ref{helicalfig1} that the $T\rightarrow0$ limit of the hairy solutions approach these domain wall solutions (the blue line). Similarly, in figure~\ref{helicalfig2}, the condensate $c_v$ and wave number $k= k_0\simeq0.256$ for the domain wall denoted by blue dots smoothly connect with the corresponding black hole solution.

To summarize, a holographic p-wave model with helical superconducting order is introduced in this subsection. As the temperature is lowered, a helical superconducting state emerges spontaneously breaking both the abelian symmetry and the three-dimensional spatial Euclidean symmetry down to Bianchi $\text{VII}_0$ symmetry.
These homogeneous, but anisotropic ground states at $T=0$ are holographically described by smooth domain wall solutions, which exhibit zero entropy density and an emergent scaling symmetry in the far IR.

Further nature of the model~\eqref{eomhelical} has been well studied in ref.~\cite{Donos:2013woa}. For example, some of the p-wave solutions can exhibit the phenomenon of pitch inversion\footnote{As the temperature is lowered, the pitch $(2\pi/k)$ first increases, becoming divergent (i.e., $k=0$) at some particular temperature, then changes sign and finally decreases in magnitude to a value $k<0$ at $T=0$.} and the symmetry of the black hole solutions is enhanced at the pitch inversion temperature. The superconducting phase can also be
$(p+ip)$ order. Depending on the mass and charge of the two-form, both the p-wave and the $(p+ip)$-wave can be thermodynamically favored. The two kinds of orders will compete with each other and there can be first order transition between them.

\section{Holographic D-wave Models}
\label{sect:dwave}
It is remarkable to see that rather simple and generic gravity models can capture many features of the phase structure of superconducting systems. Nevertheless, in order to construct more sophisticated and more realistic models one clearly needs to include additional ingredients. The focus of this part is on realising an important missing phase, i.e. d-wave superconductivity (superfluidity). The importance is self-evident since many unconventional superconductors admit either d-wave or mixed symmetry. A natural candidate for modelling the d-wave condensate is to use a charged spin two field in the bulk, instead of a charged scalar field or a vector field. Based on this approach, there are two acceptable holographic models describing the d-wave condensate in the literature.

The authors of ref.~\cite{Chen:2010mk} first constructed a minimal gravitational model by introducing a symmetric, traceless rank-two tensor field minimally coupled to a U(1) gauge field in the background of an AdS black hole. The d-wave condensate appears below a critical temperature via a second order phase transition, resulting in an isotropic  superconducting phase but no hard gap for its optical conductivity. Let us call it CKMWY d-wave model in terms of the initials of the five authors. The other effective holographic d-wave model was proposed soon after the first one with the same matter fields but with much more complex interactions~\cite{Benini:2010pr}. The phase diagram, optical conductivity, as well as fermionic spectral function were investigated in detail. With a fixed metric, this model has advantages such as being ghost-free and having the right propagating degrees of freedom. This model will be named as BHRY d-wave model for short in what follows.
\subsection{The CKMWY d-wave model}
\label{subsect:CKMWY model}
To construct a holographic d-wave model, the minimal effective bulk action including gravity, U(1) gauge field and tensor field reads~\cite{Chen:2010mk}
\begin{eqnarray}\label{wwdwave}
S =\frac{1}{2\kappa ^{2}}\int d^{4}x\sqrt{-g}\left[\mathcal{R}+\frac{6}{L^{2}}-(D_{\mu }B_{\nu \gamma
})^{^\dagger }D^{\mu }B^{\nu \gamma }-m^{2}B_{\mu \nu }^{^\dagger}B^{\mu \nu }-
\frac{1}{4}F_{\mu \nu }F^{\mu \nu }\right],
\end{eqnarray}%
where $D_\mu=\nabla_\mu-iq A_\mu$ is the covariant derivative in the black hole background, $L$ is the AdS radius that will be set to unity, and $q$ and $m^{2}$ are the charge and mass
squared of $B_{\mu \nu }$, respectively. Working in the probe limit, i.e. $q\rightarrow\infty$ with $q A_\mu$ and $qB_{\mu \nu }$ fixed, the matter part can be treated as perturbations in the 3+1 dimensional AdS black hole background~\eqref{AdSswtz}.

We would like to realize a d-wave superconductor on the boundary such that a condensate emerges on the $x-y$ plane with translation invariance and the rotational symmetry is broken down to Z(2) with the condensate changing its sign under a $\pi/2$ rotation on the $x-y$ plane. Therefore, we use an ansatz for $B_{\mu \nu }$
and $A_{\mu }$ as
\begin{eqnarray}
B_{xx}=-B_{yy}=\psi(r)\;,\quad\quad A=\phi(r)\,dt\;,
\end{eqnarray}%
with all other field components being turned off and $\psi(r)$ and $\phi(r)$ being real functions. The background geometry is fixed as AdS-Schwarzschild black hole given in~\eqref{AdSswtz}. Then the final equations of motion read
\begin{equation}
\begin{split}
\psi''+(\frac{f'}{f}-\frac{2}{r})\psi'-\left(\frac{2f'}{rf}+\frac{m^2}{f}-\frac{q^2\phi^2}{f^2}\right)\psi=0\;,\\
\phi''+\frac{2}{r}\phi'-\frac{4 q^2 \psi^2}{r^4f}\phi=0\;.
\end{split}
\end{equation}
These two equations are very similar as the case for the Abelian-Higgs model (see equations~\eqref{higsEOMs}). Therefore, it is natural to expect $\psi$ to condense spontaneously below a critical temperature. More precisely, we demand the following asymptotic form near the AdS boundary $r\rightarrow\infty$
\begin{eqnarray}
\phi=\mu-\rho/r+\cdots,\quad \psi=f_1 r^{\Delta_-}+\cdots,
\end{eqnarray}%
with $\Delta_-=\frac{1- \sqrt{17+4m^{2}}}{2}$.  Note that the expansion of $\psi$ is chosen such that the charged operator dual to $B_{\mu\nu}$ has no source, thus can acquire an expectation value proportional to $f_1$ spontaneously. According to holographic dictionary, $\mu$ is interpreted as the chemical potential, and $\rho$ as the charge density in the dual theory. The order parameter of the boundary theory can be obtained by reading the asymptotic behaviour of $B$, i.e.
\begin{equation}
\langle \mathcal{O}_{ij}\rangle =\left(
\begin{array}{cc}
f_{1} & 0 \\
0 & -f_{1}%
\end{array}%
\right),
\end{equation}
where $(i,j)$ are the indexes in the boundary coordinates $(x,y)$. In what follows, we shall keep the chemical potential $\mu$ fixed and choose $q$ to be minus one, which is the setup adopted by ref.~\cite{Chen:2010mk}.

\begin{figure}[h]
\centering
\includegraphics[scale=0.9]{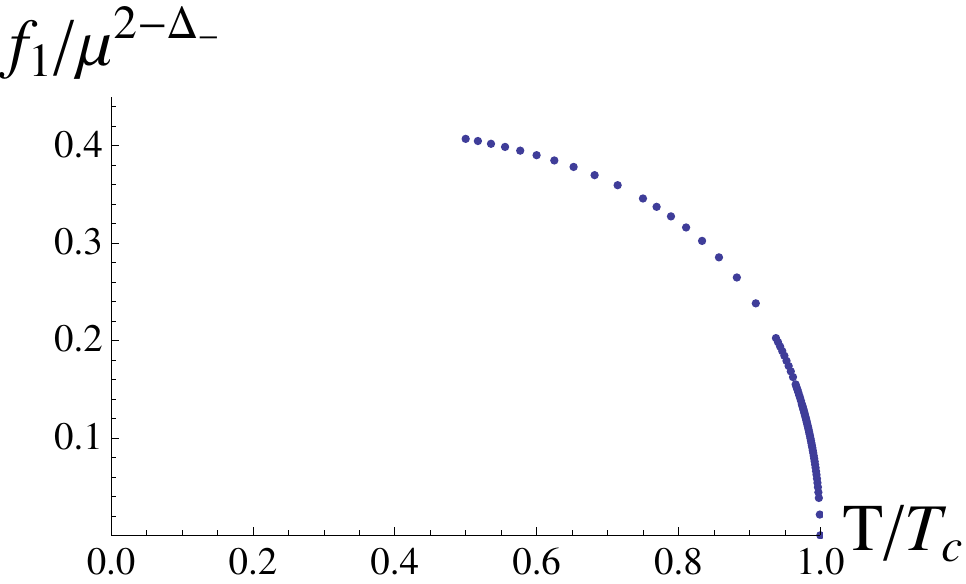}
\caption{\label{dcondenw} The d-wave condensate as a function of temperature. The condensate goes to zero at the critical temperature $T_c$. We choose $m^2=-1/4$. Used with permission from ref.~\cite{Chen:2010mk}.}
\end{figure}

The d-wave condensate as a function of temperature can be obtained numerically, which is shown in figure~\ref{dcondenw}. One can see clearly that below $T_c$, the tensor field is Higgsed to break the U(1) symmetry spontaneously in the boundary theory. Numerical calculation further ensures that the phase transition characterized by the d-wave condensate is second order with the mean field critical behaviour $f_1\sim (T_c-T)^{1/2}$. The conductivity has also been studied, which uncovered that the AC conductivity is isotropic and below $T_c$, the DC conductivity becomes infinite but has no hard gap.

\subsection{The BHRY d-wave model}
\label{subsect:BHRY model}
The approach of the CKMWY d-wave model just writes down a minimal action for the spin two field without looking in detail at the constraint equations required to get the correct number of propagating degrees of freedom. Soon, the authors of ref.~\cite{Benini:2010pr} analyzed in more detail the effective action for the spin two field and how the constraint equations could be satisfied.
The desired theory for a charged, massive spin two field in a fixed Einstein background takes the following form
\begin{equation}\label{dwBHRY}
\begin{split}
S=\frac{1}{2\kappa^2}\int d^{d+1}x\sqrt{-g}(-\frac{1}{4} F_{\mu\nu} F^{\mu\nu}+ \mathcal{L}_d),\\
\mathcal{L}_d=-|D_\rho\varphi_{\mu\nu}|^2+2|D_\mu\varphi^{\mu\nu}|^2+|D_\mu\varphi|^2
-\big[D_\mu\varphi^{\dagger\mu\nu}D_\nu \varphi+\text{h.c.}\big]-i q F_{\mu\nu} \varphi^{\dagger\mu\lambda} \varphi^\nu_\lambda\\
-m^2\big(|\varphi_{\mu\nu}|^2-|\varphi|^2\big)+2\mathcal{R}_{\mu\nu\rho\lambda} \varphi^{\dagger\mu\rho}\varphi^{\nu\lambda}-\frac{1}{d+1} \mathcal{R} |\varphi|^2\;,
\end{split}
\end{equation}
where $D_{_\mu}=\nabla_\mu-i q A_\mu$, $\varphi\equiv{\varphi^\mu}_\mu$, $\varphi_\rho\equiv D^{\mu}\varphi_{\mu\rho}$ and ${\mathcal{R}^\mu}_{\nu\rho\lambda}$ is the Riemann tensor of the background metric. The above theory is ghost-free and describes the correct number of propagating degrees of freedom. The disadvantage is that one has to be restricted to work in a fixed background spacetime that satisfies the Einstein condition $\mathcal{R}_{\mu\nu}=\frac{2\Lambda}{d-1}g_{\mu\nu}$. In the context of holographic superconductors, this restriction forces us to work in the probe approximation where the spin two field and gauge field do not influence on the metric. One such a geometry is given by the AdS-Schwarzschild black hole with a planar horizon
\begin{equation}
ds^2 = \frac{L^2}{z^2} \Big(-f(z)\,dt^2 + d \vec x_{d-1}^2 + \frac{dz^2}{f(z)} \Big), \quad f(z)=1-\Big(\frac{z}{z_h} \Big)^d \;.
\end{equation}
The black hole horizon is located at $z=z_h$, while the conformal boundary of the spacetime is located at $z=0$.
The temperature of this black hole is
\begin{equation}
T = \frac{d}{4 \pi z_h} \;.
\end{equation}

\subsubsection{The d-wave condensate}
We consider an ansatz where $\varphi_{\mu\nu}$ and $A_\mu$ depend only on the radial coordinate $z$ and only the space components of $\varphi_{\mu\nu}$ are turned on.
According to ref.~\cite{Benini:2010pr}, it is consistent to turn on a single component of $\varphi_{\mu\nu}$ and to set other components of the gauge field except for $A_t$ to be zero. Then our ansatz is
\begin{equation}\label{dhansatz}
A= \phi(z) \, dt \;, \qquad \varphi_{xy}(z)=\frac{L^2}{2z^2} \, \psi(z) \;,
\end{equation}
with all other components of $\varphi_{\mu\nu}$ set to zero, and $\phi$ and $\psi$ real.

With the above ansatz~\eqref{dhansatz}, the equations of motion for $\phi$ and $\psi$ are given by
\begin{equation}
\begin{split}
\psi'' + \left( \frac{f'}{f} - \frac{d-1}{z} \right) \psi' + \left( \frac{ q^2 \phi^2}{f^2} - \frac{m^2 L^2}{z^2 f} \right) \psi=0\;,\\
\phi'' + \frac{3-d}z \, \phi' - \frac{q^2L^2}{z^2 f} \, \psi^2 \, \phi=0\;.
\end{split}
\end{equation}
Here the prime denotes the derivative with respect to the radial coordinate $z$. To solve the above coupled equations, one demands that two fields near the boundary $z=0$ should behave as
\begin{eqnarray}
\phi=\mu-\rho z^{d-2}+\cdots, \quad \psi=\psi_{+} z^{\Delta}+\cdots,
\end{eqnarray}
where $\Delta=d/2+\sqrt{d^2+4m^2L^2}/2$. The unitary bound implies that $\Delta\geq d$ for spin two operators. Therefore, the mass of $\varphi_{\mu\nu}$ has a lower bound, i.e. $m^2\geq 0$.\footnote{In fact, the previous d-wave model does not consider this aspect.} Note that the fall-off of $\psi$ is chosen so that the dual charged operators have no deformation but can acquire expectation value spontaneously.
Up to a normalization, the coefficients $\mu$, $\rho$ and $\psi_{+}$ are interpreted as chemical potential, charge density and the expectation value of the $xy$ component for the spin two operator $\mathcal{O}_{xy}$, respectively. At the horizon, one should require $\phi(z_h)=0$ in order to keep $g^{\mu\nu}A_{\mu}A_{\nu}$ being finite at the horizon.
\begin{figure}[h!]
\centering
\includegraphics[scale=0.53]{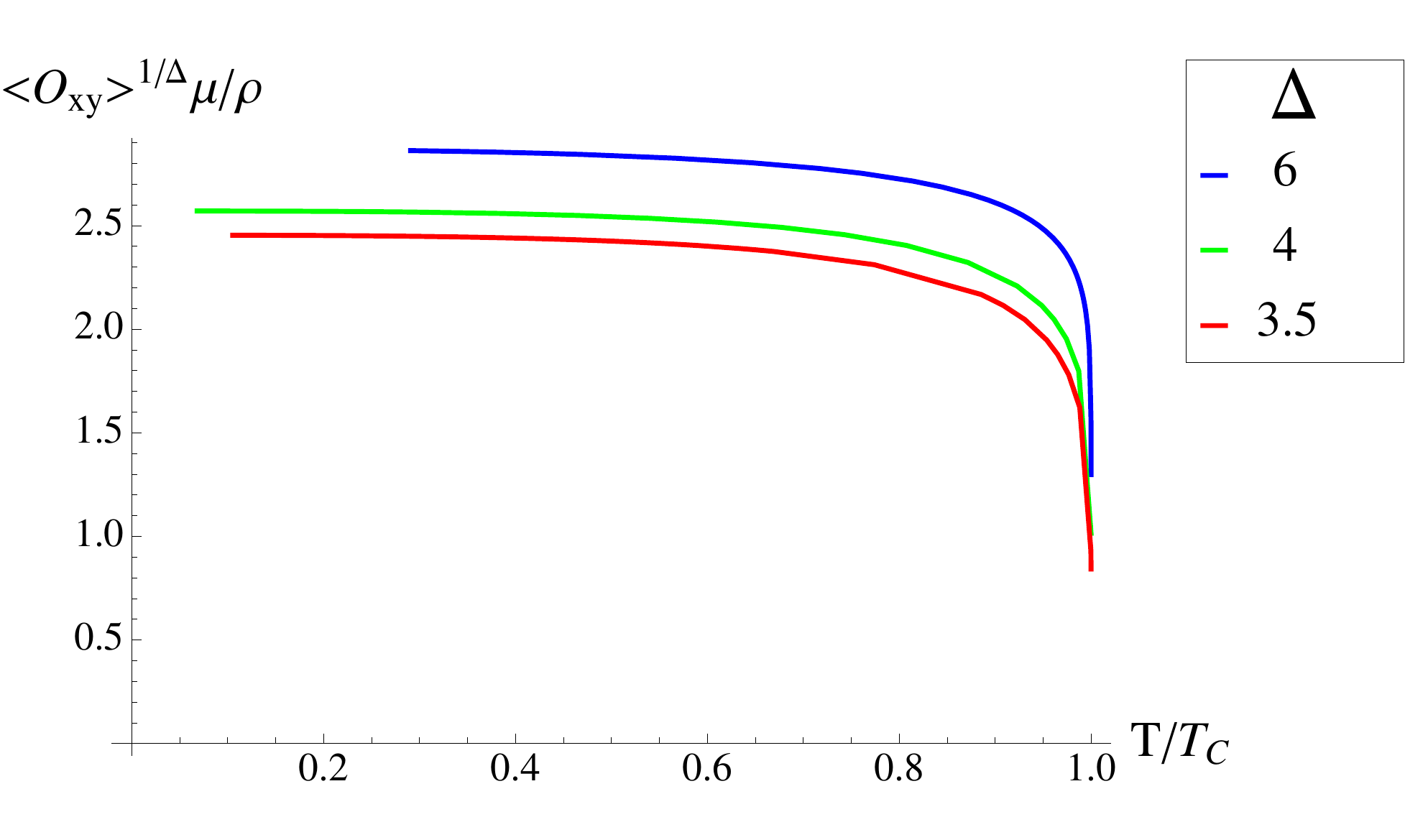}
\caption{\label{dhcondens} The condensate $\langle \mathcal{O}_{xy} \rangle$ in $d=3$ dimensional spacetime as a function of the temperature for various values of $\Delta$. The curves form top to down correspond to $\Delta=6, 4$ and $3.5$, respectively. Used with kind permission from ref.~\cite{Benini:2010pr}.}
\end{figure}

In what follows we will focus on $d=3$. The resulting boundary value problem can be solved directly, for example, by shooting method. A typical dependence of $\langle \mathcal{O}_{xy}\rangle$ on the temperature is shown in figure~\ref{dhcondens}. As we lower the temperature, the normal phase becomes unstable to developing tensor hair at a certain critical temperature $T_c$. This is a typical second order phase transition.

\subsubsection{Conductivity}
We are interested in the electromagnetic response of the condensed phase. To incorporate this feature, we extract the optical conductivity of this d-wave model by linear response theory. The conductivity tensor $\sigma _{ij}$ can be defined through
\begin{equation}
J_{i}=\sigma _{ij}\;E_{j}\;,
\end{equation}
where $i, j=x, y$. $J$ and $E$ are the electric current and electric field, respectively. To compute the conductivity in a holographic framework we turn on a source for the current $J_i$ dual to the gauge field in the bulk. Following the standard approach discussed in previous sections, we perturb the gauge field by $\delta A=e^{-i\omega t}A_{x}(r)dx$.
The bulk equations of motion couple linear fluctuations of the gauge field $A_x$ to some spin two components.
To obtain a consistent set of equations, one should also turn on time dependent fluctuations of the complexified fields $\varphi_{ty}$, $\varphi_{ty}^\dagger$, $\varphi_{zy}$ and $\varphi_{zy}^\dagger$. The linearized equations of motion for the $e^{-i\omega t}$ component of these fluctuations are~\cite{Benini:2010pr}
\begin{subequations}
\label{dhEOMlinear}
\begin{align}
\label{dhEOMAx}
0 &= A_x'' + \frac{f'}f \, A_x' + \frac{\omega^2}{f^2} \, A_x + \frac{q \psi}{2f^2} \, \big[ (\omega - 2q \phi) \varphi_{ty}^\dagger - (\omega + 2q\phi) \varphi_{ty} \big] \nonumber \\
&\quad - \frac{iq\psi}2 \, \big( {\varphi^\dagger_{zy}}' - \varphi_{zy}' \big) + \frac{iq}{2f} \, (\psi' f - \psi f') \big( \varphi_{zy}^\dagger - \varphi_{zy} \big) \ , \\
\label{dhEOMty}
0 &= \varphi_{ty}'' + \frac2z \, \varphi_{ty}' - \frac{2f + m^2L^2}{z^2 f} \, \varphi_{ty} + L^2 \, \frac{q\omega + 2q^2 \phi}{4z^2 f} \, \psi A_x 
+ \frac i2\, \big[ 2(\omega + q \phi) \varphi_{zy}' + q \phi' \varphi_{zy} \big] \ , \\
\label{dhEOMRzy}
0 &= \big[ (\omega + q\phi)^2 z^2 - m^2L^2 f \big] \, \varphi_{zy} + \frac i4\, L^2 q f \psi A_x' + \frac i2 L^2 q f \psi' A_x \nonumber \\
&\quad - i(\omega + q\phi) z^2 \varphi_{ty}' - \frac i2 \, \big[ 4(\omega + q \phi)z + q \phi'z^2 \big] \, \varphi_{ty} \ .
\end{align}
\end{subequations}
The equations for $\varphi_{ty}^\dagger$ and $\varphi_{zy}^\dagger$ can be obtained by complex conjugation and an additional transformation $\omega$ to $-\omega$ from the last two equations.
The functions $\varphi_{zy}$ and $\varphi_{zy}^\dagger$ can be eliminated from the first two equations by virtue of~\eqref{dhEOMRzy}, leaving three coupled differential equations for $A_x$, $\varphi_{ty}$ and $\varphi_{ty}^\dagger$. Since the fluctuation $A_y$ decouples from above set of fluctuations, we can conclude that the Hall conductivity $\sigma_{xy}(\omega)$ is vanishing.

The conductivity is related to the retarded Green's function for the charge current. To calculate the retarded function, one should impose causal boundary conditions on the equations of motion. As a consequence, the near-horizon modes of the gauge field and spin two field are falling into the horizon, i.e., $A_x$, $\varphi_{ty}$ and $\varphi_{ty}^\dagger$ have the behaviour as
\begin{equation}
(z_h-z)^{- i\omega z_h/ 3} \;.
\label{nearhorizon}
\end{equation}
Near the boundary $z=0$, the asymptotical behaviour for the perturbation fields $A_x$, $\varphi_{ty}$ and $\varphi_{ty}^{\dagger}$ is given by
\begin{eqnarray}
\nonumber
A_x&=&A_x^{(0)}+A_x^{(1)}z+\cdots,\\
\varphi_{ty}&=&\varphi_{ty-}z^{\Delta_-}+\varphi_{ty+}z^{\Delta_+}+\cdots,\\ \nonumber
\varphi_{ty}^{\dagger}&=&\varphi^\dagger_{ty-}z^{\Delta_-}+\varphi^\dagger_{ty+}z^{\Delta_+}+\cdots,
\end{eqnarray}
with $\Delta_{\pm}= \frac{-1\pm \sqrt{9+4m^2L^2}}{2}$. Here $\varphi_{ty-}$ and $\varphi^\dagger_{ty-}$ are identified as the source terms, while $\varphi_{ty+}$ and $\varphi^\dagger_{ty+}$ are the normalizable fluctuations. Since the presence of source terms $\varphi_{ty-}$ and $\varphi^\dagger_{ty-}$ will also source the U(1) current, one should
look for solutions where the source in the series expansion of $\varphi_{ty}$ and $\varphi_{ty}^{\dagger}$ should vanish. Finally, one can obtain the conductivity in the $x$ direction as
\begin{equation}
\sigma_{xx}=\frac{A_x^{(1)}}{i \omega A_x^{(0)}}\;.
\end{equation}
\begin{figure}[hbt]
\centering
\includegraphics[scale=0.65]{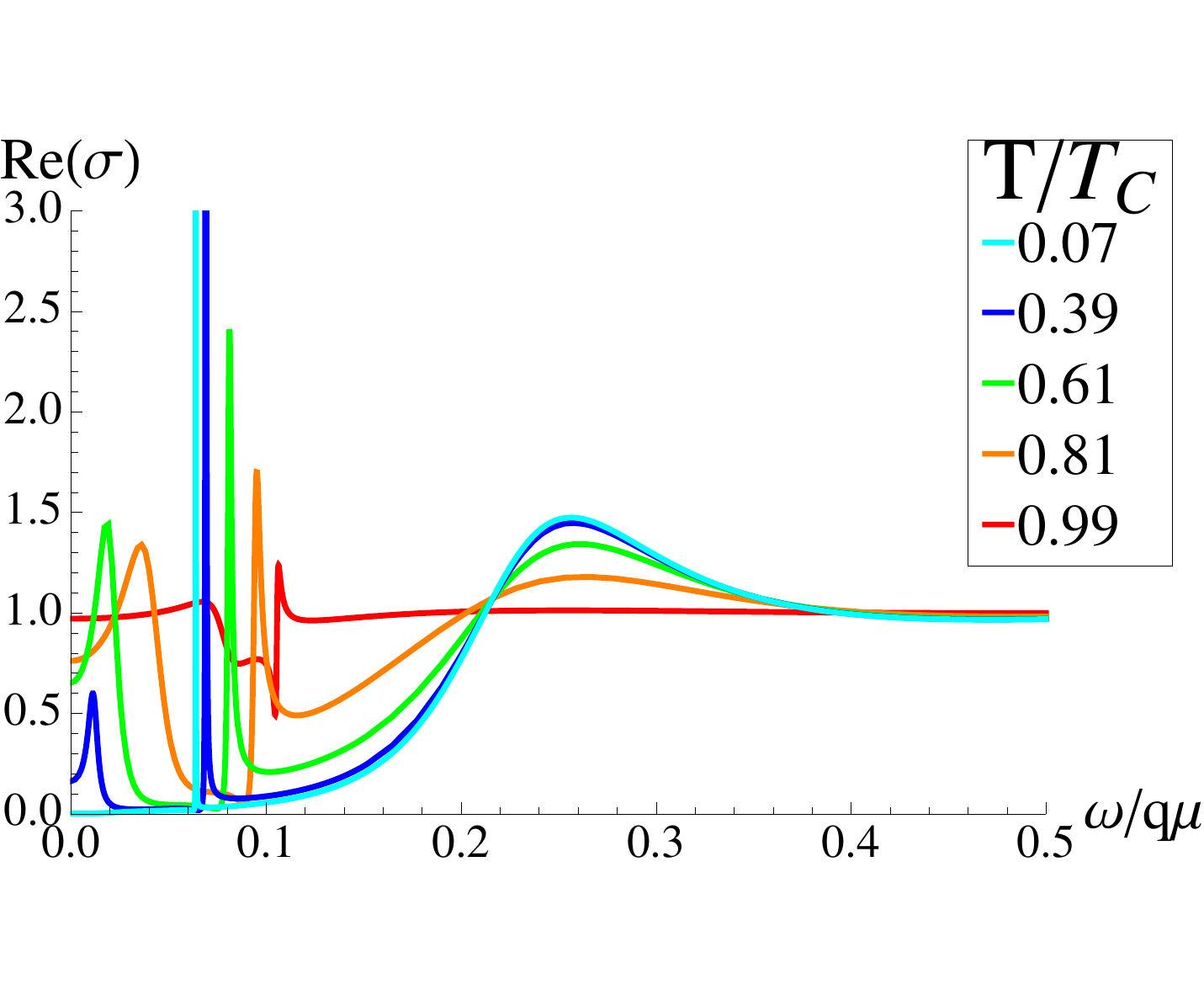}
\caption{\label{dhconductivity} The real part of the conductivity as a function of frequency for a $\Delta=4 $ condensate. Used with permission from ref.~\cite{Benini:2010pr}.}
\end{figure}

In order to obtain the conductivity in the $y$ direction, we look at the effect of a $\pi/2$ rotation of the condensate in equations~\eqref{dhEOMlinear}. This rotation operation flips the sign of $\psi$, which is equivalent to flipping the sign of $A_x$ by viewing of the equations of motion. Changing the sign of $A_x$ is equivalent to flipping the sign of both the electric field (the $A_x^{(0)}$ term) and the current (the $A_x^{(1)}$ term), thus conductivity will be unaffected under such a change of sign. This implies that the conductivity is proportional to the identity matrix. In fact, the isotropy of conductivity is a consequence of the symmetries that $\sigma_{ij}$ has in the d-wave case. An isotropic conductivity for a d-wave superconductor can be also produced in an explicit microscopic model~\cite{isosigma}.

The numerical results for the conductivity are shown in figure~\ref{dhconductivity} for a d-wave condensate of conformal dimension $\Delta=4$.  As the temperature is lowered one can observe a spike in the conductivity, which is a signal of a bound state. This spike is localized at smaller values of $\omega$ as the temperature is lowered. A second spike in the conductivity appears to disappear as the temperature is decreased. One can also see that $\text{Re}[\sigma(\omega)]$ does not vanish even for arbitrary small frequency $\omega$, so there is no hard gap in the dual boundary superconducting phase.

The fermionic spectral function in this holographic superconductor with a d-wave condensate has been well studied in ref.~\cite{Benini:2010qc}. It was showed that, with a suitable bulk Majorana coupling, the Fermi surface is anisotropically gapped. At low temperatures the gap shrinks to four nodal points, while at high temperatures the Fermi surface is partially gapped generating four Fermi arcs. The $(d+id)$ condensate for the BHRY model was investigated in ref.~\cite{Chen:2011ny}, in which the existence of fermi arcs is confirmed and a non-vanishing Hall conductivity is obtained in the absence of a magnetic field.

Although both d-wave models we reviewed above can  be used to study the properties of a superconducting phase transition with a d-wave order parameter in a dual strongly
interacting field theory, the construction is not ideal. For example, the BHRY d-wave model can only work in the probe limit. However, it is well known that including the back reaction of matter fields would lead to a much richer phase structure. It will be desirable to study a consistent holographic d-wave model with back reaction.
To write down an action for a charged spin two field propagating in a curved spacetime is challenging, because it usually suffers from non-hyperbolic and non-causal behaviour of the spin two field. Apart from those two effective models, the authors of ref.~\cite{Kim:2013oba} discussed top-down models for holographic d-wave superconductors in which the order parameter is a charged spin two field in the bulk.

\section{Competition and Coexistence of Superconducting Order Parameters}
\label{sect:competition}

\begin{figure}[hbt]
\centering
\includegraphics[scale=0.85]{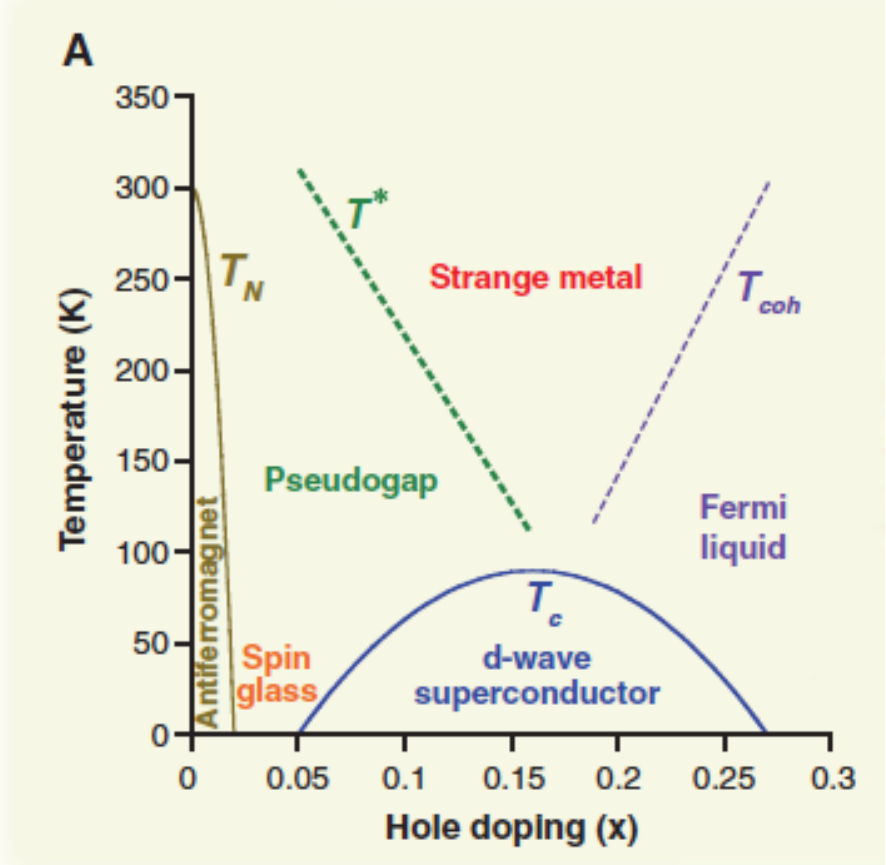}
\includegraphics[scale=0.75]{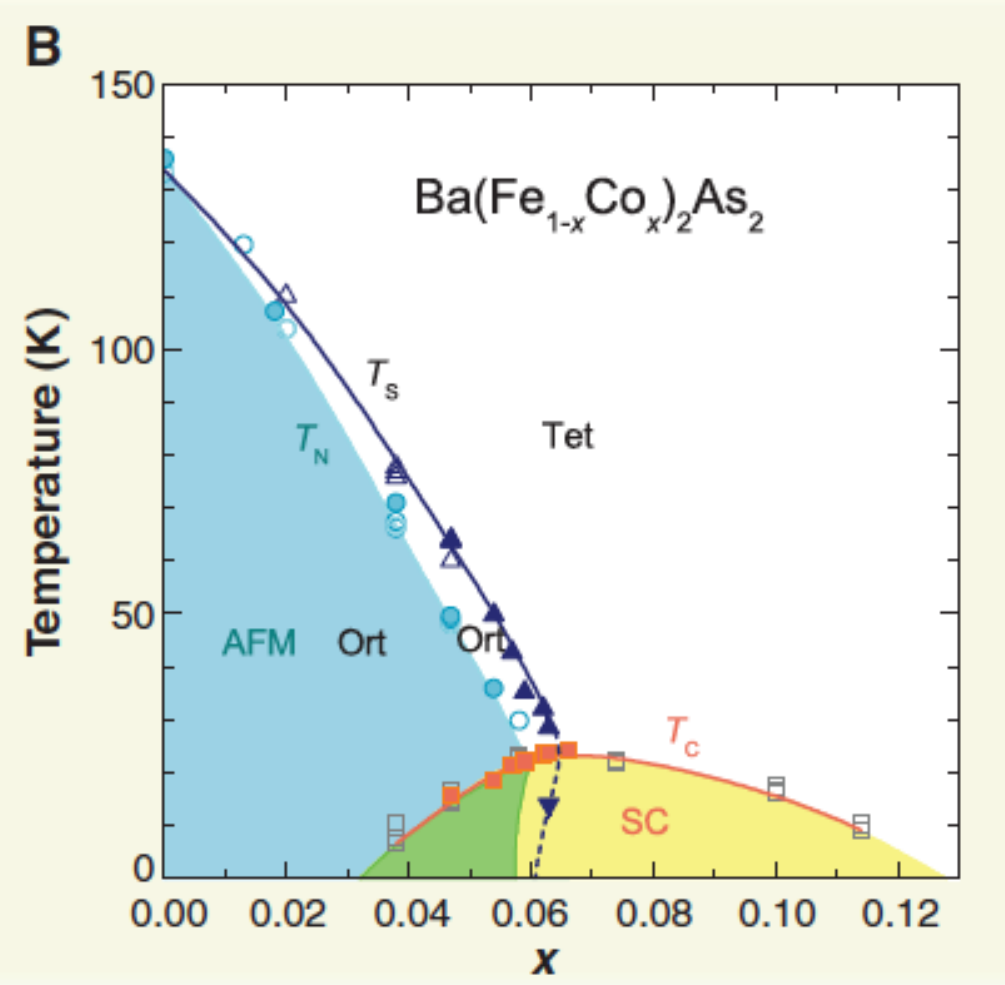}
\caption{\label{hightphase} Schematic phase diagrams of the cuprates (left) and the pnictide superconductor $Ba(Fe_{1-x}Co_x)As_2$ (right). In the right plot, the antiferromagnetic phase is labeled by AFM, the normal (tetragonal) phase is denoted by Tet, and the superconducting phase by SC. Note the similarities of the phase diagrams. Adapted with permission from ref.~\cite{Norman:2011} .}
\end{figure}

The holographic models of s-wave, p-wave and d-wave superconductors, which have scalar, vector and spin-2 order parameters respectively, have been discussed in the previous sections. These models were based on a specific setup where the dynamics in the bulk involves only a single order parameter. It is desirable to generalize the single order parameter case to multi order parameters case because  the real high $T_c$ superconducting systems indeed involve various orders, such as magnetic ordering and superconductivity, see figure~\ref{hightphase} and, e.g., refs.~\cite{Berg2009,Zaanen1012}. The holographic correspondence provides us a convenient way to investigate the interaction for these orders by simply introducing dual fields in the bulk as well as appropriate couplings among them. Following this strategy, several works on the competition of multi order parameters in the holographic superconductor models have already been made~\cite{Cai:2013wma,Nie:2013sda,Amado:2013lia,Li:2014wca,Nishida:2014lta,Basu:2010fa,Musso:2013ija,Liu:2013yaa,Amoretti:2013oia,Wen:2013ufa,Donos:2012yu}. In the following, we will review the competition between two s-wave orders~\cite{Cai:2013wma}, the competition between s-wave order and p-wave order~\cite{Nie:2013sda,Amado:2013lia} and the competition between s-wave order and d-wave order~\cite{Li:2014wca} one by one. The first case concerns the competition between two orders with the same symmetry and the last two cases are to study the competition of orders with different symmetry. The phase diagrams are also drawn for the corresponding models.

\subsection{Competition and coexistence of two s-wave orders }
\label{ss}
Historically, Ginzburg-Landau theory has proved to be an extraordinarily valuable phenomenological tool in understanding single-component superconductors. Its generalization to the two-component Ginzburg-Landau model (TCGL) was constructed, and its applicability to the two-band systems studied in refs.~\cite{Silaev:2012,Shanenko:2011,Vagov:2012}. Upon switching on the interband coupling between the two components, this model can describe the phenomenon of the two gaps in materials such as MgB$_{2}$ ($s_{++}$)~\cite{Carlstrom:2010,Buzea:2001} and iron pnictides ($s_{+-}$)~\cite{iron-based1,iron-based2,iron-based3}. Applying the multi-band Ginzburg-Landau theories to the gravity side, the holographic multi-band superconductor model can be realized involving some competing scalar fields coupled to a single gauge field. Such system exhibits rich phase structure. Next, we will describe the holographic model concretely and show results explicitly.

\subsubsection{The holographic model}

Let us start a holographic superconductor model with $N$ scalar hairs in $(3+1)$ dimensional anti-de Sitter spacetime. The action reads~\cite{Cai:2013wma}
\begin{equation}\label{action0}
S =\frac{1}{2\kappa^2}\int d^4 x
\sqrt{-g}[\mathcal{R}+\frac{6}{L^2}-\frac{1}{4}F_{\mu\nu} F^{\mu \nu}+\sum_{k=1}^{N}(-|\nabla\psi_k-ie_kA\psi_k|^2-m_k^2|\psi_k|^2)-\mathcal{V}_{intact}],
\end{equation}
where $e_k$ and $m_k$ ($k=1,2,...,N$) are the charge and mass of the scalar field $\psi_k$, respectively. The term $\mathcal{V}_{intact}$ denotes the possible interaction among bulk matter fields. Here one can perform a rescaling of the type $A_\mu\rightarrow\frac{1}{e_2}A_\mu, \psi_k\rightarrow\frac{1}{e_2}\psi_k$ to set the charge of the scalar field $\psi_2$ to unity. We are interested in the dynamics and mutual interaction among different orders. Here we limit ourselves to the case with $N=2$. The concrete model we will study is described by the following action
\begin{equation}\label{action}
\begin{split}
S =\frac{1}{2\kappa^2}\int d^4 x
\sqrt{-g}[\mathcal{R}+\frac{6}{L^2}+\frac{1}{e_2^2}\mathcal{L}_m],\\
\mathcal{L}_m=-\frac{1}{4}F_{\mu\nu} F^{\mu \nu}-|D_1\psi_1|^2-m_1^2|\psi_1|^2-|D_2\psi_2|^2-m_2^2|\psi_2|^2,
\end{split}
\end{equation}
where we have defined $D_{1\mu}=\nabla_\mu-i\frac{e_1}{e_2}A_\mu$ and $D_{2\mu}=\nabla_\mu-iA_\mu$. The parameter $e_2$ controls the strength of the back reaction and $e_1/e_2$ is the effective charge of $\psi_1$ or the ratio of two scalar charges.

The hairy black hole solution is assumed to take the following metric form
\begin{equation}\label{metric}
ds^2=-f(r)e^{-\chi(r)}dt^2+\frac{dr^2}{f(r)}+r^2(dx^2+dy^2),
\end{equation}
together with homogeneous  matter fields
\begin{equation}\label{matter}
\psi_1=\psi_1(r),\quad \psi_2=\psi_2(r),\quad A=\phi(r)dt.
\end{equation}
The horizon $r_h$ is determined by $f(r_h)=0$ and the temperature of the black hole is given by
\begin{equation}\label{temp}
T=\frac{f'(r_h)e^{-\chi(r_h)/2}}{4\pi}.
\end{equation}
One can use the U(1) gauge symmetry to set $\psi_1$ to be real. After using the $r$ component of Maxwell's equations we can also safely choose $\psi_2$ to be real.
We will work in the unites where $L=1$. Then, the independent equations of motion in terms of the above ansatz are deduced as follows
\begin{equation}\label{eomsss}
\begin{split}
\psi_1''+(\frac{f'}{f}-\frac{\chi'}{2}+\frac{2}{r})\psi_1'+(\frac{e_1^2}{e_2^2}\frac{\phi^2e^{\chi}}{f^2}-\frac{m_1^2}{f})\psi_1=0, \\
\psi_2''+(\frac{f'}{f}-\frac{\chi'}{2}+\frac{2}{r})\psi_2'+(\frac{\phi^2e^{\chi}}{f^2}-\frac{m_2^2}{f})\psi_2=0, \\
\phi''+(\frac{\chi'}{2}+\frac{2}{r})\phi'-\frac{2}{f}(\frac{e_1^2}{e_2^2}\psi_1^2+\psi_2^2)\phi=0,\\
\frac{f'}{f}+\frac{r}{2e_2^2}(\psi_1'^2+\psi_2'^2)+\frac{re^{\chi}\phi'^2}{4e_2^2f}+\frac{r}{2e_2^2f}(m_1^2\psi_1^2+m_2^2\psi_2^2)+\frac{re^{\chi}\phi^2}{2e_2^2f^2}
(\frac{e_1^2}{e_2^2}\psi_1^2+\psi_2^2)\\-\frac{3r}{f}+\frac{1}{r}=0 ,\\
\chi'+\frac{r}{e_2^2}(\psi_1'^2+\psi_2'^2)+\frac{re^{\chi}\phi^2}{e_2^2f^2}(\frac{e_1^2}{e_2^2}\psi_1^2+\psi_2^2)=0,
\end{split}
\end{equation}
where a prime denotes the derivative with respect to $r$.

The gravity background describing the normal phase is just the AdS Reissner-Nordstr\"om black hole with a planar horizon
\begin{equation}\label{normalss}
\phi(r)=\mu(1-\frac{r_h}{r}),\quad \psi_1(r)=\psi_2(r)=0,\quad f(r)=r^2(1-\frac{r_h^3}{r^3})+\frac{r_h^2}{4r^2}\frac{\mu^2}{e_2^2}(1-\frac{r}{r_h}),
\end{equation}
where $r_h$ is the black hole horizon and $\mu$ is the chemical potential of the black hole.
\subsubsection{Phase transition}
The two band model is controlled by four model parameters, i.e., $m_1^2$, $m_2^2$, $e_2$, and $e_1/e_2$. Here we will choose $m_1^2=0$ and $m_2^2=-2$. One may expect that the model admits three different superconducting phases. The first superconducting phase corresponds to the case with $\psi_1\neq0$ and $\psi_2=0$ (Phase-\uppercase\expandafter{\romannumeral1}). The second superconducting phase corresponds to the case with $\psi_2\neq0$ and $\psi_1=0$ (Phase-\uppercase\expandafter{\romannumeral2}). The third superconducting phase admits the region where both scalars condense simultaneously.
\begin{figure}[h]
\centering
\includegraphics[scale=0.59]{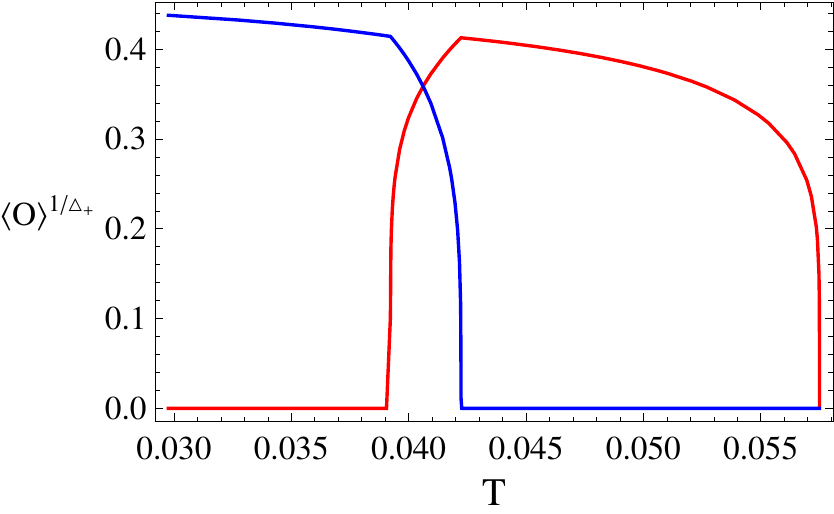}
\includegraphics[scale=0.59]{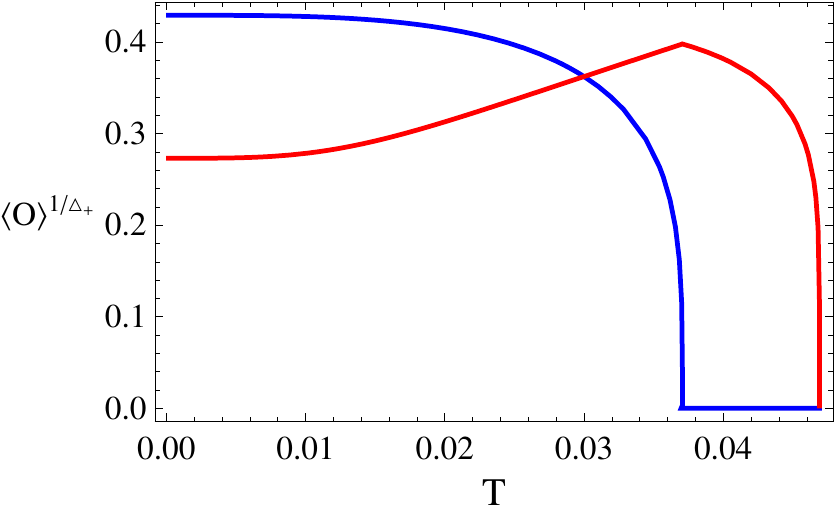}
\includegraphics[scale=0.59]{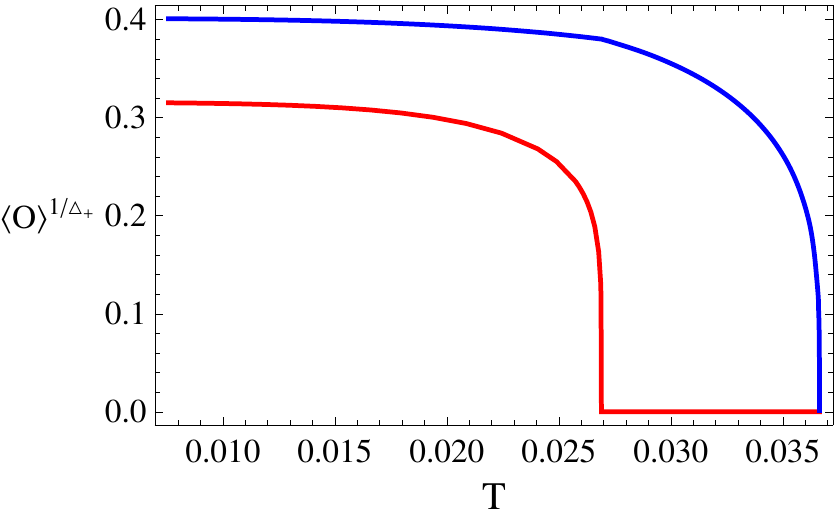}
\caption{\label{condensateABC} The condensate as a function of temperature for three kinds of the coexisting phase. The red curve is for $\psi_1$, while the blue one is for $\psi_2$. The condensate for Phase-C with $e_1/e_2=1.95$ and $e_2=4$ is shown in the left plot. The condensate for Phase-A with $e_1/e_2=1.95$ and $e_2=2$ in the middle plot and for Phase-B with $e_1/e_2=1.9$ and $e_2=1.5$ in the right plot. Three plots were taken from ref.~\cite{Cai:2013wma}.}
\end{figure}

The numerical results are shown in figure~\ref{condensateABC}. The model admits three kinds of the coexisting phase. The first kind is shown in the left panel of figure~\ref{condensateABC}. As we lower temperature, the scalar $\psi_1$ first condenses at $T_c$ where the superconducting phase transition happens; when we continue lowering temperature to a certain value, say $T_2$, the scalar $\psi_2$ begins to condense, while the condensate of $\psi_1$ decreases, resulting in the phase with both orders; if one further lowers temperature, the first condensate quickly goes to zero at a temperature $T_3$; when temperature is lower than $T_3$, there exists only the condensate of $\psi_2$. This superconducting phase is denoted by Phase-C and it is the case uncovered in the probe limit in ref.~\cite{Basu:2010fa}. The second kind of the coexisting phase is presented in the middle plot. It is different from the first one in that the coexisting phase survives even down to a low temperature. We denote this case by phase-A. Depending on the back reaction, the inverse of phase-A is also true: the condensate of $\psi_1$ emerges following the condensate of $\psi_2$, and then both orders are always present. This case is labeled as Phase-B drawn in the right plot.

To determine whether those above three coexistence phases are thermodynamically favored in their own parameter spaces, one should calculate the free energy of the system for each phase. Working in grand canonical ensemble, the chemical potential is fixed. In gauge/gravity duality the grand potential $\Omega$ of the boundary thermal state is identified with temperature times the on-shell bulk action with Euclidean signature. The Euclidean action must include the Gibbons-Hawking boundary term for a well-defined Dirichlet variational principle and further a surface counter term for removing divergence. Note that we are considering a stationary problem, the Euclidean action is related to the Minkowski one by a minus sign as
\begin{equation}
\begin{split}
-2\kappa^2 S_{Euclidean}=\int d^4 x
\sqrt{-g}[&\mathcal{R}+\frac{6}{L^2}+\frac{1}{e_2^2}\mathcal{L}_m]+\int_{r\rightarrow\infty} d^3x
\sqrt{-h}(2\mathcal{K}-\frac{4}{L})\\
+&\frac{1}{e_2^2}\int_{r\rightarrow\infty} d^3x\sqrt{-h}(\frac{{\triangle_1}_+-3}{L}\psi_1^2+\frac{{\triangle_2}_+-3}{L}\psi_2^2),
\end{split}
\end{equation}
where $h$ is the determinant of the induced metric on the boundary $r\rightarrow\infty$, and $\mathcal{K}$ is the trace of the extrinsic
curvature. By using of the equations of motion~\eqref{eomsss} and the asymptotical expansion of matter and metric functions near the AdS boundary, the grand potential $\Omega$ can be expressed as
\begin{equation}\label{grand1}
\frac{2\kappa^2\Omega}{V_2}=\varepsilon,
\end{equation}
where $V_2=\int dx dy$ and the constant $\varepsilon$ is from the asymptotical expansion of $f=r^2+\varepsilon/r+\cdots$. For the normal phase given in~\eqref{normalss}, one has $\varepsilon=-r_h^3-\frac{r_h}{4}\frac{\mu^2}{e_2^2}$.

\begin{figure}[h]
\centering
\includegraphics[scale=0.55]{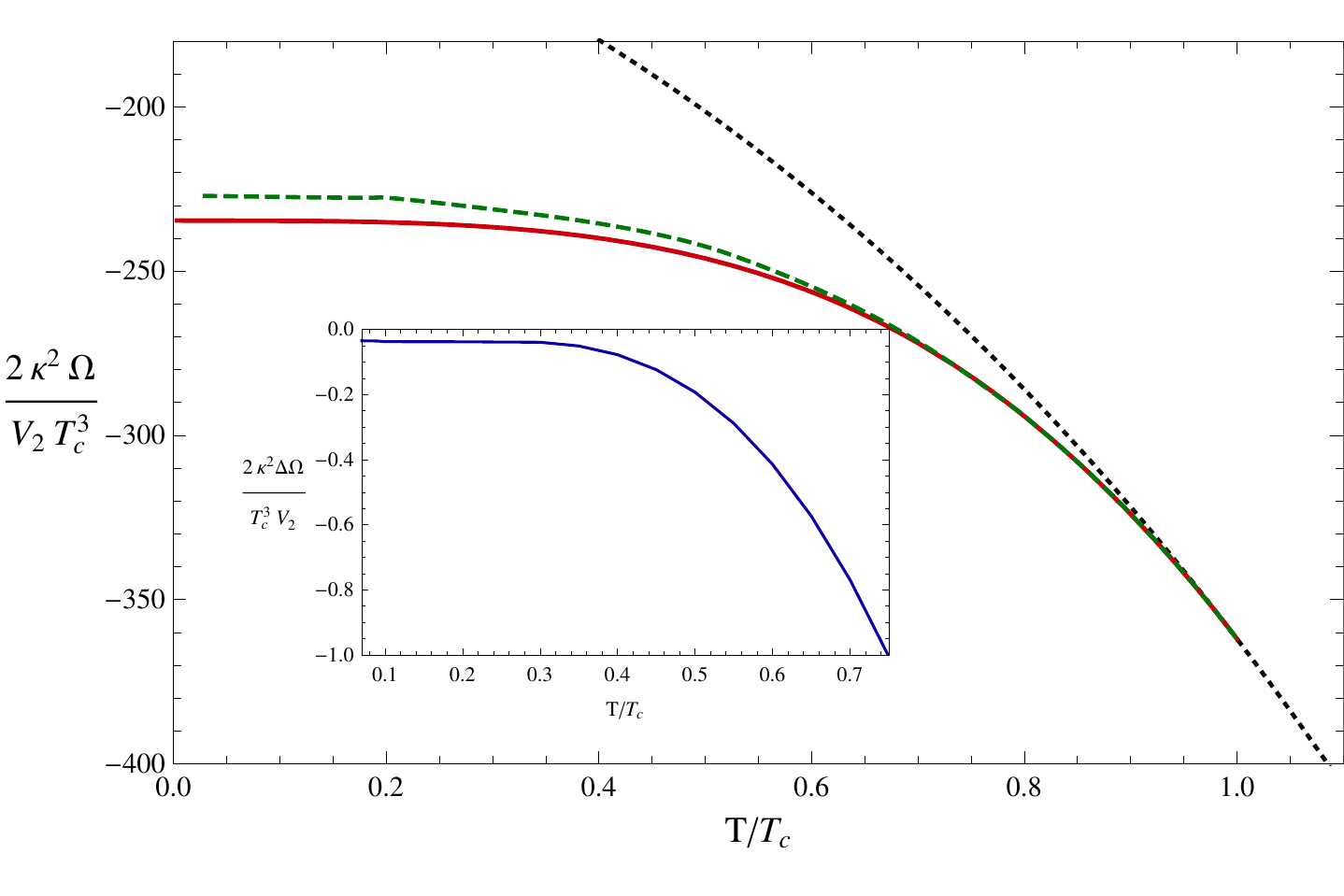}
\includegraphics[scale=0.55]{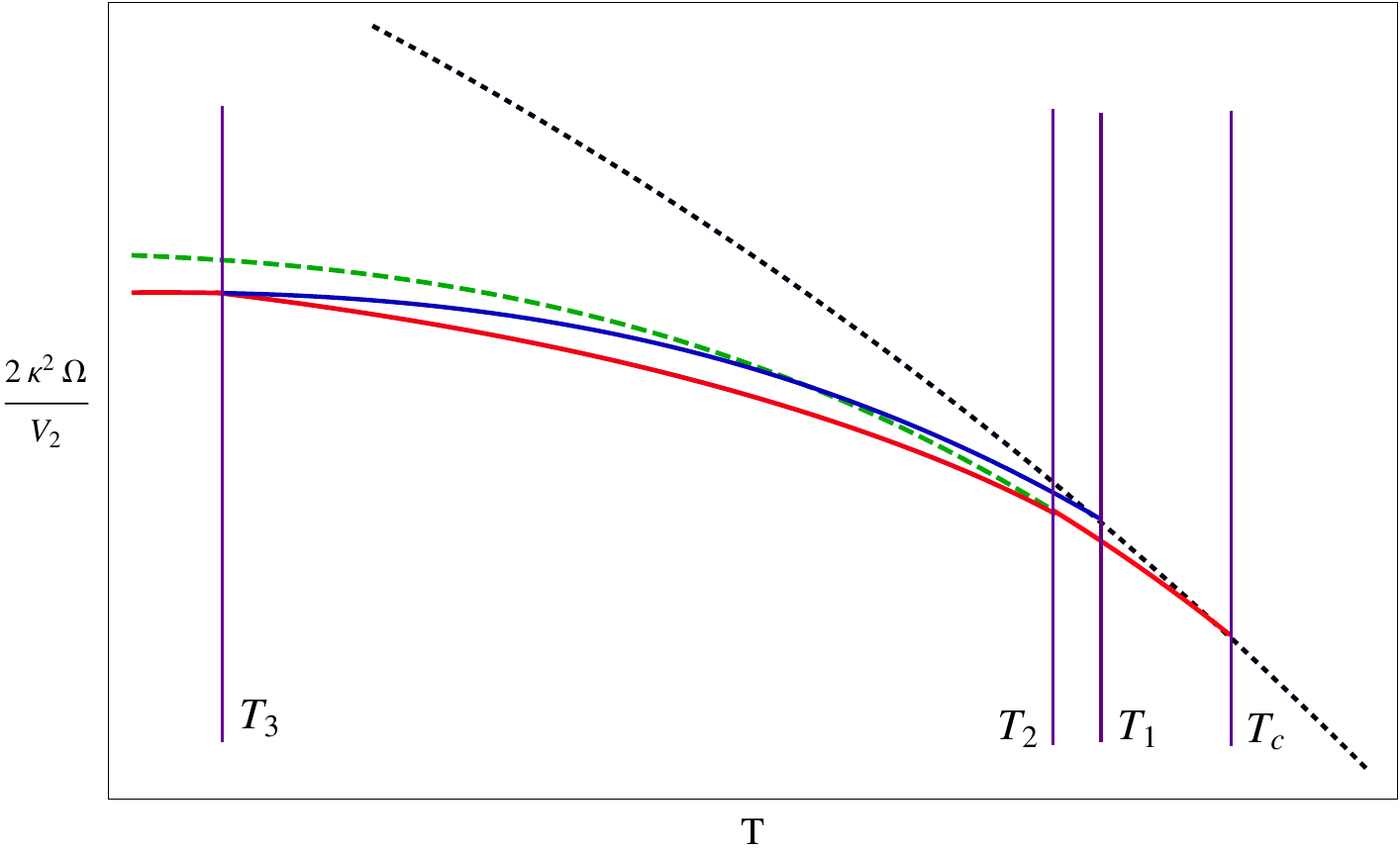}\caption{\label{freeAss} The free energy as a function of temperature for Phase-A(C) labeled as the solid red curve. The equations of motion also admit three other types of solutions, i.e., the normal phase (dotted black curve), Phase-\uppercase\expandafter{\romannumeral1} (dashed green curve) and Phase-\uppercase\expandafter{\romannumeral2} (solid blue curve). The curve in the insert of the left plot is the difference of free energy between Phase-A and Phase-\uppercase\expandafter{\romannumeral2}. One can see phase-A(C) indeed has the lowest free energy. The plots were taken from ref.~\cite{Cai:2013wma}.}
\end{figure}

\begin{figure}[h]
\centering
\includegraphics[scale=1.0]{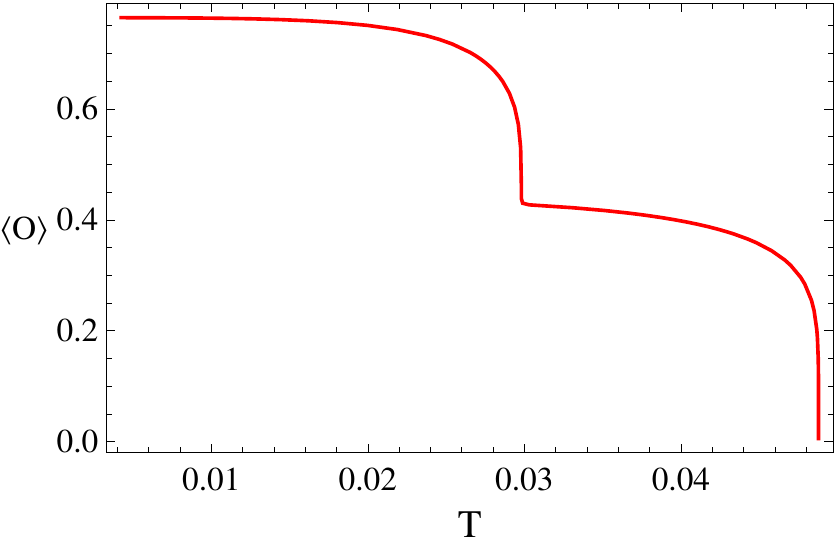} \caption{\label{condensateAA} The total condensate as a function of temperature for Phase-A. We set $e_1/e_2=2$ and $e_2=2$. The two special points at $T_c\simeq0.0488\mu$ and $T\simeq0.0298\mu$ correspond to the superconducting critical point and the point at which $\psi_2$ begins to emerge in Phase-A, respectively. The figure was taken from ref.~\cite{Cai:2013wma}.}
\end{figure}

The free energy corresponding to phase-A(C) is drawn in figure~\ref{freeAss}. From each plot, phase-A(C) does have the lowest free energy, indicating that once phase-A(C) appears, it is thermodynamically favored. But for phase-C, there is only a narrow window admitting the two orders to coexist. Comparing phase-B with phase-A, the only difference is that $\psi_2$ condenses before $\psi_1$. From  figure~\ref{freeAss}, one can also see there exist two transition points in both cases. One is the critical superconducting phase transition and the other within the superconducting phase, which indicates the fact that our system is multi-band. In order to see this clearly, one can define the total condensate as $\langle\mathcal{O}\rangle=\langle O_{1+}\rangle^{1/\Delta_{1+}}+\langle O_{2+}\rangle^{1/\Delta_{2+}}$, and draw $\langle\mathcal{O}\rangle$ as a function of temperature in figure~\ref{condensateAA}. As one lowers temperature, $\langle\mathcal{O}\rangle$ emerges at the critical superconducting phase transition point, then at a certain temperature inside the superconducting phase it has a sudden increase, where the condensate of the other $\psi$ appears. Such a behaviour is very reminiscent of the one in the real multi-band superconductor.

\subsubsection{Conductivity}
\label{sect:conduc}
\begin{figure}[h]
\centering
\includegraphics[scale=1.1]{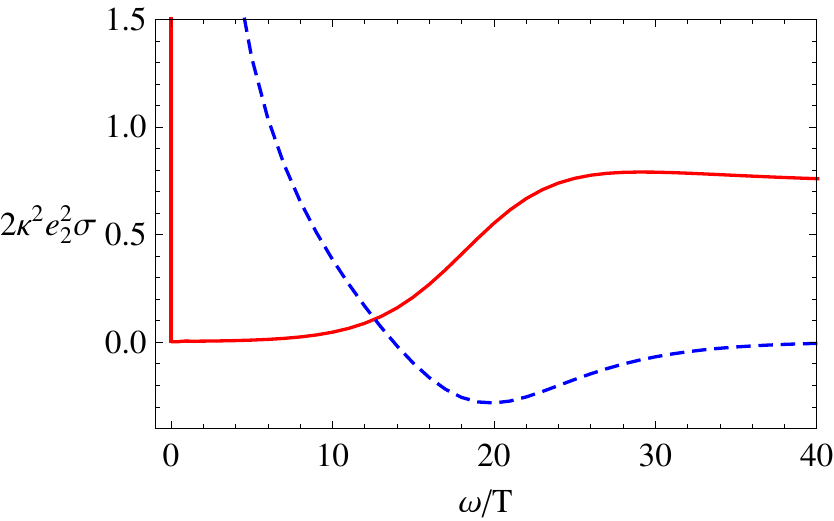}\ \ \ \
\caption{\label{conductivityss} The optical conductivity as a function of frequency at temperature $T=0.0273\mu$ for Phase-A. The red solid line is the real part of the conductivity, while the blue dashed line is the imaginary part of the conductivity.  Here  the parameter $e_1/e_2=1.95$ and $e_2=2$ are taken. There is a delta function at the origin for the real part of the conductivity. Figure taken from ref.~\cite{Cai:2013wma}.}
\end{figure}

  In order to ensure the system is indeed in a superconducting state, one should  calculate the conductivity $\sigma$. Since now the back reaction is included, one has to consider the fluctuations of $A_x$ and $g_{tx}$. Assuming both perturbations have a time dependence of the form $e^{-i\omega t}$,  the final equation of motion to calculate the conductivity can be obtained as
\begin{equation}\label{Axeom}
A_x''+(\frac{f'}{f}-\frac{\chi'}{2})A_x'+[(\frac{\omega^2}{f^2}-\frac{\phi'^2}{e_2^2f})e^{\chi}-\frac{2}{f}(\frac{e_1^2}{e_2^2}\psi_1^2+\psi_2^2)]A_x=0.
\end{equation}
Since the conductivity is related to the retarded two-point function of the U(1) current, i.e, $\sigma=\frac{1}{i\omega}G^R(\omega,k=0)$, one imposes the ingoing boundary condition near the horizon
\begin{equation}\label{ingoing}
A_x=(r-r_h)^{-\frac{i\omega}{4\pi T}}[a_0+a_1(r-r_h)+a_2(r-r_h)^2+\cdots],
\end{equation}
with $a_0, a_1, a_2$ being constants. The gauge field $A_x$ near the boundary $r\rightarrow\infty$ falls off as
\begin{equation}\label{axbound}
A_x=A^{(0)}+\frac{A^{(1)}}{r}+\cdots.
\end{equation}
According to the AdS/CFT dictionary, the retarded Green function can be read as $G^R=\frac{1}{2\kappa^2 e_2^2}\frac{A^{(1)}}{A^{(0)}}$, from which one can obtain the conductivity
\begin{equation}\label{conduc}
\sigma(\omega)=\frac{1}{i\omega}G^R(\omega,k=0)=\frac{1}{2\kappa^2 e_2^2}\frac{A^{(1)}}{i\omega A^{(0)}}.
\end{equation}
The optical conductivity as a function of frequency in the region with two order parameters is presented in  figure~\ref{conductivityss}. One can see clearly that the optical conductivity in two band model behaves qualitatively similar to the  model with only one scalar order discussed in figure~\ref{sconduc}. In addition, from the Kramers-Kronig relations, one can conclude that the real part of the conductivity has a Dirac delta function at $\omega=0$ since the imaginary part has a pole, i.e., Im$[\sigma(\omega)]\sim\frac{1}{\omega}$.

\subsubsection{Phase diagram}
\begin{figure}[h]
\centering
\includegraphics[scale=0.85]{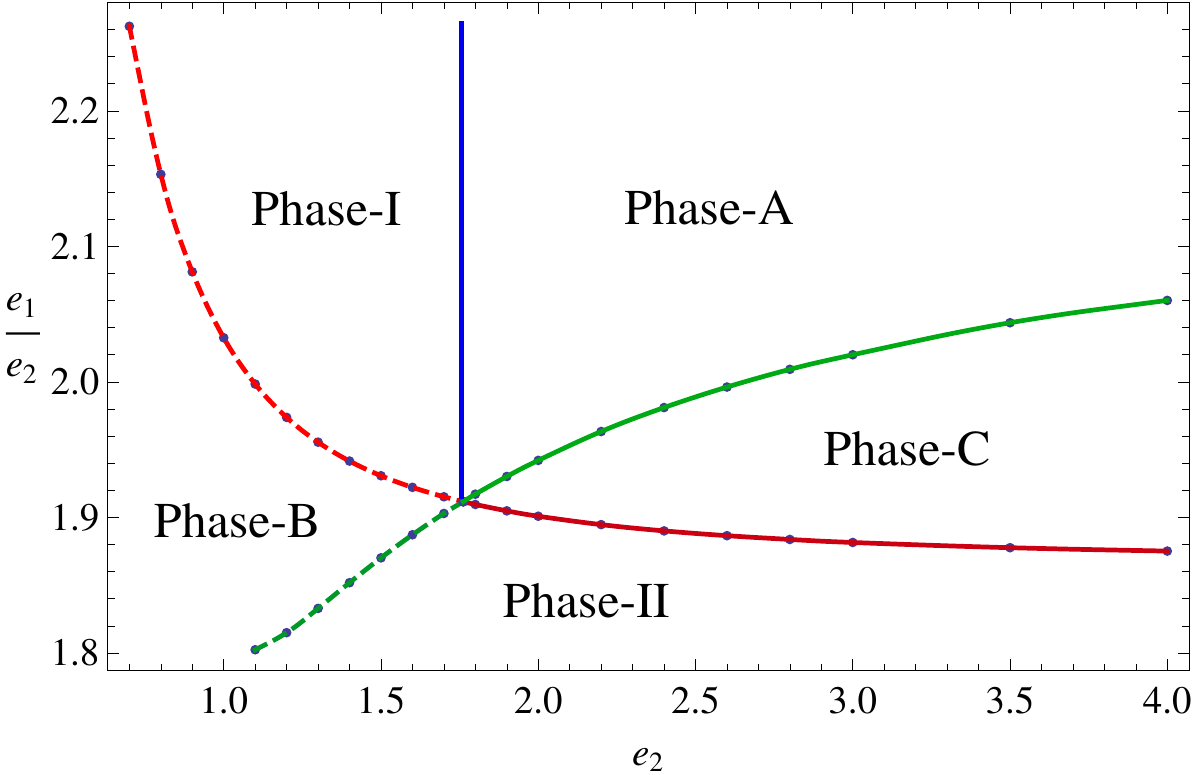}
\caption{\label{diagramss} The full phase diagram for the five superconducting phases. Depending on $e_1/e_2$ and $e_2$, the phase diagram is divided into five parts. The most thermodynamically favored phase in each part is labeled. This figure was taken from ref.~\cite{Cai:2013wma}.}
\end{figure}

Constructing the parameter space is helpful to learn in which region the superconducting orders can coexist. One  can complete this task  by just turning the problem as an eigenvalue problem. Focus on the concrete model discussed in this paper, i.e., $m_1^2=0$ and $m_2^2=-2$, a good starting point is to find the critical valve of the ratio $e_1/e_2$ such that $T$ is a critical temperature at which $\psi_1$ begins to vanish or emerge. At such a temperature, $\psi_1$ is very small and can be treated as a perturbation on the background where only $\psi_2$ condenses
\begin{equation}\label{conduc}
-\psi_1''-(\frac{f'}{f}-\frac{\chi'}{2}+\frac{2}{r})\psi_1'+\frac{m_1^2}{f}\psi_1=\frac{e_1^2}{e_2^2}\frac{\phi^2e^{\chi}}{f^2}\psi_1,
\end{equation}
where $\{\phi, f, \chi\}$ are functions describing the hairy AdS black hole with only $\psi_2$ non-vanishing. Imposing the appropriate boundary conditions, this equation can be considered as an eigenvalue problem with positive eigenvalue $e_1^2/e_2^2$. The full phase diagram for the five superconducting phases is shown in figure~\ref{diagramss}.

From figure~\ref{diagramss},  one  has as many as five superconducting phases in the model apart from the normal phase. Depending on the model parameters $e_1/e_2$ and $e_2$, each phase can be most thermodynamically stable in some region of parameter space. As one increases the strength of the back reaction, the region for Phase-C with the coexisting behaviour of two order parameters only in a narrow window is gradually forced to shrink and finally vanishes at $e_2^{critical}$, while the regions for Phase-A and Phase-B where both order parameters always present enlarge. In this sense, one can conclude that the gravity which provides an equivalent attractive interaction between the holographic order parameters tends to make the coexistence of two orders much more easy rather than more difficult.

In this subsection~\ref{ss},  a holographic superconductor model with more than one order parameter in four dimensions has been studied, where each complex scalar field in the bulk is minimally coupled to a same U(1) gauge field. This can be interpreted as a holographic multi-band superconductor model. Concretely, we have discussed the two-band case with mass squares $m_1^2=0$ and $m_2^2=-2$ for two bulk scalar fields $\psi_1$ and $\psi_2$, respectively. Depending on the strength of the back reaction $1/e_2^2$ and the relative charge ratio $e_1/e_2$  of the two scalar fields, the model admits as many as five different superconducting phases. Three of them, denoted by Phase-A, Phase-B and Phase-C, exhibit the coexistence region of two order parameters. More specifically, for Phase-C, as one lowers the temperature, the second scalar $\psi_2$ condenses following $\psi_1$  will completely suppress the condensate of the first order, i.e., $\psi_1$ will go to zero finally. The condensate behaviours in Phase-A and Phase-B are similar. One of the two orders condenses first, and once the other begins to condense, both always coexist. However, this model is limited to the competition of the order parameters with the same asymmetry. Therefore it is quite interesting to study the holographic models with superconducting order parameters with different spins. This will be done in the following subsections.
\subsection{Competition between  s-wave and p-wave orders}
\label{sp}
In this subsection, we will study  two holographic superconductor models with both s-wave and p-wave condensed. One is proposed in ref.~\cite{Nie:2013sda}, where the authors built a holographic superconductor model with a scalar triplet charged under an SU(2) gauge field in the bulk. The other holographic s+p model in ref.~\cite{Amado:2013lia} consists of a scalar doublet charged under an U(2) gauge field living in a planar Schwarzschild black hole geometry.  The discussions for both models are limited to the probe limit case.
\subsubsection{The holographic s+p superconductor with a scalar triplet charged under an SU(2) gauge field}
To realize the s-wave and p-wave superconductivity in one model, we first consider a real scalar triplet charged in an SU(2) gauge field in the gravity side. The full action is~\cite{Nie:2013sda}
\begin{eqnarray}
\begin{split}
&S =\frac{1}{2 \kappa_g ^2}\int d^{d+1}x \sqrt{-g} (R-2\Lambda)+S_M,\\
&S_M=\frac{1}{g_c^2}\int d^{d+1}x \sqrt{-g}(-D_\mu \Psi^{a} D^\mu \Psi^a-\frac{1}{4}F^a_{\mu\nu}F^{a\mu\nu}-m^2 \Psi^a\Psi^a),
\end{split}
\end{eqnarray}
where $\Psi^a$ is an SU(2) charged scalar triplet in the vector representation of the SU(2) gauge group, and
\begin{equation}
D_\mu\Psi^a=\partial_\mu \Psi^a+\epsilon^{abc}A^b_\mu\Psi^c.
\end{equation}
$F^a_{\mu\nu}$ is the gauge field strength which is the same as~\eqref{supstrength} and reads
\begin{equation}
F^a_{\mu\nu}=\partial_\mu A^a_\nu-\partial_\nu A^a_\mu +\epsilon^{abc} A^b_\mu A^c_\nu.
\end{equation}
$g_c$ is the Yang-Mills coupling constant as well as the SU(2) charge of $\Psi^a$. One can redefine the fields $A^a_{\mu}$ and $\Psi^a$ to get the standard expression where the coupling $g_c$ appears in the derivative operator $D_\mu$. Here we limit ourselves to the case of probe limit. This limit can be realized consistently by taking the limit $g_c\rightarrow\infty$.

In the probe limit, we consider the $d+1$ dimensional AdS black brane as the background with metric
\begin{equation}
\label{metric_sp}
ds^2=-f(r)dt^2+\frac{1}{f(r)}dr^2+r^2dx_i dx^i.
\end{equation}
$x^i$s are the coordinates of a $d-1$ dimensional Euclidean space.
The function $f(r)$ is
\begin{equation}\label{fEinstein}
f(r)=r^2\left(1-\frac{r_h^d}{r^d}\right),
\end{equation}
with $r_h$ the horizon radius. Here the AdS radius $L$ has been set to be unity. The temperature of the black brane is related to $r_h$ as
\begin{equation}\label{TemperatureE}
T=\frac{d }{4\pi}r_h.
\end{equation}
This is just the temperature of dual field theory in the AdS boundary.

Let us consider the following ansatz  for the matter fields
\begin{eqnarray}
\Psi^3=\Psi_3(r),\quad A^1_t=\phi(r),\quad A^3_x=\Psi_x(r),
\end{eqnarray}
with all other field components being turned off. In this ansatz, we take $A^1_\mu$ as the electromagnetic U(1) field. With this ansatz, the equations of motion of matter fields in the AdS black brane background read
\begin{eqnarray}\label{eoms}
\nonumber
\phi''+\frac{d-1}{r}\phi' -\Big(\frac{2  \Psi_3^2}{f}+\frac{\Psi_x^2}{r^2 f}\Big)\phi&=&0, \\
\Psi_x''+\Big(\frac{d-3}{r}+\frac{f'}{f}\Big)\Psi_x'+\frac{\phi^2}{f^2}\Psi_x&=&0, \\ \nonumber
\Psi_3''+\Big(\frac{d-1}{r}+\frac{f'}{f}\Big)\Psi_3'-\Big(\frac{m^2}{f}-\frac{\phi^2}{f^2}\Big)\Psi_3&=&0.
\end{eqnarray}
One can see $\Psi^3$ and $\Psi_x$ are not coupled in their equations of motion, but they are both coupled to the same U(1) electromagnetic field. In this model,  thus one can easily realize the s-wave and p-wave superconductivity consistently.

\begin{figure}[h!]
\centering
\includegraphics[width=9.0cm] {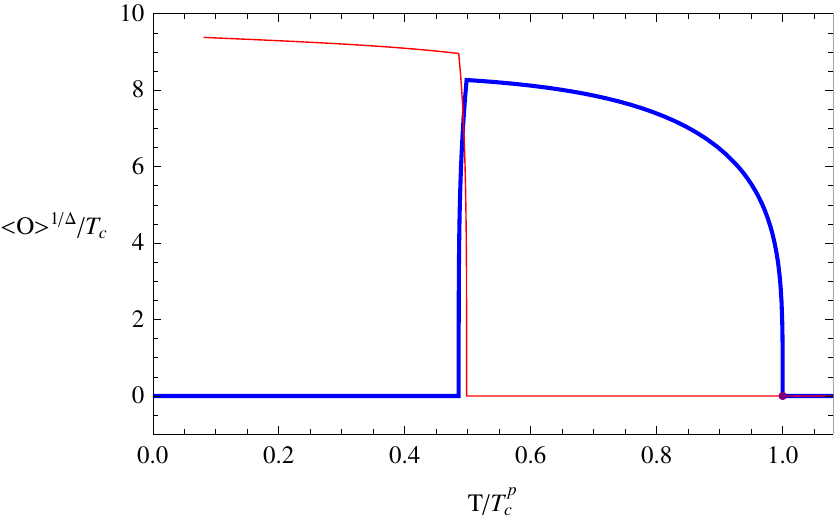}
\caption{\label{SPcondensation} Condensate of the operators in the s+p coexisting phase. The blue curve is for the condensate of the p-wave operator, while the red curve is for the condensate of the s-wave operator. The figure was taken from ref.~\cite{Nie:2013sda}.}
\end{figure}

\begin{figure}
\includegraphics[width=7.5cm] {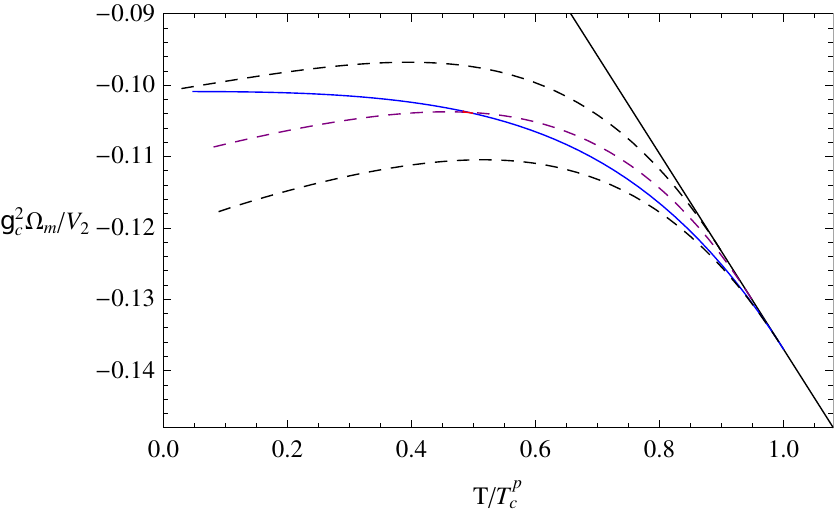}
\includegraphics[width=8.1cm] {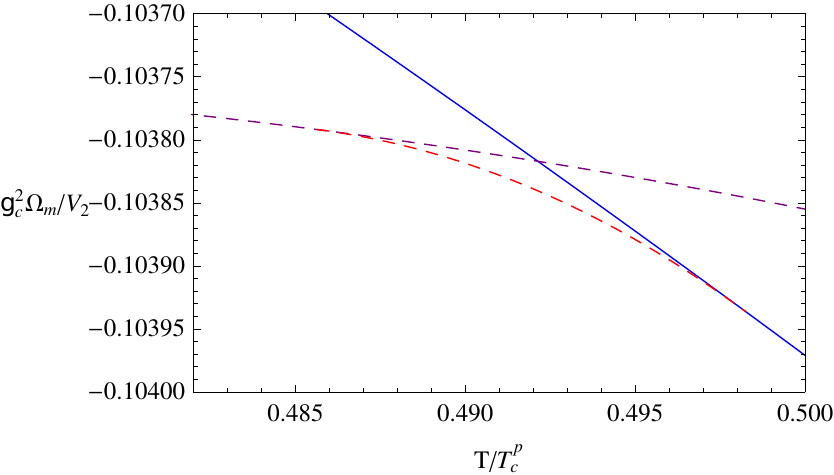}
\caption{\label{FreeEnergysp} (Left) The Gibbs free energy versus temperature for various phases.  The black solid  curve is for the normal phase, the blue solid curve is for the p-wave phase, and the dashed lines from bottom to top are for the s-wave phase with operator dimension $\Delta=\Delta_{cI}$, $(6+\sqrt{3})/4$, and $\Delta_{cII}$, respectively. (Right) The Gibbs free energy in the region near the intersection point of the p-wave curve and the s-wave curve with $\Delta=(6+\sqrt{3})/4$. The plots were taken from ref.~\cite{Nie:2013sda}.}
\end{figure}

We take the case with $\Delta=\Delta_{eg}=(6+\sqrt{3})/4$ as an example. The condensate behaviour for the coexisting phase is drawn in the left plot of figure~\ref{SPcondensation}. We can see that the $s+p$ coexisting phase starts from the p-wave phase and ends with the pure s-wave condensate phase. Based on the calculation of free energy shown in figure~\ref{FreeEnergysp}, we confirm that the s+p coexisting phase indeed has the lowest free energy and is thus thermodynamically favored in the temperature region. Thus the potential first order phase transition from the pure p-wave phase to the pure s-wave phase is replaced by the phase transitions from the p-wave phase to the s-wave phase through an s+p coexisting phase. And all the three phase transitions are continuous ones, and are of characteristic of second order phase transition within the numerical accuracy.

\begin{figure}[h!]
\centering
\includegraphics[width=16cm] {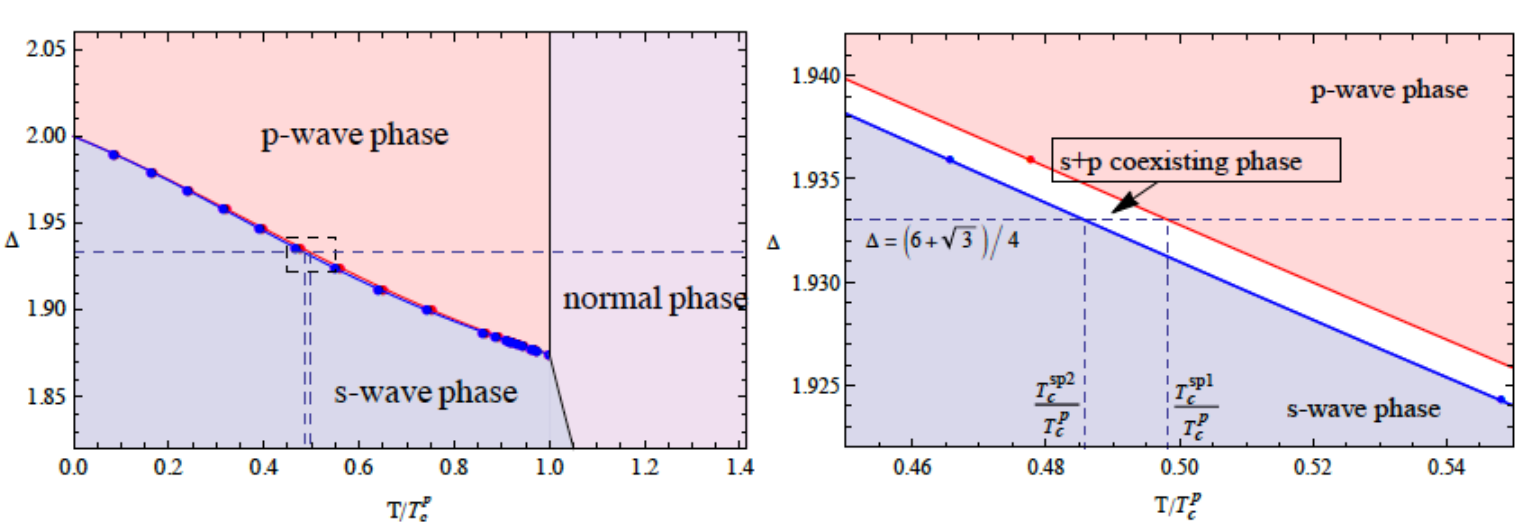}
\caption{\label{phasediagram}The $\Delta-T$ phase diagram. The normal phase, the s-wave phase, d-wave phase and the coexisting phase are colored differently. The right plot is an enlarged version of the coexisting region. The figures were taken from ref.~\cite{Nie:2013sda}.}
\end{figure}

From figure~\ref{FreeEnergysp}, we see that the Gibbs free energy curves of the s-wave and p-wave phases have an intersection when $\Delta_{cI}<\Delta<\Delta_{cII}$. The s+p coexisting phase just exists in this interval.  By computing the values of $T_c^{sp1}$ and $T_c^{sp2}$ and getting the relations $T_c^{sp1}(\Delta)$ and $T_c^{sp2}(\Delta)$ in the region $\Delta_{cI}<\Delta<\Delta_{cII}$ , a phase diagram of the holographic model on the $\Delta$--$T$ plane can be shown in figure~\ref{phasediagram}. We can see from the figure that the system contains four kinds of phases known as the normal phase, the s-wave phase, the p-wave phase and the s+p coexisting phase. The s+p coexisting phase is favored in the area between the blue line and the red line. The region for the s+p coexisting phase is very narrow in the phase diagram. This is similar to the situation of the coexisting phase with two s-wave orders in the probe limit studied in ref.~\cite{Basu:2010fa}.  However, for the latter, the region with the coexisting phase is enlarged with the full back reaction~\cite{Cai:2013wma}. This would be due to the additional interaction between the two scalar fields in the bulk through gravity and  this interaction reduces the repellency between the two condensates.

Very recently, the back reaction effect was included in this model~\cite{Nie:2014qma}, which showed a rich phase structure and various condensate behaviours such as the ``n-type" and ``u-type" ones. The phase transitions to the p-wave phase or s+p coexisting phase become first order in strongly back reacted cases. The phase diagrams similar as figure~\ref{phasediagram} in different strength of back reaction were constructed, indicting that the region for the s+p coexisting phase is enlarged with a small or medium back reaction parameter, while is reduced in the strongly back reacted case.

\subsubsection{The holographic s+p superconductor with a scalar doublet charged under a U(2) gauge field}
In this sector, we consider a holographic s+p model consisting of a scalar doublet charged
under a U(2) gauge field living in a $(3+1)$-dimensional Schwarzschild-AdS black brane geometry. The action for the matter sector  reads~\cite{Amado:2013lia}
\begin{equation}
S=\int d^4 x \sqrt{-g}\left(-\frac{1}{4}\tilde{F}^{\mu\nu}_{c}\tilde{F}_{\mu\nu}^c-m^2\Psi^\dagger \Psi-(D^\mu\Psi)^\dagger D_\mu\Psi\right)\,,
\label{accionsp}
\end{equation}
with
\begin{eqnarray}
&&\Psi=\sqrt{2} \begin{pmatrix}
\lambda \\
\psi
\end{pmatrix}\;,\quad
D_{\mu}=\partial_{\mu}-iA_{\mu}\;,\quad
A_{\mu}=A_{\mu}^{c}T_c\,, \\ \nonumber
&&
 T_0=\frac{1}{2}\mathbb{I}\;,\quad
 T_i=\frac{1}{2}\sigma_i\,.
\end{eqnarray}
The system lives in the Schwarzschild-AdS background~\eqref{metric_sp}. Considering the following consistent ansatz for the fields
\begin{equation}
A^{(0)}_0= \Phi(r)\,,\quad   A^{(3)}_0 = \Theta(r)\,,\quad A^{(1)}_1 = w(r)\,,\quad \psi=\psi(r)\,,
\end{equation}
with all functions being real-valued, the resulting equations of motion read
\begin{equation}\label{pu2eom}
\begin{split}
\psi''+\left(\frac{f'}{f}+\frac{2}{r}\right)\psi'+\left(\frac{(\Phi-\Theta)^2}{4f^2}-\frac{m^2}{f}-\frac{w^2}{4r^2f}\right)\psi=0\,,\\
\Phi''+\frac{2}{r} \Phi'-\frac{\psi^2}{f}(\Phi-\Theta)=0\,,\\
\Theta''+\frac{2}{r}\Theta'+\frac{\psi^2}{f}(\Phi-\Theta)-\frac{w^2}{r^2f}\Theta=0\,,\\
w''+\frac{f'}{f}w'+\frac{\Theta^2}{f^2}w-\frac{\psi^2}{f}w=0\,.
\end{split}
\end{equation}
In what follows we choose the scalar to have $m^2 = -2$ and thus the corresponding dual operator  has mass dimension 2.

The UV asymptotic behaviour of the fields, corresponding to the solution of equations~\eqref{pu2eom} in the limit $r\to\infty$, is given by
\begin{equation}
\begin{split}
&\Phi=\mu-\rho/r+\cdots\,,\\
&\Theta=\mu_3-\rho_3/r+\cdots\,,\\
&w = w^{(0)}+w^{(1)}/r+\cdots\,,\\
&\psi = \psi^{(1)}/r+\psi^{(2)}/r^2+\cdots\,,
\end{split}
\end{equation}
where in the dual field theory side, $\mu$ and $\rho$ are respectively the chemical potential and charge density corresponding
to the overall $U(1)\subset U(2)$ generated by $T_0$, whereas $\mu_3$ and $\rho_3$ are the chemical potential and
charge density corresponding to the $U(1)\subset SU(2)$ generated by $T_3$. $\psi^{(1)}$ is the source of a scalar operator of dimension 2,
while $\psi^{(2)}$ is its expectation value. Finally $w^{(0)}$ and $w^{(1)}$ are the source and vacuum expectation value of the current operator
$J_x^{(1)}$.  Notice that in a background where $w(r)$
condenses the $SU(2)\subset U(2)$ is spontaneously broken, and moreover spatial rotational symmetry is spontaneously broken too.

We are looking for solutions of the equations~\eqref{pu2eom} where $\psi$, $w$, or both acquire non-trivial profiles. First we will switch on a chemical potential $\mu$ along the overall U(1), while requiring that the other chemical potential $\mu_3$ remains null. Therefore the UV boundary conditions are
\begin{equation}
\psi^{(1)}=0\,,\quad w^{(0)}=0\,,\quad \mu_3 =0\,.
\label{uvconds}
\end{equation}
In the IR regularity requires $A_t$ to vanish at the black hole horizon. So far, the holographic multi-component superfluid model has been realized.

\begin{figure}[h!]
\includegraphics[width=7.8cm]{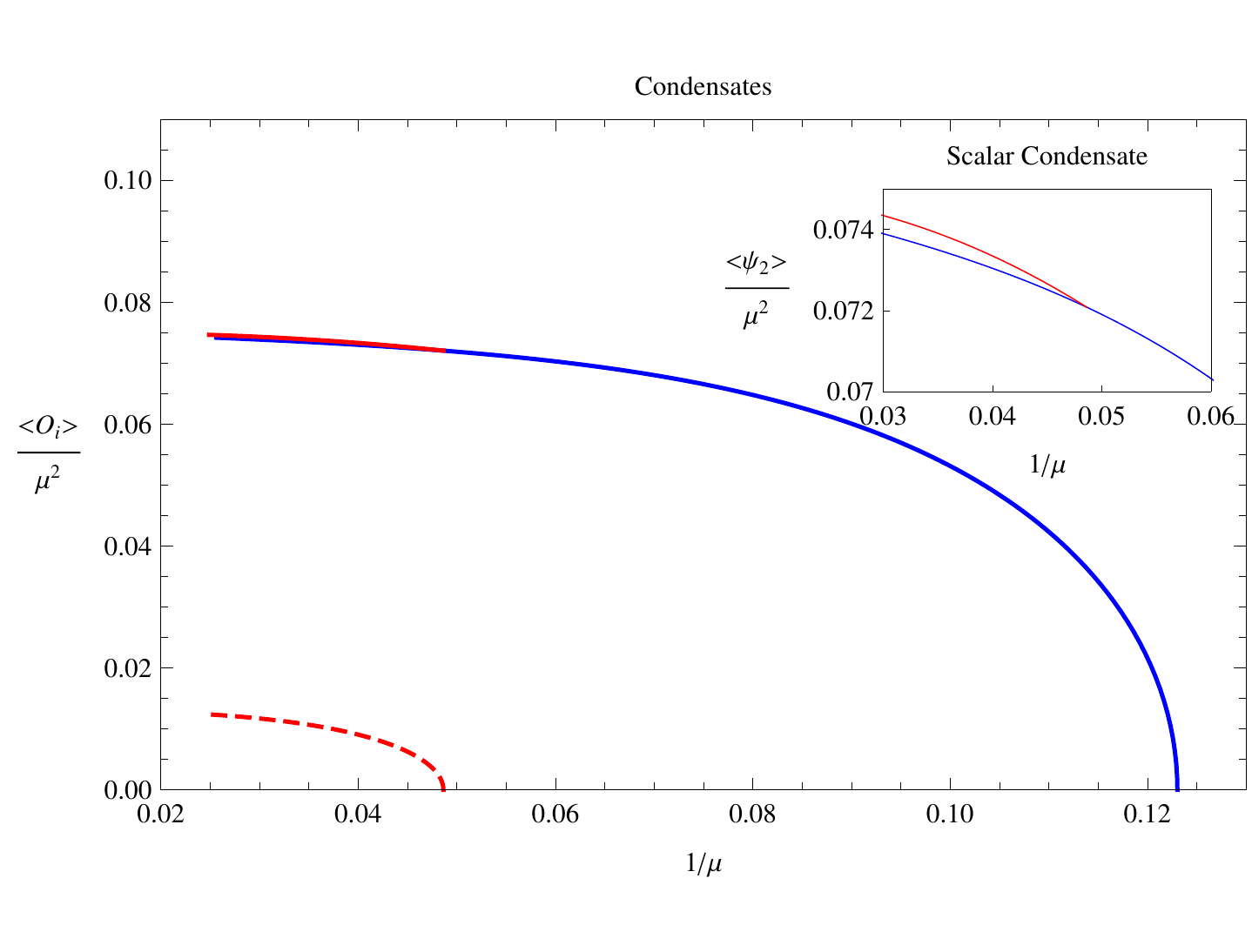}
\includegraphics[width=7.9cm]{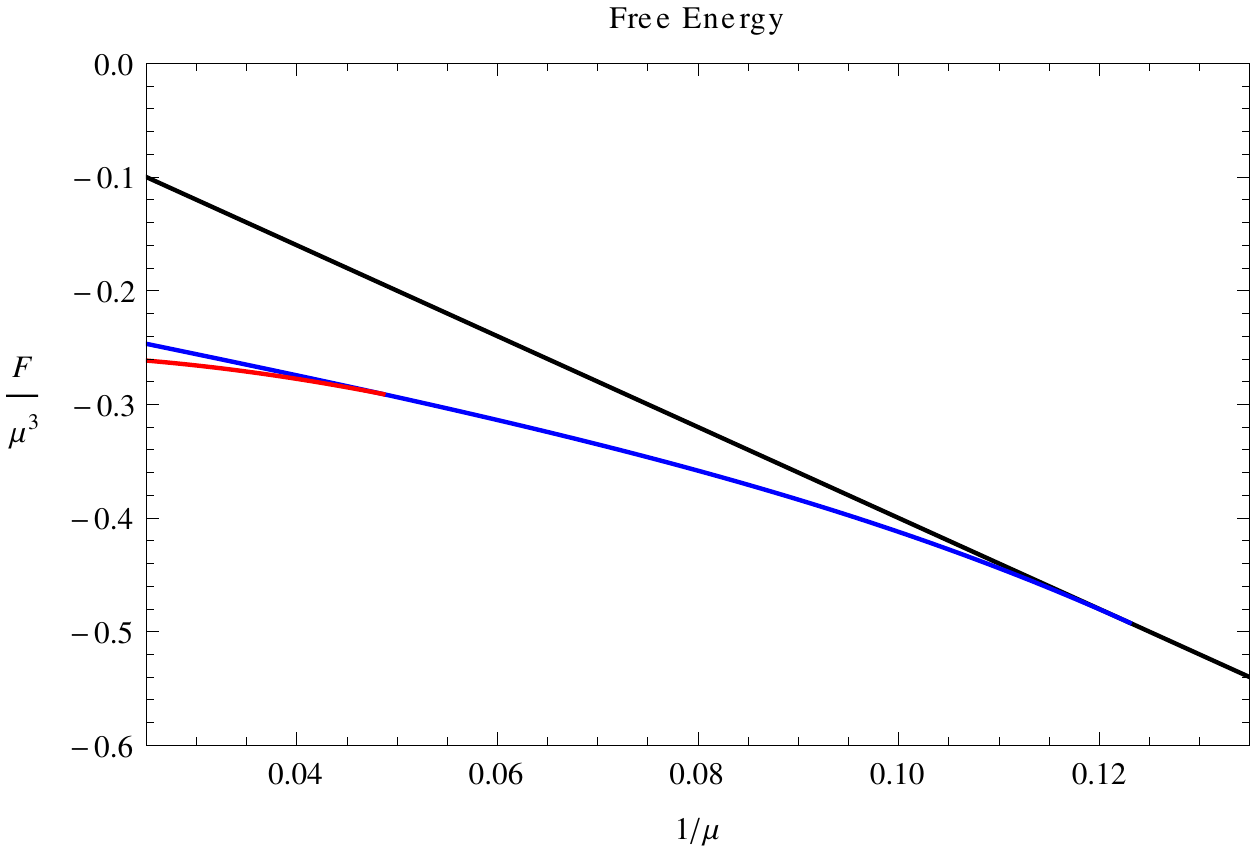}
\caption{\label{condensatespp} Left: Condensates $ \psi^{(2)}$ (solid) and $w^{(1)}$ (dashed) as a function of $1/\mu$ in the s-wave (blue) and s+p-wave (red) phases. The p condensate appears at $\mu_{sp}$ such that $\mu_s/\mu_{sp}\simeq0.395$. The inset zooms in on the plot of $ \psi^{(2)}$ to show the difference in the scalar condensate between the s (blue) and the s+p (red) solutions. Right: Free energy of the different solutions versus $1/\mu$: normal phase in black, s-wave phase in blue, and s+p-wave phase in red. Reprinted with permission from ref.~\cite{Amado:2013lia}.}
\end{figure}

In the left plot of figure~\ref{condensatespp} the condensates $\langle \mathcal{O}_2 \rangle \sim \psi^{(2)}$ and $\langle J_x^{(1)}\rangle\sim w^{(1)}$ are plotted as a function of the chemical potential. Notice that the solution where both condensates coexist extends down to as low $1/\mu$ as where the decoupling limit is trustable. And the free energy for the different solutions is shown in the right plot of figure~\ref{condensatespp}. At small chemical potential only the normal phase solution exists. At $\mu=\mu_s \simeq 8.127$ there is a second order phase transition to the s-wave solution. If one keeps increasing $\mu$, at $\mu_{sp}\simeq 20.56$ there is a second order phase transition from the s-wave phase to the s+p-wave phase. The system stays in the s+p-wave phase for $\mu>\mu_{sp}$.

Next, we relax the condition $\mu_3 =0$ and study the phase diagram of the system as a function of $\mu$ and $\mu_3/\mu$. Notice that turning on a second chemical potential means to explicitly break $U(2)\rightarrow U(1)\times U(1)$. The system can now be interpreted as a holographic dual to an unbalanced mixture. The UV boundary conditions now read
\begin{equation}
\psi^{(1)}=0\,,\quad w^{(0)}=0\,.\label{uvconds2}
\end{equation}
\begin{figure}[h!]
\begin{center}
\includegraphics[scale=0.65]{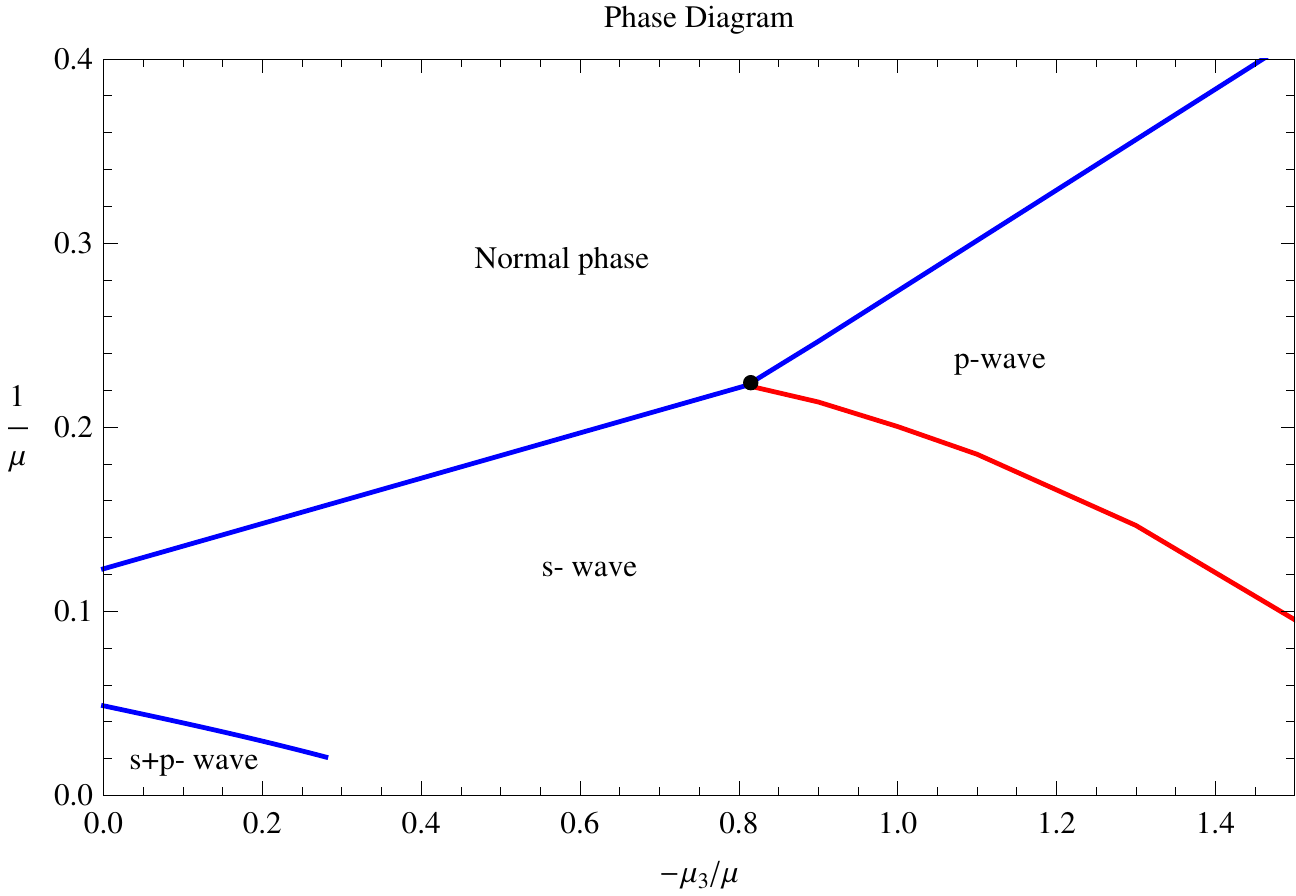}
\caption{\label{phasediagram_2} Phase diagram of the unbalanced system as a function of $1/\mu$ and
$\mu_3/\mu$. Second order phase transitions are denoted by blue lines, whereas the red line corresponds to a
first order phase transition. Reprinted with permission from ref.~\cite{Amado:2013lia}.}
\end{center}
\end{figure}
In this case, by computing the free energy of the different solutions, the phase diagram of the system as a function of $1/\mu$ and $\mu_3/\mu$ is plotted in figure~\ref{phasediagram_2}. For small values of $\mu_3/\mu$, the solution where both condensates coexisting extends down to as lower $1/\mu$ as where we can trust the decoupling limit. As $|\mu_3|/\mu$ gets larger, the transition to the s+p-wave phase happens at a higher value of $\mu$.  For $|\mu_3|/\mu$ is large enough, the p-wave phase is preferred at intermediate values of $\mu$. Therefore, as $\mu$ is increased above a critical value $\mu_p$ the system goes from the normal to the p-wave phase through a second order phase transition. If $\mu$ is increased even further a first order phase transition takes the system from the p-wave to the s-wave phase. The tricritical point where the normal, s-wave and p-wave phases meet happens at $1/\mu\simeq 0.223$ and $|\mu_3|/\mu\simeq 0.815$, the p-wave solution is never energetically preferred for $|\mu_3|/\mu< 0.815$.

In summary, in this subsection we have reviewed the competition between s-wave and p-wave order through two holographic superconductor models. The first model is realized the s+p superconductor with a scalar triplet charged under an SU(2) gauge field and the other is constructed with a scalar doublet charged under a U(2) gauge field. The s+p coexisting phase exists in both models. In the first model, the s+p coexisting phase is narrow and one condensation tends to kill the other. This competing behaviour is similar to the case shown in the condensed matter system~\cite{Pallab:2014}. However, in the second case, the condensates feed on different charge densities and the coexisting phase survives down to a low temperature. Therefore, it should be noted that the competing scenario is model dependent. In next subsection, we will study the competition between the scalar field and the tensor field, i.e., the competition between the s-wave and d-wave orders.

\subsection{Competition between s-wave order and d-wave order}
\label{sd}
In section~\ref{sect:dwave}, we have mentioned two acceptable holographic models describing the d-wave condensation. The CKMWY d-wave model is reviewed in subsection~\ref{subsect:CKMWY model} and the BHRY d-wave model in subsection~\ref{subsect:BHRY model}. In order to realize the condensation of s-wave order and d-wave order in one holographic model, one can simply combine the Abelian-Higgs model with a d-wave model. Thus, one could have two holographic models with s-wave order and d-wave order. Here we will discuss  the competition between s-wave order and d-wave order for both  d-wave models in the probe limit where one neglects the back reaction of matter fields to the background geometry~\cite{Li:2014wca}. The phase structures are given and the behaviours of the thermodynamic quantities for the s+d coexisting phase are also studied. The coexisting phase does appear in both models and is thermodynamically favored.
\subsubsection{The s-wave + BHRY d-wave model}
\label{sect:superfluid}
To study the competition between s-wave and d-wave orders, let us first start with the holographic model by combining the  Abelian-Higgs s-wave model and BHRY d-wave model. The holographic model with a scalar field $\psi_1$, a symmetric tensor field $\varphi_{\mu\nu}$ and a U(1) gauge field $A_\mu$ is described by the following action~\cite{Li:2014wca}:~\footnote{Ref.~\cite{Nishida:2014lta} also discussed the following model, but with a coupling between the scalar field and the tensor field, and studied the phase structure in terms of the coupling parameter and temperature with fixed charges of two orders. In the folloowing discussion, there is no direct interaction between scalar and tensor fields and the model parameter is the ratio of two fields. Note that in paper \cite{Nishida:2014lta}, when the coupling $\eta= 0$, there also exists a coexisting phase under the model parameters $m^2_1 = -2, m^2_2 = 0$ and $q_2 = 1.95$. Both results are consistent with each other in that case.}

\begin{equation}\label{BHRY}
\begin{split}
S=\frac{1}{2\kappa^2}\int d^{4}x\sqrt{-g}(- \frac{1}{4} F_{\mu\nu} F^{\mu\nu} -|D\psi_1|^2-m_1^2|\psi_1|^2+ \mathcal{L}_d),\\
\mathcal{L}_d=-|\tilde{D}_\rho\varphi_{\mu\nu}|^2+2|\tilde{D}_\mu\varphi^{\mu\nu}|^2+
|\tilde{D}_\mu\varphi|^2
-\big[\tilde{D}_\mu\varphi^{\dagger\mu\nu}\tilde{D}_\nu \varphi+\text{h.c.}\big]-i q_2 F_{\mu\nu} \varphi^{\dagger\mu\lambda} \varphi^\nu_\lambda\\
-m_2^2\big(|\varphi_{\mu\nu}|^2-|\varphi|^2\big)+2R_{\mu\nu\rho\lambda} \varphi^{\dagger\mu\rho}\varphi^{\nu\lambda}-\frac{1}{4} R |\varphi|^2,
\end{split}
\end{equation}
where $D_{\mu} = \nabla_\mu - i q_1 A_\mu$ and $\tilde{D}_\mu = \nabla_\mu - i q_2 A_\mu$, $\varphi\equiv{\varphi^\mu}_\mu$, $\varphi_\rho\equiv g^{\mu\lambda}\tilde{D}_{\lambda}\varphi_{\mu\rho}$ and ${R^\mu}_{\nu\rho\lambda}$ is the Riemann tensor of the background metric. $\psi_1$ is the scalar order and $\psi_{\mu\nu}$ is the tensor order. The parameters $q_1$ and $q_2$ are the charges of the scalar and the tensor fields, respectively. One can perform a rescaling to set the charge $q_1$ of the scalar to be unity. Then the phase structure of this theory is determined by the ratio $q_2/q_1$ by fixing the mass square of the scalar field $m_1^2$ and the mass square of the tensor field $m_2^2$. We shall set $q_1=1$ without loss of generality in the following discussion.

Working in the probe limit, we choose the background metric to be the 3+1 dimensional AdS-Schwarzschild black hole with planar horizon~\eqref{metric_sp}. And we consider the following ansatz
\begin{equation}
\label{ansatz_1sd}
A_\mu \, dx^\mu =  \phi(r) \, dt  \;, \quad \psi_1=\psi_1(r)\quad
\varphi_{xy}=\varphi_{yx} = \frac{r^2}{2} \, \psi_2(r) \;,
\end{equation}
with $\phi(r)$, $\psi_1(r)$ and $\psi_2(r)$ all real functions.

With the above ansatz~\eqref{ansatz_1sd}, the equations of motion for $\phi$, $\psi_1$ and $\psi_2$ are given by
\begin{eqnarray}\label{EOMs}
\begin{split}
\phi'' + \frac{2 \phi'}{r} - \frac{2}{f}\phi \psi_1^2- \frac{q_2^2} {f} \phi \psi_2^2=&0, \\
\psi_1'' + \frac{f'}{f} \psi_1'+ \frac{2}{r} \psi_1' + \frac{\phi^2}{f^2} \psi_1- \frac{m_1^2}{f} \psi_1=&0, \\
\psi_2'' + \frac{f'}{f} \psi_2' + \frac{2}{r} \psi_2' + \frac{ q_2^2 \phi^2}{f^2} \psi_2- \frac{m_2^2}{f} \psi_2=&0.
\end{split}
\end{eqnarray}
Here the prime denotes the derivative with respect to $r$. With this model at hand, we can study the competition mechanism between the s-wave order and d-wave order.
It is easy to see that equations~\eqref{EOMs} have a symmetry
\begin{equation}\label{symmetry}
m_1^2\leftrightarrow m_2^2,\ q_2\rightarrow 1/q_2, \ \phi\rightarrow q_2 \phi,\  \psi_1\rightarrow q_2\psi_2 /\sqrt{2},\ \ \psi_2\rightarrow\sqrt{2} q_2\psi_1.
\end{equation}
Under this symmetry transformation, the role of s-wave and d-wave would interchange each other. Without loss of generality, here we  focus on the case $m_1^2<m_2^2$.

Before solving the set of coupled equations~\eqref{EOMs} numerically, we make a briefly qualitative analysis on the possible phases for such a model. Following ref.~\cite{Basu:2010fa}, we rephrase the equations for the s-wave and d-wave as a potential problem. The evolution equations for s-wave and d-wave in equations~\eqref{EOMs} can be rewritten as follows
\begin{eqnarray}
\begin{split}
\frac{d^2}{d y^2}\tilde{\psi_1}-\tilde{V}_{1eff}(y)\tilde{\psi_1}&=&0, \\
\frac{d^2}{d y^2}\tilde{\psi_2}-\tilde{V}_{2eff}(y)\tilde{\psi_2}&=&0,
\end{split}
\end{eqnarray}
where $dy=-\frac{dz}{z^2 f}$ with $z=1/r$ and $V_{1eff}(z)=-f^2(\frac{\phi^2}{f^2}-\frac{m_1^2}{f}+\frac{f_{,z}}{f}z^3)$ and $V_{2eff}(z)=-f^2(\frac{q_2^2 \phi^2}{f^2}-\frac{m_2^2}{f}+\frac{f_{,z}}{f}z^3)$. Now in terms of the new variable $y$, the equations of motion for s-wave and d-wave are rephrased as a potential problem on a semi infinite line, i.e., $y\in [0, \infty)$. Our qualitative discussion is based upon the lemma proven in ref.~\cite{Basu:2010fa}. For the case $q_2^2<1$, no matter which gauge field configuration we choose, we always have $V_{1eff}<V_{2eff}$. Therefore the phase structure of the system is the same as that of s-wave holographic superconductor with a single scalar. While, for the case $q_2^2\geq1$, one may expect that the d-wave field with large charge $q_2$ will always dominate. However, the potential $V_{1eff}$ diverges like $\frac{1}{y^2}$ near the boundary $y=0$ when we lower the temperature. Therefore, lowering the temperature possibly makes the mass dependent potential more important and hence the s-wave tends to dominate. We will confirm this with the following numerical calculation.

\begin{figure}[h!]
\centering
\includegraphics[scale=1]{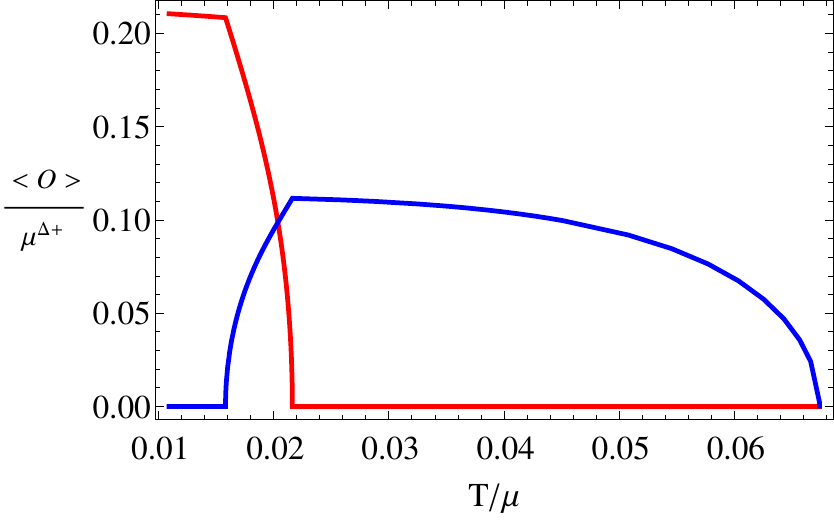}\ \ \ \
 \caption{\label{co} Condensate of the operators in the s+d coexisting phase. The blue curve is for the condensate of the d-wave operator, while the red curve is for the s-wave operator. The figure was taken from ref.~\cite{Li:2014wca}.}
\end{figure}

\begin{figure}[h!]
\centering
\includegraphics[scale=0.9]{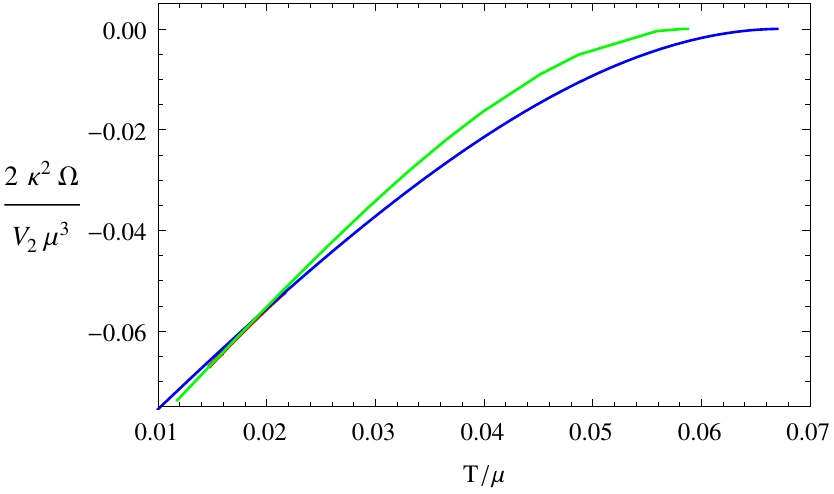}\ \ \ \
\includegraphics[scale=0.91]{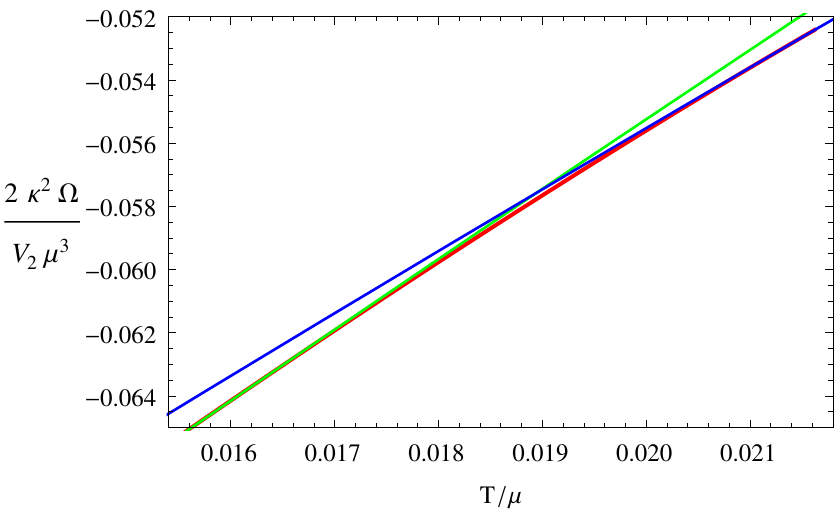}\ \ \ \
 \caption{\label{free} The left plot shows the difference of Gibbs free energy  between the superconducting phase and the normal phase. The blue curve is for the d-wave phase, the green line is for the s-wave phase, while the red curve is for the s+d coexisting phase.  The right plot is an enlarged version of the left one to show the s+d phase more clearly. Figures taken from ref.~\cite{Li:2014wca}.}
\end{figure}

Here we set the mass square $m_1^2=-2$ and $m_2^2=7/4$ and we take $q_2=2.66$ as a typical example. Our numerical results confirm that the model does admit the coexistence region of two orders with different symmetry, which is drawn in figure~\ref{co}. We find that the s+d coexisting phase starts from the d-wave phase and ends with the pure s-wave condensate phase. The calculation of the free energy confirms that the coexisting s+d phase is thermodynamically favored as shown in figure~\ref{free}.

\begin{figure}[h!]
\centering
\includegraphics[scale=0.9]{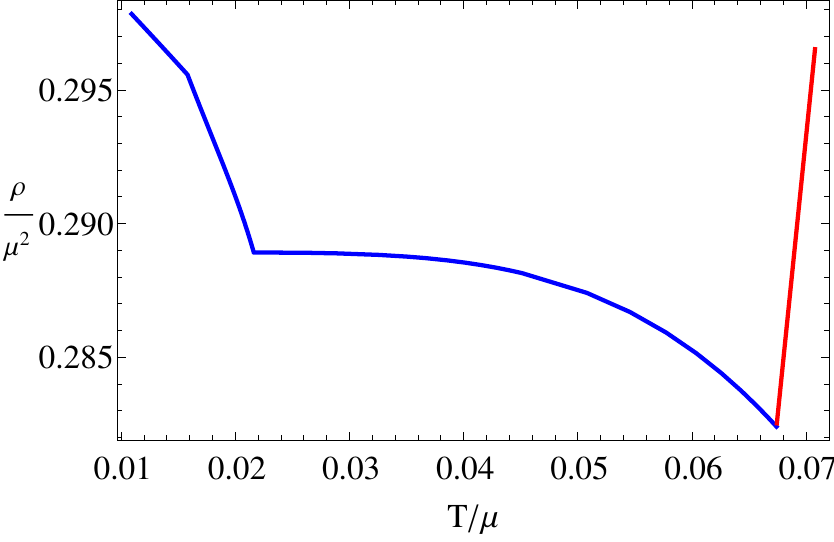}
\includegraphics[scale=0.88]{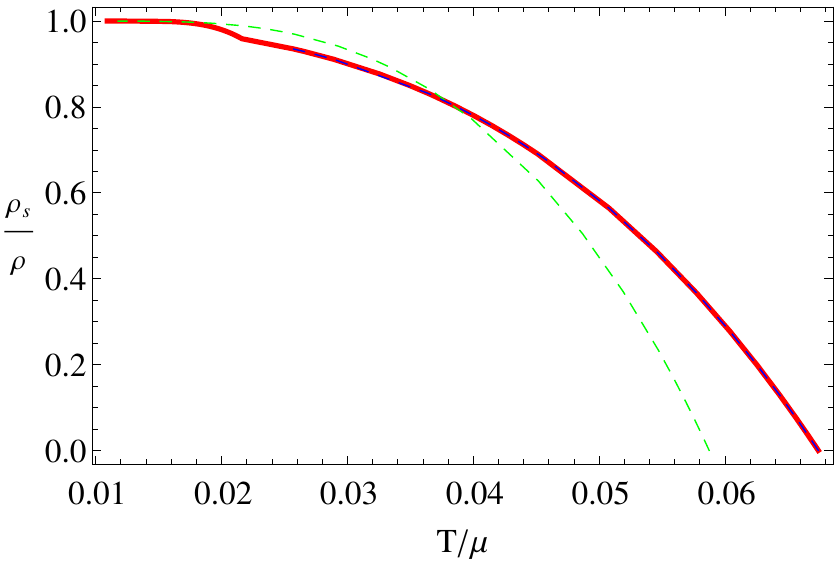}
 \caption{\label{ratio_1} Left: The total charge density as a function of temperature. Right: The ratio of the superconducting charge density over the total charge density $\rho_s/\rho$ versus temperature. The plots were taken from ref.~\cite{Li:2014wca}.}
\end{figure}

Based on the above discussions, it can be seen clearly that there exist three particular points at which the derivative of the charge density with respect to temperature is discontinuous, indicating a second order phase transition. The one with the highest temperature is the critical point for the superconducting phase transition, while the remaining two points are inside the superconducting phase, indicating the appearance and disappearance of coexisting phase. We can also see the signal of phase transition from the behaviour of the total charge density as a function of temperature and the ratio $\rho_s/\rho$ versus temperature shown in figure~\ref{ratio_1}, where $\rho_s$ is the superconducting charge density $\rho_s=\rho-\rho_n$ and $\rho_n$ is the normal charge density carried by the black hole.

\begin{figure}[h]
\centering
\includegraphics[scale=0.9]{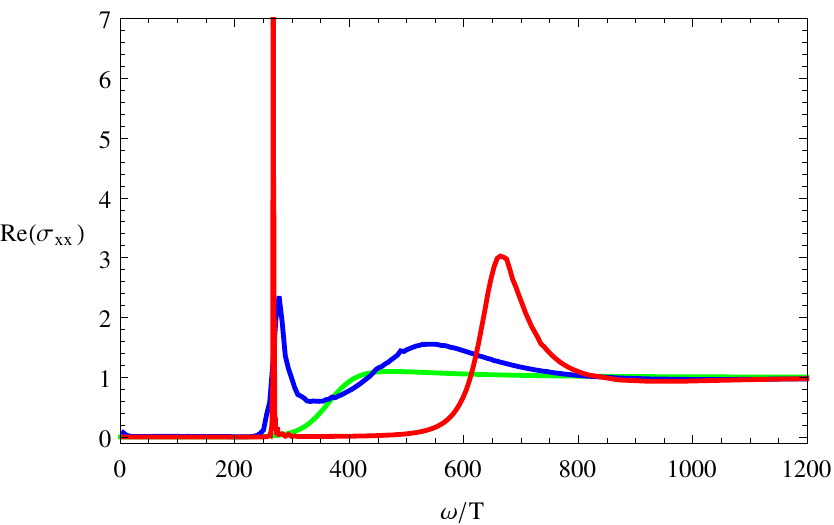}
\includegraphics[scale=0.92]{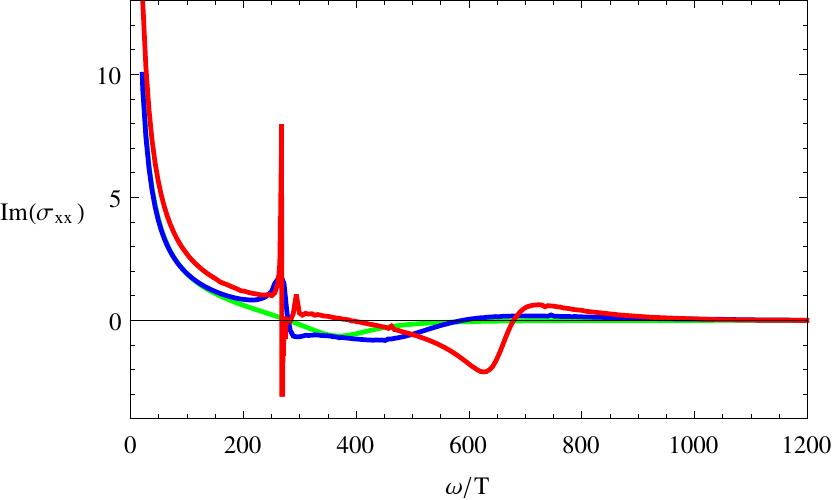}
 \caption{\label{aa} The real part (left) and imaginary part (right) of the conductivity as a function of frequency at temperature $T=0.018\mu$. The red curve is for the s+d coexisting phase, the green line is for the pure s-wave phase and the blue curve for the pure d-wave phase. Plots taken from ref.~\cite{Li:2014wca}. }
\end{figure}

In order to ensure the system is indeed in a superconducting state, we calculate the optical conductivity $\sigma(\omega)$, which corresponds to the red line shown in figure~\ref{aa}. We see that much more interesting phenomena happen in the low frequency region. Unlike the s-wave case which only has a bump at $\omega/T \simeq 400$ in figure~\ref{aa}, for pure d-wave condensate, apart from a much more obvious bump at $\omega/T \simeq 500$, Re($\sigma_{xx}$) has an additional spike at a lower frequency. This spike may indicate the existence of a bound state~\cite{Benini:2010pr}. One can see clearly that such peak becomes much more sharper in the s+d coexisting state, thus the bound state is enhanced due to the additional condensate of s-wave order.

\begin{figure}[h!]
\centering
\includegraphics[scale=1]{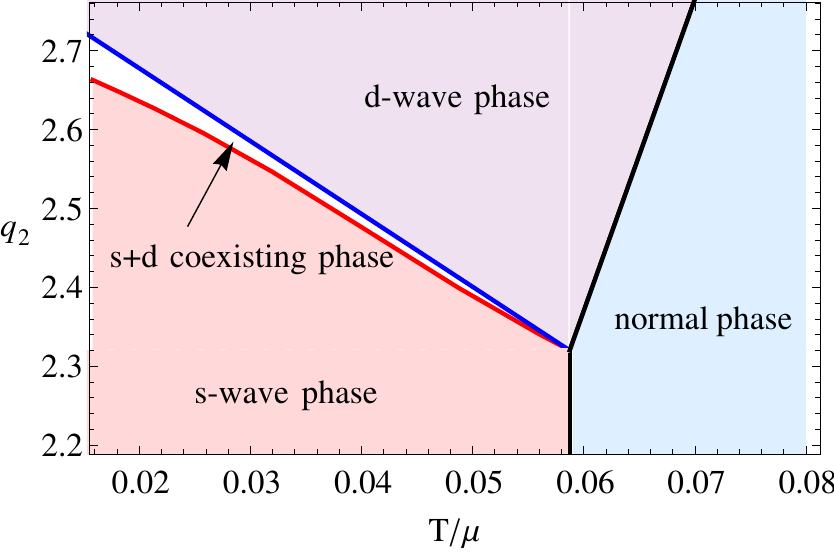}\ \ \ \
 \caption{\label{phase_1} The $q_2$-$T$ phase diagram. The four phases are colored differently and we label the most thermodynamically favoured phase in each region. The figure was taken from ref.~\cite{Li:2014wca}.}
\end{figure}

To be complete, we give the phase diagram with $m_1^2=-2$ and $m_2^2=7/4$  shown in figure~\ref{phase_1}, which can tell us in which region the coexisting phase appears. From figure~\ref{phase_1}, we see that the coexisting phase exists only in a narrow region in the phase diagram. We denote the critical temperature for a single s-wave or d-wave starting to condense as $T_{cs}$ and $T_{cd}$. If we set the charges of the s-wave and d-wave fields to unity, then $T_{cs}/\mu \simeq 0.0588$ and $T_{cd}/\mu\simeq0.0253$. We see that
\begin{itemize}
\item  In the regime $q_2<T_{cs}/T_{cd}\simeq 2.323$, the s-wave dominates the system and there is no condensation of the d-wave order.
\item  As $q_2$ increases beyond $2.323$, the s+d phase appears, which emerges from the d-wave phase and ends with a pure s-wave.
\item If we continue increasing $q_2$ to the case $q_2>1.155 T_{cs}/T_{cd}\simeq 2.683$, the s-wave order never condenses and the resulting phase diagram is the same as that of model with only d-wave order.
\end{itemize}

Finally, we try to give a qualitative explanation on the mechanism through which the condensation of one order affects the dynamics of the other order. Note that here the back reaction is not taken into account. Thus the two fields interact only through their effect on the gauge field once one or both has (have) condensed. Through looking at the gauge field we may give some insight into the competing mechanics between two orders.
\begin{figure}[h!]
\centering
\includegraphics[scale=1]{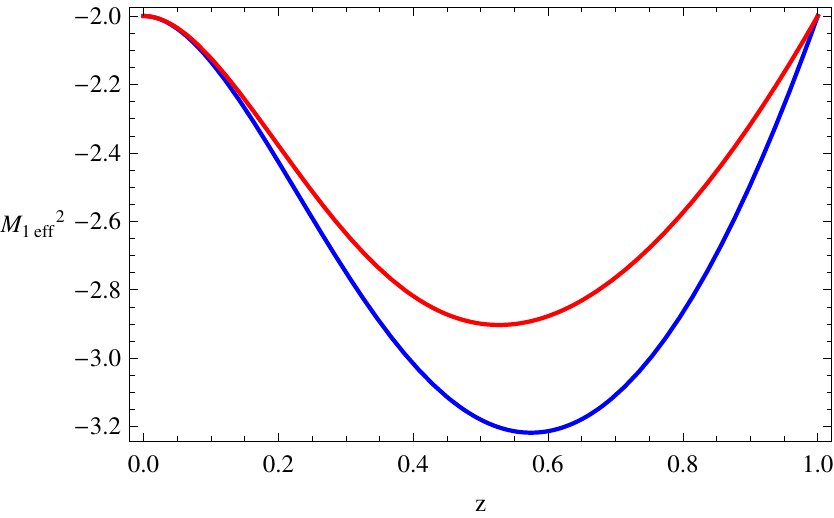}\ \ \ \
 \caption{\label{eff_mass} The blue curve is the effective mass square of s-wave without the condensation of d-wave. The red curve is the effective mass square of the s-wave under the condensation of d-wave. It can be seen clearly that the effective mass of s-wave increases after the condensation of d-wave. This figure was taken from ref.~\cite{Li:2014wca}.}
\end{figure}
\begin{itemize}
\item First, after the d-wave order condenses, if one keeps lowering the temperature and reaches the critical temperature at which the pure s-wave would condense, this condensation does not happen. This is due to the fact that the condensation of the d-wave increases the effective mass of the s-wave, thus prevents the instability of the s-wave to happen, which can be seen from figure~\ref{eff_mass}. This reflects the competition between s-wave and d-wave.
\item However,  further decreasing the temperature, the condensation of s-wave does happen. This is due to the fact that the effective mass of the s-wave is lowered and ultimately even if the condensation of the d-wave depleted the gauge potential, the background with only d-wave order becomes unstable.
\item Finally, the condensate of the s-wave order kills the first one. This should be due  to the effective mass square of the s-wave being lower.
\end{itemize}
It should be noted that this phenomenon is model dependent. This narrow coexistence region of two superconducting orders and the fact that one condensate can eventually kill the other also happen for two s-wave orders in ref.~\cite{Basu:2010fa} and the s+p case in ref.~\cite{Nie:2013sda}. The competition diagram is similar to the competition between the conventional s-wave and the triplet Balian-Werthamer or the B-phase pairings in the doped three dimensional narrow gap semiconductors, such as $\mathrm{Cu}_x\mathrm{Bi}_2\mathrm{Se}_3$ and $\mathrm{Sn}_{1-x}\mathrm{In}_x\mathrm{Te}$ in the condensed matter system~\cite{Pallab:2014}. Although in ref.~\cite{Pallab:2014} the competition is apparently between a s-wave order and a p-wave order, d-wave and p-wave are similar in some circumstances, for example, their excitations of the normal component can be probed using low frequency photons.

\subsubsection{The s-wave + CKMWY d-wave model}
With the same strategy, in this subsection we study the competition between s-wave order and d-wave order in the model combining the Abelian-Higgs s-wave model~\cite{Hartnoll:2008vx} with the CKMWY d-wave model~\cite{Chen:2010mk}. The full action including a U(1) gauge field $A_\mu$, a complex scalar field $\psi_1$ and a symmetric, traceless tensor field $B_{\mu\nu}$ takes the following form~\cite{Li:2014wca}
\begin{equation}\label{CKMWY}
S=\frac{1}{2\kappa^2}\int d^4
x\sqrt{-g}(-\frac{1}{4}F_{\mu\nu}F^{\mu\nu}-|D\psi_1|^2-m_1^2|\psi_1|^2+\tilde{\mathcal{L}}_d),
\end{equation}
with
\begin{equation}
\tilde{\mathcal{L}}_d=-g^{\mu\lambda}(\tilde{D}_{\mu}B_{\nu\gamma})^{\dagger}\tilde{D}_{\lambda}B^{\nu\gamma}-m_2^2B_{\mu\nu}^{\dagger}B^{\mu\nu}.
\end{equation}
Here $D_\mu = \nabla_\mu - i q_1 A_\mu$ and $\tilde{D}_\mu = \nabla_\mu - i q_2 A_\mu$.
In the probe limit, matter fields can be treated as perturbations in the 3+1 dimensional AdS black hole background~\eqref{metric_sp}. Let us consider the following ansatz
\begin{equation}
\psi_1=\psi_1(r),\quad B_{xx}=-B_{yy}=\psi_2(r),\quad A_t=\phi(r)dt,
\end{equation}
with all other field components being turned off and $\psi_1(r)$, $\psi_2(r)$ and $\phi(r)$ being real functions. Then the explicit equations of motion read
\begin{equation}\label{EOM_2}
\begin{split}
\phi''+\frac{2}{r}\phi'-\frac{4 q_2^2 \psi_2^2}{r^4f}\phi-\frac{2q_1^2 \psi_1^2}{f}\phi=0,\\
\psi_1''+(\frac{f'}{f}+\frac{2}{r})\psi_1'+\frac{q_1^2\phi^2}{f^2}\psi_1-\frac{m_1^2}{f}\psi_1=0,\\
\psi_2''+(\frac{f'}{f}-\frac{2}{r})\psi_2'+\frac{q_2^2\phi^2}{f^2}\psi_2-\frac{2f'}{rf}\psi_2-\frac{m_2^2}{f}\psi_2=0.
\end{split}
\end{equation}
\begin{figure}[h!]
\centering
\includegraphics[scale=0.85]{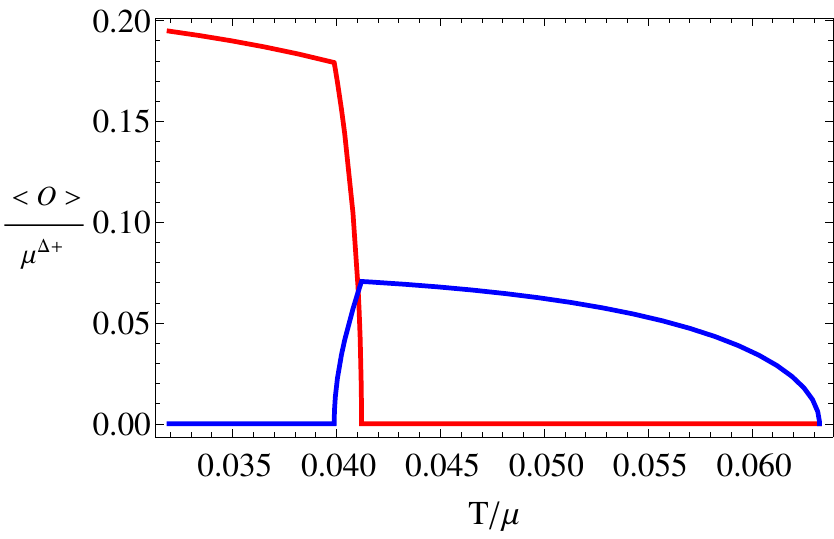}\ \ \ \
\includegraphics[scale=0.85]{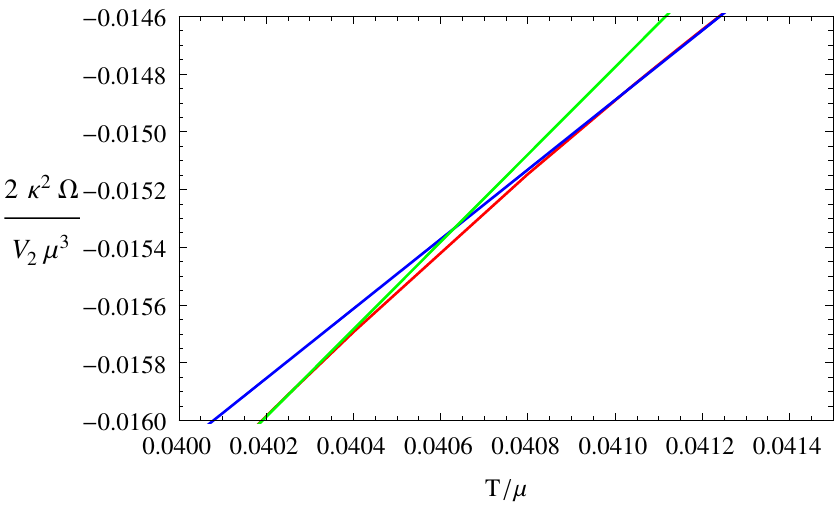}\ \ \ \
 \caption{\label{condensate_2}  The let plot  shows the condensation in the s+d coexisting phase. The right plot shows  the differences of Gibbs free energy between superconducting phases and the normal phase. Here the blue line stands for the d-wave phase, the green one for the s-wave phase and the red one for
the s+d coexisting phase. The figure is reproduced from ref.~\cite{Li:2014wca}.}
\end{figure}

The numerical results are shown in figure~\ref{condensate_2} for the case with $q_1=1$, $q_2=1.34$, $m_1^2=-2$ and $m_2^2=-13/4$. As seen, a new phase with both s-wave order and d-wave order coexistence can appear near $T^{cross}$ and this s+d coexisting phase has the lowest free energy and is thus thermodynamically preferred to the s-wave phase and d-wave phase. In more detail, as we lower the temperature of the system, it first undergoes a phase transition from the normal phase to the pure d-wave phase at $T_c^d$. Then at $T_c^{sd1}$, a new phase transition occurs, and the system goes into an s+d coexisting phase. Finally the system undergoes the third phase transition from the s+d coexisting phase to a pure s-wave phase at $T_c^{sd2}$. Note that all the three phase transitions are second order.

\begin{figure}[h]
\centering
\includegraphics[scale=0.9]{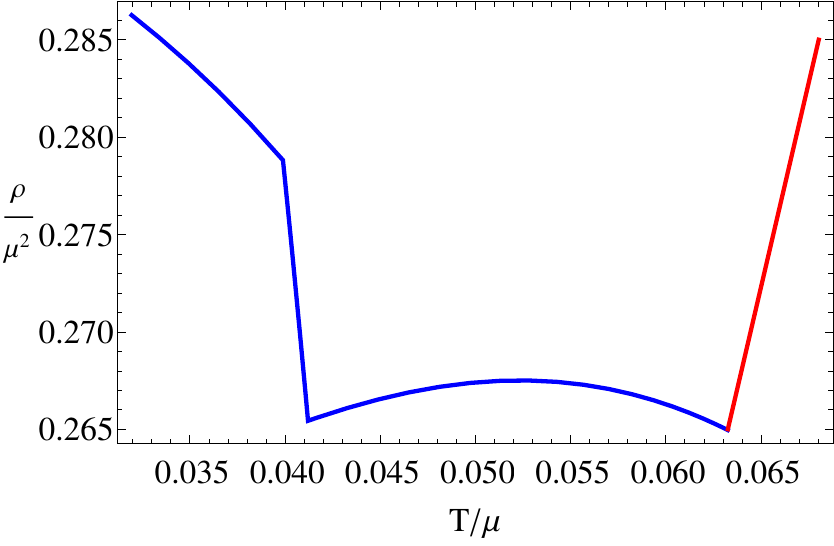}\ \ \ \
\caption{\label{charge_2} The total charge density as a function of the temperature. The red curve is for the normal phase, while the blue one corresponds to the superconducting phase. There are three special temperatures at which the derivatives of charge density with the temperature are discontinuous. Figure taken from ref.~\cite{Li:2014wca}.}
\end{figure}

The feature of the phase transitions can also be seen clearly from the charge density as the function of temperature in figure~\ref{charge_2}.  One can see that the charge density with respect to temperature is continuous, but its derivative is discontinuous at three special points, indicating three second order phase transitions. These features are the same as those for the model in the previous subsection. But there is a little difference in the behaviour of the total charge density for the d-wave phase. In the s-wave + BHRY d-wave model, the total charge density changes monotonously with the temperature, while it behaves non-monotonous in the present case.

\begin{figure}[h]
\centering
\includegraphics[scale=0.9]{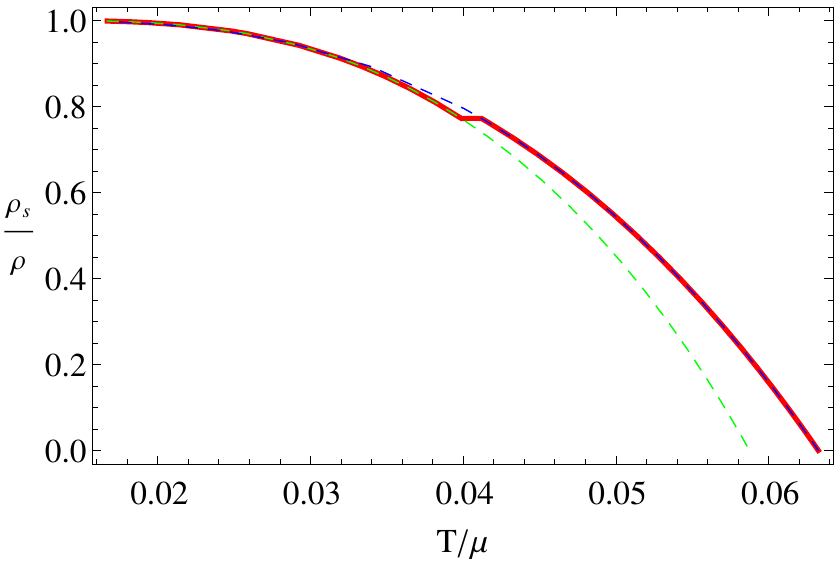}\ \ \ \
\caption{\label{ratio_2} The ratio of the superconducting charge density over the total charge density, $\rho_s/\rho$, with respect to the temperature. The red curve describes the ratio $\rho_s/\rho$ when the system transfers from the d-wave phase to s-wave phase through the s+d coexisting phase. The green dashed blue curve is for the ratio $\rho_s/\rho$ of the pure s-wave phase and the blue dashed curve is the ratio for the pure d-wave phase. The figure was taken from ref.~\cite{Li:2014wca}.}
\end{figure}

The information of the phase transitions can also be revealed via the behaviour of the ratio $\rho_s/\rho$ with respect to the temperature.
From figure~\ref{ratio_2}, one can see that the ratio $\rho_s/\rho$ also has a small kink in the region of the coexisting phase. Comparing figure~\ref{ratio_1} with figure~\ref{ratio_2}, we see that in the former case, the green dashed curve for the pure s-wave phase intersects with the blue dashed curve for the pure d-wave phase. In contrast, the green dashed curve in figure~\ref{ratio_2} is always lower than the blue dashed curve. Therefore, as one lowers the temperature, the ratio $\rho_s/\rho$ in the s+d coexisting phase increases for the former \eqref{BHRY}, while it decreases for the latter~\eqref{CKMWY}. The authors of ref.~\cite{Nie:2013sda} investigated an s+p coexisting phase and found the decrease of the ratio $\rho_s/\rho$ in the coexisting phase, similar to figure~\ref{ratio_2}. They suggested that it might be an experimental signal of the phase transition from a single condensate phase to a coexisting phase. Nevertheless,  the results here uncover that the ratio $\rho_s/\rho$ versus temperature is model dependent.

\begin{figure}[h!]
\centering
\includegraphics[scale=0.88]{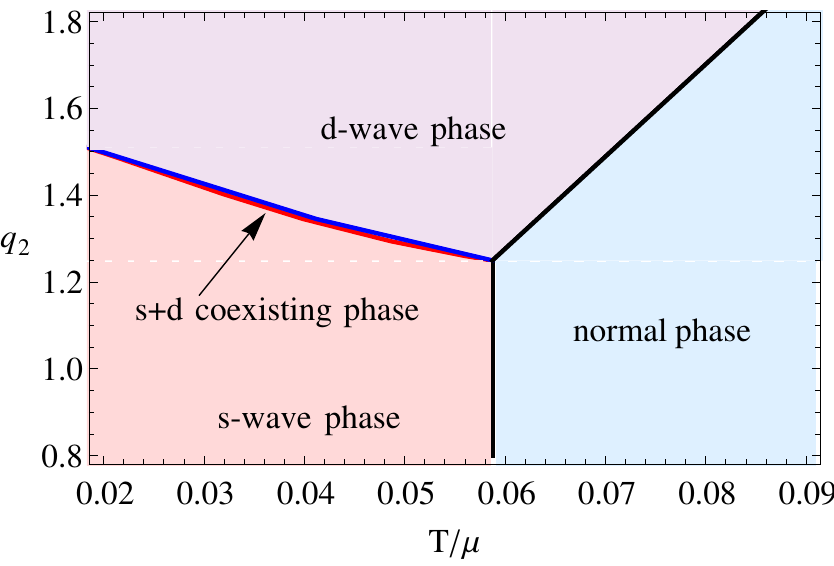}\ \ \ \
\includegraphics[scale=0.88]{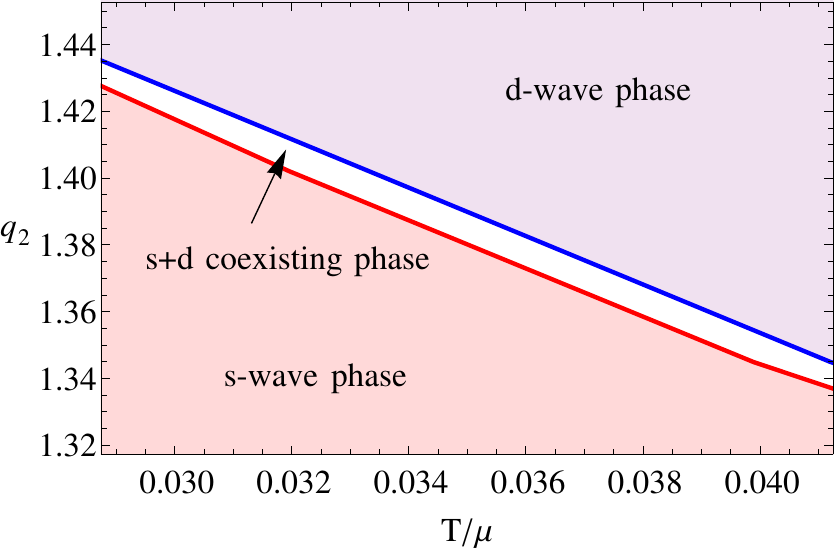}\ \ \ \
 \caption{\label{phase_2} The $q_2$-$T$ phase diagram with $m_1^2=-2$ and $m_2^2=-\frac{13}{4}$. The most thermodynamically favored phase in each part is labeled. The s+d coexisting phase exists only in a narrow region. The right plot is an enlarged version for the coexisting region in order to see this more clearly. Plots taken from ref.~\cite{Li:2014wca}.}
\end{figure}

The phase diagram for the model~\eqref{CKMWY} with $m_1^2=-2$ and $m_2^2=-13/4$ in the $q_2-T$ plane is shown in figure~\ref{phase_2}. As the s-wave + BHRY d-wave model,  the system also contains four kinds of phases known as the normal phase, s-wave phase, d-wave phase and  s+d coexisting phase. The normal phase dominates in the high temperature region, the s-wave phase dominates in the lower temperature region with small $q_2$ below the red curve, and the d-wave phase dominates in the higher temperature zone with large $q_2$ above the blue curve. The s+d coexisting phase is favoured in the area between the red and blue curves. The region for the s+d coexisting phase is very narrow in the phase diagram, which indicates that the s-wave and d-wave phases generally repel each other, but they can coexist in a very small range of temperature.

Comparing the two holographic setups, i.e., the model~\eqref{BHRY} and the model~\eqref{CKMWY}, one can see some common features as follows.
\begin{itemize}
\item The s+d coexisting phase does exist in a region of the model parameter $q_2/q_1$. Once the coexisting phase appears, it is always thermodynamically favoured, compared to the pure s-wave and pure d-wave superconducting phases, which can be seen from the free energy in figure~\ref{free} and figure~\ref{condensate_2}.
\item All phase transitions are second order in these two holographic models.
\item One can see from figure~\ref{phase_1} and figure~\ref{phase_2} that the phase structure is very similar for both models. The region for the s+d coexisting phase is very narrow in the phase diagram, indicating that the s-wave and d-wave phases generally repel each other.
\end{itemize}

There exist also  some differences in the two models. For suitable model parameters in the first model, as the temperature is lowered, the s-wave order condenses inside the d-wave order resulting in the coexisting phase with both orders. However, when the scalar order condenses the first one starts to disappear, and finally only the s-wave condensate is left for sufficiently low temperatures. If one changes the model parameter $m_1^2\leftrightarrow m_2^2$, the inverse is also true: the condensate of d-wave order emerges following the condensate of s-wave order, and then the d-wave condensate finally kills the s-wave order. Those two kinds of coexisting phase are one to one correspondence. In contrast, in the second model, one sees only  the first kind of the coexisting phase. What's more, for the first model, the ratio $\rho_s/\rho$ increases in the s+d coexisting phase as the temperature is lowered, while it decreases in the second case. This gives an obvious evidence that the ratio $\rho_s/\rho$ versus temperature is model dependent.

\section{Coexistence and Competition of Magnetism and Superconductivity}
\label{sect:M&S}
The novel paired mechanism makes p-wave superconductor have many features which are different from the traditional knowledge coming from s-wave superconductor both in theories and experiments. In the usual picture,  superconductivity and magnetism are incompatible with each other. Especially, ferromagnetic phase, a spontaneously magnetized phase which has nonzero magnetic moment without external magnetic field and appears when the temperature is lower than a critical one called ``Curie temperature", can not coexist with superconductivity at a sample.\footnote{However, under special conditions superconductivity may coexist with antiferromagnetic order, where neighboring electron spins arrange in an antiparallel configuration. Since antiferromagnets don't have net magnetism, we won't involve them here.} This is rooted in the microscopic theory of superconductivity from BCS theory. However, this understanding is broken by p-wave superconductor. The discovery of the superconducting ferromagnet\footnote{We will use ``superconducting ferromagnet" to denote the materials whose Curie temperature is higher than superconducting transition temperature and ``ferromagnetic superconductor"  to denote the opposite case.} materials, such as UGe$_2$~\cite{Lonzarich},  URhGe~\cite{Aoki}, UCoGe~\cite{Huy} and ZrZn$_2$~\cite{Pfleiderer}, came as a big surprise. In this material, superconductivity is realized well below the Curie temperature, without expelling the ferromagnetic order.

The nature of superconducting state in ferromagnetic materials is currently under debate. For a review of phenomenological theory of ferromagnetic unconventional superconductors with spin-triplet Cooper pairing of electrons, one can see refs.~\cite{D.I.Uzunov,MacHida,Nevidomskyy}. However, the microscopic theory about the coexistence of magnetism and superconductivity in strongly interacting heavy electrons is either too complex or insufficiently developed to describe the complicated behaviour. So it is still a fascinating thing to find a suitable theory to describe the coexistence and competition of the ferromagnetism and superconductivity in strong correlated system.

Holographic frame to discuss the coexistence and competition between spontaneously magnetic order phase and superconductivity initiated  in refs.~\cite{Amoretti:2013oia,Iqbal:2010eh}. Because of lacking an individual model to describe spontaneously magnetization and the time reversal symmetry broken, these models cannot give complete features of this topic. A very new idea proposed in ref.~\cite{Cai:2014oca} tries to give an independent model describing spontaneously magnetization in holographic frame, which opens a new direction. We will introduce the main results in this framework. For more details, one can refer to refs.~\cite{Cai:2014oca,Cai:2014dza,Cai:2015mja}.

\subsection{The holographic model for ferromagnetism/paramagnetism phase transition}
Before going on the topic of coexistence of ferromagnetism and superconductivity, let's first review how to build a holographic ferromagnetism which is independent on superconductivity in ref.~\cite{Cai:2014oca}. This model is realized by adding a real antisymmetric field into Einstein-Maxwell theory in a (3+1) dimensional  AdS spacetime,
\begin{equation}\label{LM2}
S=\frac1{2\kappa^2}\int d^4x\sqrt{-g}(\mathcal{R}+\frac6{L^2}-F^{\mu\nu}F_{\mu\nu}+\lambda^2 L_M)£¬
\end{equation}
where
\begin{equation}\label{LM3}
L_M=-\frac14\nabla^\mu M^{\nu\tau}\nabla_\mu M_{\nu\tau}-\frac{m^2}4M^{\mu\nu}M_{\mu\nu}-\frac12M^{\mu\nu}F_{\mu\nu}-\frac J8 V(M_{\mu\nu}).
\end{equation}
Here $2\kappa^2=16\pi G$ and $G$ is the Newtonian gravitational constant, $\lambda$ and $J$~are two constants, $m$ is the mass of the real tensor field $M_{\mu\nu}$, $A_\mu$ is the gauge potential of U(1) gauge field. The antisymmetric tensor field $M_{\mu\nu}$~is the effective polarization tensor of the U(1) gauge field strength~$F_{\mu\nu}$ with the self-interaction $V(M_{\mu\nu})$ which should be expanded as the even power of $M_{\mu\nu}$. The probe limit corresponds to $\lambda\rightarrow0$. Under this limit, the equation for polarization field decouples from the gauge field and gravity field,
\begin{equation}\label{eomFM}
  \nabla^2M_{\mu\nu}-m^2M_{\mu\nu}-J{M_\mu}^\delta {M_\delta}^\tau {M_\tau}_\nu-F_{\mu\nu}=0,
\end{equation}
with the dyonic Reissner-Nordstr\"om (RN) background~\cite{Cai},
\begin{eqnarray}\label{geom}
  ds^2 &=& r^2(-f(r)dt^2+dx^2+dy^2)+\frac{dr^2}{r^2f(r)},\nonumber \\
   f(r) &=&1-\frac{1+\mu^2+B^2}{r^3}+\frac{\mu^2+B^2}{r^4},\\
   A_\mu &=& \mu(1-1/r)dt+Bxdy. \nonumber
\end{eqnarray}
Here the horizon radius has been scaled to be unitary. If we only care about the magnetic part of polarization field, then a self-consistent ansatz for polarization field is $M_{\mu\nu}=-p(r)dt\wedge dr+\rho(r)dx\wedge dy$. Taking this ansatz into equation~\eqref{eomFM}, we have
\begin{equation}\label{rhop}
\begin{split}
\rho''+\frac{f'\rho'}f-\left( \frac{2f'}{rf}+\frac4{r^2}+\frac{m^2}{r^2f}\right)\rho+\frac{J\rho^3}{r^6f}-\frac{B}{r^2f}=0,\\
p''+\left(\frac{f'}{f}+\frac4r\right)p'-\left(\frac2{r^2}+\frac{m^2}{r^2f}\right)p-\frac{Jp^3}{r^2f}-\frac{\mu}{r^4f}=0,
\end{split}
\end{equation}
where a prime denotes the derivative with respect to $r$. It is interesting to see that these two equations decouple from each other in this case, which makes it to be possible that we can neglect the dynamic of $p(r)$ if we only care about the dynamic of magnetism. At the horizon, we need to impose a regular boundary condition. Near AdS boundary, the linearized equations have following asymptotic solutions for $\rho(r)$,
\begin{equation}\label{asym2}
\rho\sim \rho_+r^{(1+\delta)/2}+\rho_-r^{(1-\delta)/2}-\frac{B}{4+m^2},\\
\end{equation}
with $\delta=\sqrt{17+4m^2}$. In order to make the theory self-consistent and spontaneous condensation appear, we need following restriction on parameters,
\begin{equation}\label{m2}
-4<m^2<-\frac32, ~~~ \text{and}~~\rho_+=0.
\end{equation}
According to the action~\eqref{LM2} one can derive the magnetic moment from polarization field, which reads
\begin{equation}\label{QN1}
N=-\frac{1}{2}\int_1^\infty dr\frac{\rho}{r^2}.
\end{equation}
Here we have set the constant $\lambda=1$ in this expression for convenience. This integration converges only when $\rho_+=0$. In the case without external magnetic field, i.e. $B=0$, if there is a solution such that $\rho(r)\neq0$, the magnetic moment then is nonzero, which gives a ferromagnetic phase for dual boundary. Because the action~\eqref{LM2} implies transformation for $\rho(r)$ such as $\rho(r)\rightarrow-\rho(r)$ under the time reversal transformation, the condensed phase of $\rho$ gives a time reversal symmetry broken spontaneously, which is necessary for magnetic ordered phase.

\subsection {Ferromagnetism and p-wave superconductivity}
Once the two independent models for ferromagnetic phase transition and p-wave superconductor are in hand, we can combine them to discuss the possibility of coexistence. For example, we can combine the Einstein-Maxwell-complex vector theory for p-wave superconductor with ferromagnetic model. The complete action reads~\cite{Cai:2014dza}
\begin{equation}\label{action1}
S=\int d^4x\sqrt{-g}\left[\mathcal{R}+\frac{6}{L^2}-F_{\mu\nu} F^{\mu \nu}+\lambda^2(\mathcal{L}_{\rho}+\mathcal{L}_{M}+\mathcal{L}_{\rho M})\right],
\end{equation}
with
\begin{equation}\label{LrhoM}
\begin{split}
&\mathcal{L}_\rho=-\frac{1}{2}\rho_{\mu\nu}^\dagger\rho^{\mu\nu}-m_1^2\rho_\mu^\dagger\rho^\mu+iq\gamma \rho_\mu\rho_\nu^\dagger F^{\mu\nu}-V_\rho,\\
&\mathcal{L}_M=-\frac14\nabla^\mu M^{\nu\tau}\nabla_\mu M_{\nu\tau}-\frac{m_2^2}4M^{\mu\nu}M_{\mu\nu}-\frac{1}2M^{\mu\nu}F_{\mu\nu}-V_M,\\
&\mathcal{L}_{\rho M}=-i\alpha\rho_\mu\rho_\nu^\dagger M^{\mu\nu},\\
&V_\rho=-\frac{\Theta}2 \rho_{[\mu}\rho_{\nu]}^\dagger\rho^\mu\rho^{\dagger\nu}.
\end{split}
\end{equation}
Here $\alpha\neq0$ and $\Theta$ are two coupling constant. $L_M$ is the Lagrangian for polarization field which is just as the same as \eqref{LM3}. $L_\rho$ is the Lagrangian for complex vector field, which is similar to the one we discussed before. However, there is an additional term $V_\rho$ which describes the magnetic moment  interaction of complex vector field. This term is irrelevant for the previous section where we only care about superconductivity but is relevant when we care about spontaneous magnetization.

Under the probe limit $\lambda\rightarrow0$, a self-consistent ansatz of action~\eqref{action1} is,
\begin{equation}\label{VMansatz}
M_{\mu\nu}=-p(r)dt\wedge dr+h(r)dx\wedge dy,~~\rho_\mu=\rho_xdx+i\rho_ydy.
\end{equation}
Then we can get the equations of motion for complex vector field and polarization field under the background~\eqref{geom},
\begin{equation}\label{eqcomp1}
\begin{split}
h''+\frac{f'}fh'+\left(\frac{Jh^2}{r^6f}-\frac{2f'}{rf}-\frac4{r^2}-\frac{m_2^2}{fr^2}\right)h-\frac{2c\alpha \rho_x^2}{r^2f}=0,\\
\rho_x''+(\frac{f'}f+\frac2r)\rho_x'+\left(\frac{q^2\phi^2}{r^4f^2}-\frac{\Theta c^2\rho_x^2}{r^4f}-\frac{m_1^2}{fr^2}-\frac{ch\alpha}{fr^4}\right)\rho_x=0,\\
c''+\left(\frac{f'}f+\frac2r+\frac{2\rho_x'}{\rho_x}\right)c'-\frac{(1-c^2)(c\Theta\rho_x^2+\alpha h)}{fr^4}=0,\\
\end{split}
\end{equation}
where we have defined $c(r)$ as $\rho_y(r)=c(r)\rho_x(r)$. Note that the equation for $p(r)$ decouples from the others. The linearized equations near the AdS boundary give following asymptotic solutions~\footnote{The asymptotic solution of $c(r)$ depends on the source free condition of $\rho_x$. When $\rho_{x+}\neq0$, asymptotic solution of $c(r)$ becomes $c=c_++c_-r^{-\delta_1}$.}
\begin{equation}\label{adsbound}
\begin{split}
\rho_x={\rho_x}_+r^{(\delta_1-1)/2}+{\rho_x}_-r^{-(\delta_1+1)/2},\quad c=c_+r^{\delta_1}+c_-,\\
h(r)=h_+r^{(1+\delta_2)/2}+h_-r^{(1-\delta_2)/2},
\end{split}
\end{equation}
where $\delta_1=\sqrt{1+4m_1^2}$ and $\delta_2=\sqrt{17+4m_2^2}$ with $m_1^2>-1/4,~m_2^2>-4$. As the previous subsection, we should impose the condition $h_+=0$ for the polarization field and ${\rho_x}_+=c_+=0$ for the complex vector field, i.e., we require that the condensation and magnetization would happen spontaneously. The equations have solutions only when $c(r)=0,\pm1$. Because of the equivalent of $\alpha\rightarrow-\alpha$ and $c\rightarrow-c$, we assume $\alpha>0$ without loss of generality. The magnetic moment is defined as the same as~\eqref{QN1},
\begin{equation}\label{TolM}
N=-\int_{r_h}^\infty\frac h{2r^2}dr.
\end{equation}
According to  the dictionary of AdS/CFT, the expectation value of p-wave superconducting order parameter is a complex vector $\overrightarrow{P}$, whose mode is $P=\sqrt{1+c^2}|\rho_{x-}|$. Though the expression of magnetic moment density does't contain the terms of complex vector field, it is effected by $\rho_\mu$ through the mixture terms in equations~\eqref{eqcomp1}.

In the pure p-wave model, the global U(1) and spatial rotation symmetries are broken spontaneously when $\rho_x$ or $\rho_y$ is nonzero without source. Here it is also true. Moreover, there is an additional symmetry breaking. If one notes following rules for time reversal transformation,
\begin{equation}\label{timerev}
h\rightarrow-h,~~~\rho_y\rightarrow-\rho_y,
\end{equation}
then when $h\neq0$ or $\rho_y=\pm\rho_x\neq0$ (they both lead nonzero magnetic moment), the time reversal symmetry is broken spontaneously, which agrees with the fact that a spontaneously magnetized phase is with a time reversal symmetry broken spontaneously.

Because the complex vector field and polarization field can condense in low temperatures in an AdS RN black hole background respectively, this model gives a wide possibility to investigate the influence between p-wave superconductivity and spontaneous magnetization. We take $T_{sc0}$ and $T_{C0}$ as the critical temperatures of $\rho_x$ and $h$, when $\alpha=0$. Depending on the values of them, the p-wave superconudcting order or ferromagnetism will appear first. The interesting question is whether the other phase transition can still happen.

\subsection{Coexistence of superconductivity and ferromagnetism}

The first case we will consider is $T_{C0}>T_{sc0}$, i.e., the ferromagnetic phase appears first. The equation for $c$ in equations~\eqref{eqcomp1} shows that $c\neq0$ if $h\neq0$. So there isn't a phase such that $\{h<0,~\rho_x\neq0,~\rho_y=0\}$. When temperature is decreased to lower than $T_{C0}$, five kinds of phases may appear. They are phase A $\{h=\rho_x=\rho_y=0\}$, phase B $\{\rho_x=\rho_y=0,h<0\}$, phase C $\{\rho_x\neq0, h=\rho_y=0\}$, phase $D_1$ $\{\rho_x=\rho_y\neq0, h<0\}$ and phase $D_2$ $\{\rho_x=-\rho_y\neq0, h<0\}$, corresponding to normal phase, pure ferromagnetic phase, pure p-wave superconducting phase and two kinds of superconducting ferromagnetic phases, respectively.

In this case, whether the p-wave superconductivity can appear depends on the sign of interaction of magnetic moment of the complex vector field, i.e., the sign of $\Theta$. The possible phases and the physical favored phase in different temperature regions are summarized in table~\ref{Tab1}. In the case of $\Theta>0$, there is a critical temperature $T_{sc}$ between $T_{C0}$ and $T_{sc0}$, lower than which, the p-wave superconductivity can appear from ferromagnetic phase and the system will show ferromagnetism and superconductivity both. In addition, the critical temperature for superconductivity is increased rather than decreased by  spontaneous magnetization. This promotion is enhanced by increasing of interaction strength between complex vector field and antisymmetric tensor field. Numerical results imply that the magnetism and superconductivity can coexist even in the zero temperature limit. However if $\Theta<0$, the p-wave superconducting state can not  appear and the system will only be in a pure ferromagnetic state.
\begin{table}[h!]
\centering
\begin{tabular}{|c|c|c|c|}
  \multicolumn{4}{c}{Phases in the case of $T_{C0}>T_{sc0}$}\\
    \hline
    temperature & $T>T_{C0}$& $T_{sc}<T<T_{C0}$&$T<T_{sc}$\\
    \hline
    Possible& A & A,B & A, B, $D_1$, $D_2$, C(if $T<T_{sc0}$)\\
    \hline
    Physical($\Theta>0$) & A& B& $D_1$\\
    \hline
    Physical($\Theta<0$) & A&
    \multicolumn{2}{c|}{B}\\
    \hline
  \end{tabular}
  \caption{The possible and physical phases in the case of $T_{C0}>T_{sc0}$. Phase A is $\{h=\rho_x=\rho_y=0\}$. Phase B is $\{h<0,~\rho_x=\rho_y=0\}$. Phase C is $\{h=\rho_y=0,~\rho_x\neq0\}$. Phase $D_1$ is $\{h<0,~\rho_x=\rho_y\neq0\}$. Phase $D_2$ is $\{h<0,~\rho_x=-\rho_y\neq0\}$. Table taken from ref.~\cite{Cai:2014dza}.}\label{Tab1}
\end{table}

The other case is $T_{sc0}>T_{C0}$, i.e., the case where the p-wave superconducting phase appears first. When $T_{C0}<T<T_{sc0}$, the equations~\eqref{eqcomp1} show there may exist three kinds of p-wave superconducting phases. One is the usual p-wave superconducting phase C($\{h=\rho_y=0, \rho_x\neq0\}$), the other two are new superconducting phases denoted as $E_1$ with $\{h<0,\rho_x=\rho_y\neq0\}$ and $E_2$  with $\{h<0,\rho_x=-\rho_y\neq0\}$. The magnetization in two phases $E_1$ and $E_2$ is induced by the p-wave pair rather than been produced spontaneously, which is different from case in phases $D_1$ and $D_2$.

\begin{table}[h!]
  \centering
  \begin{tabular}{|c|c|c|c|}
  \multicolumn{4}{c}{Phases in the case of $T_{sc0}>T_{C0}$ and $\Theta>0$}\\
    \hline
    Temperature & $T>T_{sc0}$& $T_{C0}<T<T_{sc0}$&$T<T_{C0}$\\
    \hline
  Possible& A &  A, $E_1$, C &  A, $E_1$, C, B\\
    \hline
    Physical &A& \multicolumn{2}{c|}{$E_1$}\\
    \hline
    \multicolumn{4}{c}{ }\\
  \end{tabular}
  \\
\centering
\begin{tabular}{|c|c|c|c|}
  \multicolumn{4}{c}{Phases in the case of $T_{sc0}>T_{C0}$ and $\Theta<0$}\\
    \hline
    Temperature & $T>T_{sc0}$& $T_{C0}<T<T_{sc0}$&$T<T_{C0}$\\
    \hline
  Possible& A &  A, C &  A, C, B\\
    \hline
    Physical &A& C& B\\
    \hline
  \end{tabular}
  \caption{The possible and physical phases in the case of $T_{sc0}>T_{C0}$. Phase A is $\{h=\rho_x=\rho_y=0\}$. Phase B is $\{h<0,~\rho_x=\rho_y=0\}$. Phase C is $\{h=\rho_y=0,~\rho_x\neq0\}$. Phase $E_1$ is $\{h<0,~\rho_x=\rho_y\neq0\}$. Phases $E_2$ is $\{h<0,~\rho_x=-\rho_y\neq0\}$. The tables were taken from ref.~\cite{Cai:2014dza}.}\label{Tab2}
\end{table}

\begin{figure}[h!]
\includegraphics[width=0.5\textwidth]{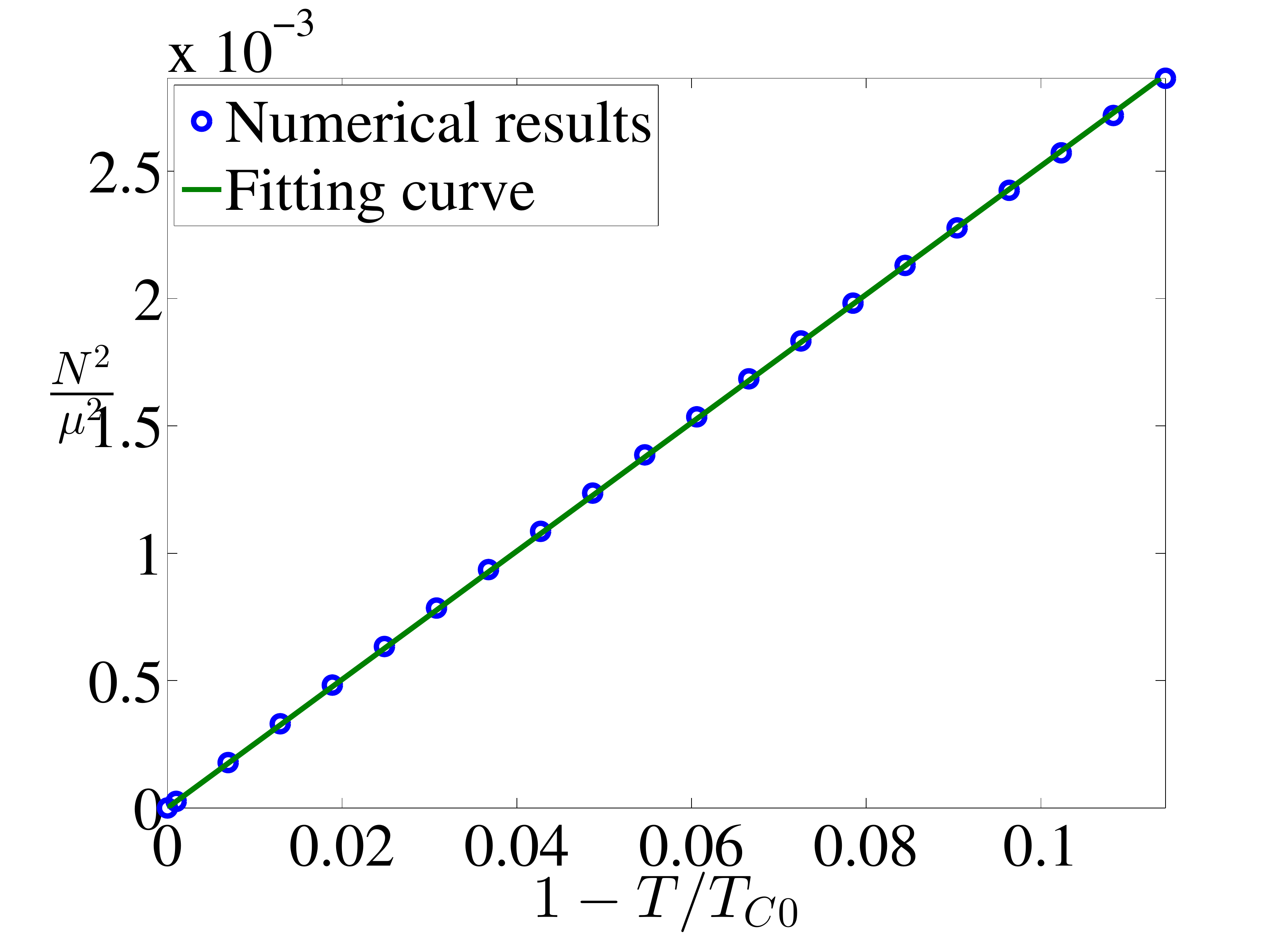}
\includegraphics[width=0.5\textwidth]{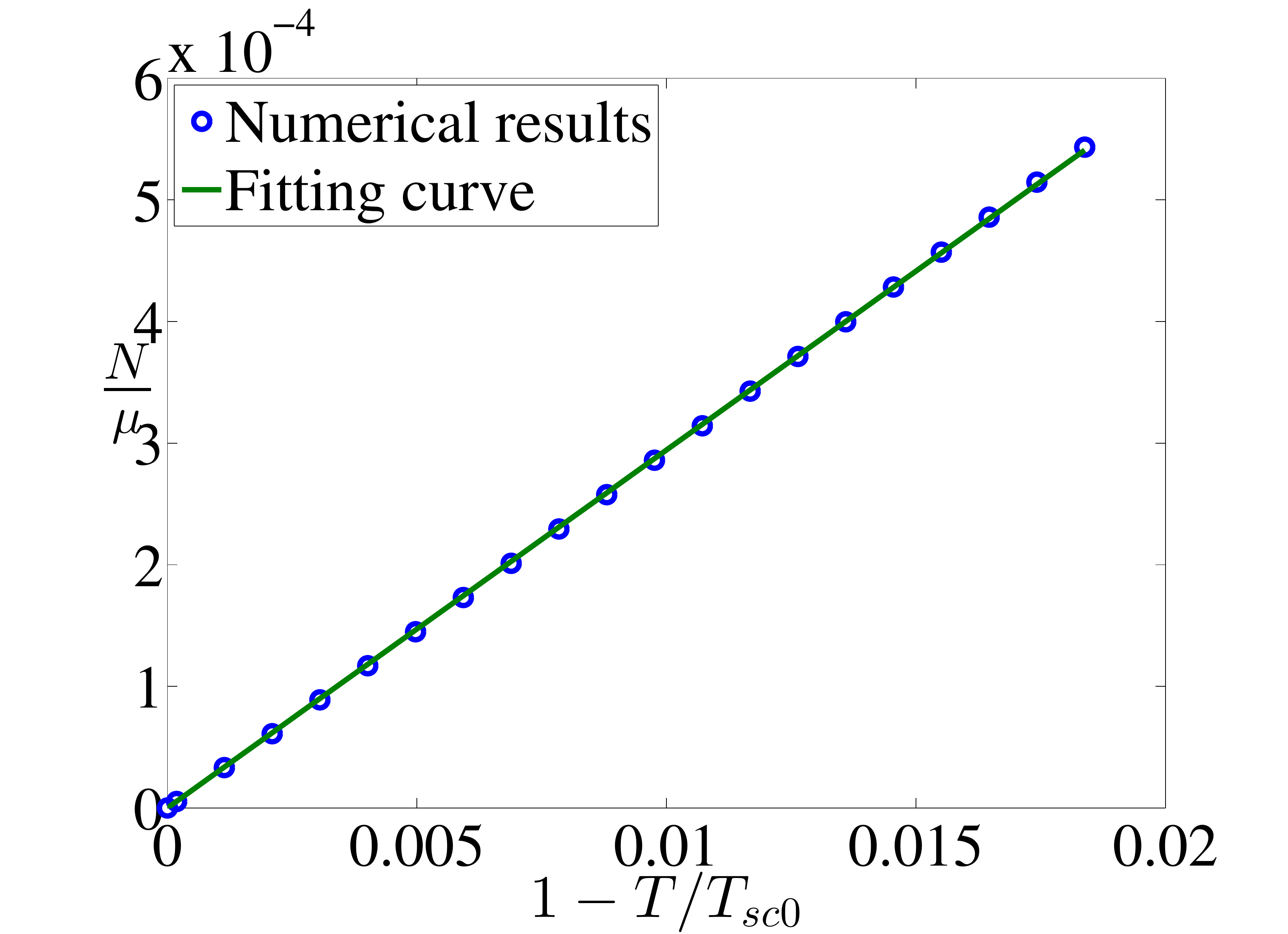}
\caption{The behaviours of $N$ near the critical temperature in the phases $D_1$(left) and $E_1$(right). Here $m_1^2=-3/16, m_2^2=-3,J=-1, \Theta=1$ and $\alpha=0.1$. In the left one, $q=1.4$. In the right one, $q=1.4$. The plots were taken from ref.~\cite{Cai:2014dza}.}
\label{TG2b}
\end{figure}

Numerical results show that situations also depend on the sign of $\Theta$. All the results are summarized  in  table~\ref{Tab2}. If $\Theta>0$,
with decreasing the temperature, the system will transit into phase $E_1$, where p-wave superconductivity with a kind of induced magnetism appears. The superconductivity and magnetism appear both, however, it should better be called a magnetic superconducting phase rather than a ferromagnetic superconducting phase, because the magnetic moment is not spontaneously produced and proportional to $T_{sc0}-T$ rather than $\sqrt{T_{sc0}-T}$ near the critical temperature (see figure~\ref{TG2b}). If $\Theta<0$, the system will be in the pure p-wave superconducting phase without magnetism if temperature is less than $T_{sc0}$. When temperature is lower than $T_{C0}$, the system will transit into the pure ferromagnetic phase from the p-wave superconducting phase. Therefore the ferromagnetism and superconductivity can not coexist if $\Theta<0$.

Tables~\ref{Tab1} and~\ref{Tab2} show that the ground state near zero temperature limit only depends on the sign of $\Theta$. These can be understood in a physical manner if we pay more attention to this phenomenological parameter in~\eqref{LrhoM}, where it was introduced to describe the self-interaction between the magnetic moments of complex vector field. The case of $\Theta>0$ means that the p-wave pair will attract the one with the same magnetic moment direction and repulse the one with the opposite magnetic moment. Under the influence of spontaneous magnetization, the magnetic moment of p-wave pair will tend to align along the direction of spontaneous magnetization. As a result, p-wave pair and spontaneous magnetization would be enhanced by each other and therefore survive. However, if $\Theta<0$, the p-wave pair will repulse the one with the same magnetic moment direction. So the p-wave pair will align without net magnetism and the system is in a pure p-wave superconducting phase in the region where superconductivity dominates. When $T<T_{C0}$, the ferromagnetism can appear, which tends to make p-wave pairs have same direction. But the p-wave pairs with same magnetic moment direction will repulse each other, which leads that the p-wave pair is not stable and will be de-paired. So the system can only be in the ferromagnetic phase.

\section{Conclusion and Discussion}
\label{sect:conclusion}
Due to the strong/weak duality characteristic  of the holographic correspondence, it provides us with a powerful tool to study the properties of strongly interacting systems by a weakly coupled gravity theory with one extra spatial dimension. Although the underlying dynamics which govern the dual field theory and the gravity are apparently different, as we have shown, in the framework of holography quantum computations in the dual (strongly coupled) field theory can be translated into classical calculations in the bulk, where one can just solve differential equations with suitable boundary conditions. Within this framework, holographic correspondence is considered as a hopeful approach to understand the properties of strongly correlated electron systems.

The bulk gravitational models that we have reviewed are some phenomenological models. In such bottom-up approach, the gravity duals were constructed using the minimal set of fields that captured the essential dynamics. They just involve gravity interacting with an effective U(1) gauge field and a charged field serving as the order parameter.  We have a lot of the degrees of freedom to choose the form of interactions as well as the value of couplings.\footnote{In principle, the arbitrary can be fixed by embedding the bulk model into some low energy effective theory of string/M theory.} Nevertheless, one has seen that those simple models would describe dual superconductors rather well. Some interesting features have been uncovered. Let us take holographic p-wave models as an example. For the SU(2) Yang-Mills model~\eqref{su2action}, the conductivities are strongly anisotropic in a manner which is suggestive of a gap with nodes. The low-lying excitations of the normal state have a relaxation time growing rapidly as the temperature is lowered, which agrees with the absence of impurity scattering. For the second model~\eqref{vecaction}, it has been found that the vector condensate can be induced by an applied magnetic field, and the condensation of the charged vector operator forms a vortex lattice structure in the spatial directions perpendicular to the magnetic field. Going beyond the probe approximation, the model displays a rich phase structure. In terms of temperature and chemical potential, the complete phase diagrams have been constructed for the conducting phase, insulating phase and their corresponding superconducting phases and some new phase boundaries are revealed. The Maxwell-vector model is a generalization of the SU(2) model in the sense that the vector field has a general mass and gyromagnetic ratio. The third model~\eqref{eomhelical} realizes a p-wave superconducting phase by involving a charged two-form in the bulk. The p-wave states exhibit a helical structure and some of them display the phenomenon of pitch inversion as the temperature is decreased. The ground state of the condensed phase has zero entropy density and exhibits an emergent scaling symmetry in the IR.

It is clear that the key ingredient in constructing a gravitational dual of a superconductor is to find an instability which breaks a U(1) symmetry, e.g., at low temperatures and causes a condensate to form spontaneously. One may ask whether those phenomenological bulk duals of superconductors are just a Ginzburg-Landau description. The answer is exactly no. Let us stress two key differences. First, the instability in the Ginzburg-Landau model must be put in by hand, while it arises naturally in holographic setup. Second, the Ginzburg-Landau model is only valid near the transition point, whereas the gravitational description can characterize the whole dynamics. For a given bulk action, scanning through values of model parameters corresponds to scanning through many different dual field theories. In that sense, a simple holographic model has a kind of universality, i.e., the results may be true for a large class of dual field theories, quite insensitive to the details of their dynamics. Another confusion is that we realized the spontaneous breaking of a continuous U(1) symmetry in $(2+1)$ dimensions at finite temperature, in apparent contradiction to the Coleman-Mermin-Wagner theorem. The cure is that the large $N$ limit evades the theorem as fluctuations are suppressed. It would be interesting to discuss the effect of bulk quantum corrections which correspond to $1/N$ corrections in the dual field theory~\cite{Anninos:2010sq}.
Finally, although the hair breaks a local U(1) symmetry in the bulk, according to the dictionary, the dual system consists of a condensate breaking a global U(1) symmetry. On the other hand, the onset of superconductivity is characterized by the condensation of a composite charged operator spontaneously breaking U(1) gauge symmetry. So strictly speaking, what one has realized is a dual theory of superfluid~\cite{Herzog:2008he,Brihaye:2011vk,Arean:2010wu,Wu:2014bba} rather than superconductor. However, in the limit that the U(1) symmetry is ``weakly gauged" one can still view the dual theory describing a superconductor.~\footnote{In fact, most of the condensed matter theories do not include dynamical photons, as their effects are usually small. For example, in the BCS theory electromagnetic field is often introduced as an external field. The possibility of introducing dynamical gauge fields in holographic superconductors was discussed in refs.~\cite{Domenech:2010nf,Gao:2012yw}.}

Throughout this brief summary we have been mainly concerned with static and homogeneous case and focused on some basic aspects.
This is a rapidly devolving field, due to the limitation of length, we are not able to give more details for many interesting developments, such as introduction of momentum dissipation (to break translational symmetry)~\cite{Flauger:2010tv,Horowitz:2013jaa,Dias:2013bwa,Zeng:2014uoa,Koga:2014hwa,Ling:2014laa,Arean:2013mta,Erdmenger:2015qqa,Kim:2015dna}, construction of holographic Josephson Junction~\cite{Horowitz:2011dz,Wang:2011rva,Siani:2011uj,Wang:2011ri,Wang:2012yj,Rozali:2013pla,Cai:2013sua,Takeuchi:2013kra,Kiritsis:2011zq,Li:2014xia}, and investigation on dynamics for far-from equilibrium state~\cite{Murata:2010dx,Bhaseen:2012gg,Sonner:2014tca,Bai:2014tla,Adams:2012pj,Gao:2012aw,Li:2013fhw,Garcia-Garcia:2013rha,Chesler:2014gya,Du:2014lwa}. The analysis for the most part has been done numerically. To explore the properties of holographic superconductors using analytical techniques can be found, for example, in refs.~\cite{Herzog:2010vz,Siopsis:2010uq,Zeng:2010zn,Cai:2011ky,Momeni:2011iw,Zeng:2012zza,Pan:2012jf,Gangopadhyay:2012gx,Huang:2013sca,Banerjee:2012vk,Momeni:2013bca,Lu:2013tza}. Optimistically the growing literature based on holographic duality might shed some light on the understanding of mysterious phenomena and eventually microscopic origins of strongly correlated superconductivity.

The applications of the holographic correspondence are still going on. It was written by G.~T.~Horowitz and J.~Polchinski~\cite{Horowitz:2006ct} that we find it difficult to believe that nature does not make use of it, but the precise way in which it does so remains to be discovered. In addition to holographic superconductors, the holographic approach has been used to understand some other aspects of condensed matter physics, including (non-)Fermi liquids~\cite{Lee:2008xf,Liu:2009dm,Cubrovic:2009ye,Faulkner:2009wj,Faulkner:2010tq}, quantum Hall effect~\cite{Davis:2008nv,Fujita:2009kw,Bergman:2010gm}, strange metals~\cite{Hartnoll:2009ns,Faulkner:2010da,Kim:2010zq,Davison:2013txa}, topological insulators~\cite{HoyosBadajoz:2010ac,Ryu:2010fe,Karch:2010mn}, Hubbard model~\cite{Fujita:2014mqa} and so on. A major application using holographic duality is to describe quantum chromodynamics (QCD), especially for the quark gluon plasma produced in particle accelerators. It is referred as AdS/QCD or holographic QCD, which has been widely studied~\cite{Polchinski:2001tt,BoschiFilho:2002ta,Erlich:2005qh,deTeramond:2005su,Babington:2003vm,Kruczenski:2003uq,Gubser:2006bz,Shuryak:2005ia,Gursoy:2008bu,Herzog:2006ra,Colangelo:2010pe,Chen:2009kx,Huang:2007fv,vanderSchee:2014qwa}. Another emerging subject is the fluid/gravity correspondence, which translates problems in fluid dynamics into problems in general relativity~\cite{Bhattacharyya:2008jc,Rangamani:2009xk,Bredberg:2011jq}. Readers who are interested in those exciting achievements are encouraged to consult those relevant references.

\section*{Acknowledgements}
This work was supported in part by the National Natural Science Foundation of China ( No.11035008, No.11375247, No.11205226 and No.11435006 ).
L Li was supported in part by European Union's Seventh Framework Programme under grant agreements (FP7-REGPOT-2012-2013-1) no 316165, the EU-Greece program ``Thales" MIS 375734 and was also co-financed by the European Union (European Social Fund, ESF) and Greek national funds through the Operational Program ``Education and Lifelong Learning" of the National Strategic Reference Framework (NSRF) under ``Funding of proposals that have received a positive evaluation in the 3rd and 4th Call of ERC Grant Schemes".
\appendix

\end{document}